\documentclass[aps,prc,preprint,tightenlines,superscriptaddress,showpacs,%
amssymb,byrevtex,nofootinbib,preprintnumbers]{revtex4-1}

\usepackage[utf8]{inputenc}
\usepackage[colorlinks=true,linkcolor=blue,urlcolor=blue,citecolor=blue,anchorcolor=blue]{hyperref}

\usepackage{graphicx,color}
\usepackage{amsmath}
\usepackage{mathrsfs}
\usepackage{dcolumn}
\usepackage{epsfig,epstopdf}
\usepackage{bm}

\usepackage{enumitem}

\usepackage[capitalise]{cleveref}
\crefrangelabelformat{equation}{(#3#1#4--#5#2#6)}

\newcommand{\Tr}{\mathrm{Tr}}

\renewcommand{\vec}{\mathbf}

\newcommand{\UCAS}{School of Physical Sciences, University of Chinese Academy of \\ Sciences (UCAS), Beijing 100049, China}
\newcommand{\CSSM}{Special Research Centre for the Subatomic Structure
  of Matter (CSSM),\\Department of Physics, University of
  Adelaide, Adelaide, South Australia 5005, Australia}
\newcommand{\DESY}{Deutsches Elektronen-Synchrotron DESY, 22603
  Hamburg, Germany}

\begin{document}
\title{Towards high partial waves in lattice QCD with a dumbbell-like operator}

\author{Jia-Jun Wu}
\affiliation{\UCAS}
\author{Waseem Kamleh}
\affiliation{\CSSM}
\author{Derek B. Leinweber}
\affiliation{\CSSM}
\author{Yan Li}
\affiliation{\UCAS}
\author{Gerrit Schierholz}
\affiliation{\DESY}
\author{Ross D. Young}
\affiliation{\CSSM}
\author{James M. Zanotti}
\affiliation{\CSSM}
\collaboration{CSSM-QCDSF Collaboration}

\begin{abstract}

An extended two-hadron operator is developed to extract the spectra
of irreducible representations (irreps) in the finite volume. 
The irreps of the group for the finite volume system are projected
using a coordinate-space operator. 
The correlation function of this operator is computationally efficient
to extract lattice spectra of the specific irrep. 
In particular, this new formulation only requires propagators to be
computed from two distinct source locations, at fixed spatial
separation.
We perform a proof-of-principle study on a $24^3 \times 48$ lattice
volume with $m_\pi\approx 900$~MeV by isolating various spectra of 
the $\pi\pi$ system with isospin-2 including a range of total momenta
and irreps. 
By applying the L\"uscher formalism, the phase shifts of $S$-, $D$- and $G$-wave $\pi\pi$ scattering with isospin-2 are extracted from the spectra.
\end{abstract}

\preprint{ADP-21-21/T1159}

\maketitle

\section{Introduction}
\label{intro}

The numerical simulation of quark and gluon fields on a finite
lattice enables a study of the hadron spectrum and strong interactions
of QCD via first principles.
Recently, there has been tremendous progress in lattice QCD
calculations of the hadron spectrum and interactions (see Refs.~\cite{Liu:2016kbb,
  Briceno:2017max, Padmanath:2019wid} for recent reviews).
From the energy levels of lattice QCD, there is a clear strategy for
how to extract scattering information for two-body systems, such as the
$\pi\pi$ system \cite{Dudek:2012xn,Dudek:2010ew, Dudek:2012gj,
  Beane:2011sc,Bulava:2016mks}.
In order to map out the energy-dependence of the scattering phase shifts
various methods have been developed to access more finite-volume energy levels,
such as the variational analysis for the excited energy
eigenvalues~\cite{Michael:1985ne,Luscher:1990ck,Blossier:2009kd,Mahbub:2009nr}, moving systems~\cite{Dudek:2012xn,
  Dudek:2012gj}, and twisted boundary conditions~\cite{Hudspith:2020tdf,Chen:2014afa}.
An important requirement to isolate distinct partial waves is the need
to distinguish the energy levels in different irreducible
representations (irreps), such as done in Ref.~\cite{Dudek:2012gj}. 

With improved control of orbital motion, there is potential for
lattice QCD to provide insight into the phenomenology of high-spin
systems. 
For instance, the relatively narrow dibaryon resonance observed by
WASA-at-COSY \cite{Adlarson:2011bh,Adlarson:2014pxj} suggests
significant coupling to the $np$ $G$-wave amplitude.
There is also potential to shed light on the nuclear $A(y)$ puzzle
\cite{Witala:1991emr,Tornow:1998zz,Huber:1998hu} or the dynamics
underlying Regge trajectories
\cite{Anisovich:2000kxa,Ebert:2009ub,Meyer:2004jc}.
While each of these objectives will (ultimately) also require advances in many-body systems on
the lattice \cite{Hansen:2019nir,Mai:2021lwb}, the physics of many-body channels
can often be suppressed at large unphysical quark masses on the
lattice---such as the recent high-$J$ study of Ref.~\cite{Johnson:2020ilc}.

The study of high angular momentum systems is an ongoing challenge in
lattice QCD.
On the cubic finite volume of a 4-dimensional lattice, the relevant
symmetry group is a subgroup of the octahedral group ($O_h$)---or the
relevant little group when considering systems at finite momenta.
Importantly, the full SO(3) group of the infinite volume physical theory is broken,
and consequently, numerical investigations are limited to the discrete symmetry of the lattice.
The issue of partial-wave mixing, and influence on discrete spectra, has been investigated theoretically and numerically in previous work,
e.g.~Refs.~\cite{Luscher:1990ux, Luscher:1986pf,Rummukainen:1995vs,Kim:2005gf,Fu:2011xz,
  Gockeler:2012yj, Guo:2012hv, Luu:2011ep, Leskovec:2012gb, Briceno:2012yi,Briceno:2014oea,Berkowitz:2015eaa,Wu:2015evh,Li:2019qvh,Li:2021mob,Amarasinghe:2021lqa}.
In this paper, we introduce a novel operator construction, designed to
provide an efficient method to isolate different lattice irreps in 
a two-hadron system.
The method relies upon constructing an operator that corresponds to a
``dumbbell'' in coordinate space, where the two body operator
is the product of two single particle operators separated by a fixed
distance.
The construction shares similarities with the ``cube''
source employed in Ref.~\cite{Berkowitz:2015eaa}.
In our case, as will be shown, we sum over the rotations of the
dumbbell at the sink in order to project correlation functions onto
the desired irrep.
The two distinguishing features of this method are that only the total
momentum of the two-hadron system is fixed and only two point-source
Dirac matrix inversions are required.
As an exploratory exercise, we study the isospin-2 $\pi\pi$ system,
for which several alternative methods have already been explored
\cite{Dudek:2010ew, Dudek:2012gj, Beane:2011sc}.
We demonstrate that we are able to successfully determine energy
levels of the various representations with different total momenta.
In Section \ref{sec:II}, the two-hadron operator and correlation
functions for specific irreps are constructed.
In Section \ref{sec:III}, a lattice-QCD calculation for isospin-2 $\pi^-\pi^-$
scattering is presented with lattice size $24^3\times48$ and the
energy levels for various irreps with different total momentum 
are extracted. These energy levels are then used to a determine
the phase shifts of $\pi^-\pi^-$ scattering.
Finally, results are summarised in Section \ref{sec:sum}.

\section{Formalism}
\label{sec:II}

\subsection{Operators in coordinate space}

\begin{figure}[tbp]
\begin{center}
\includegraphics[width=0.7\columnwidth]{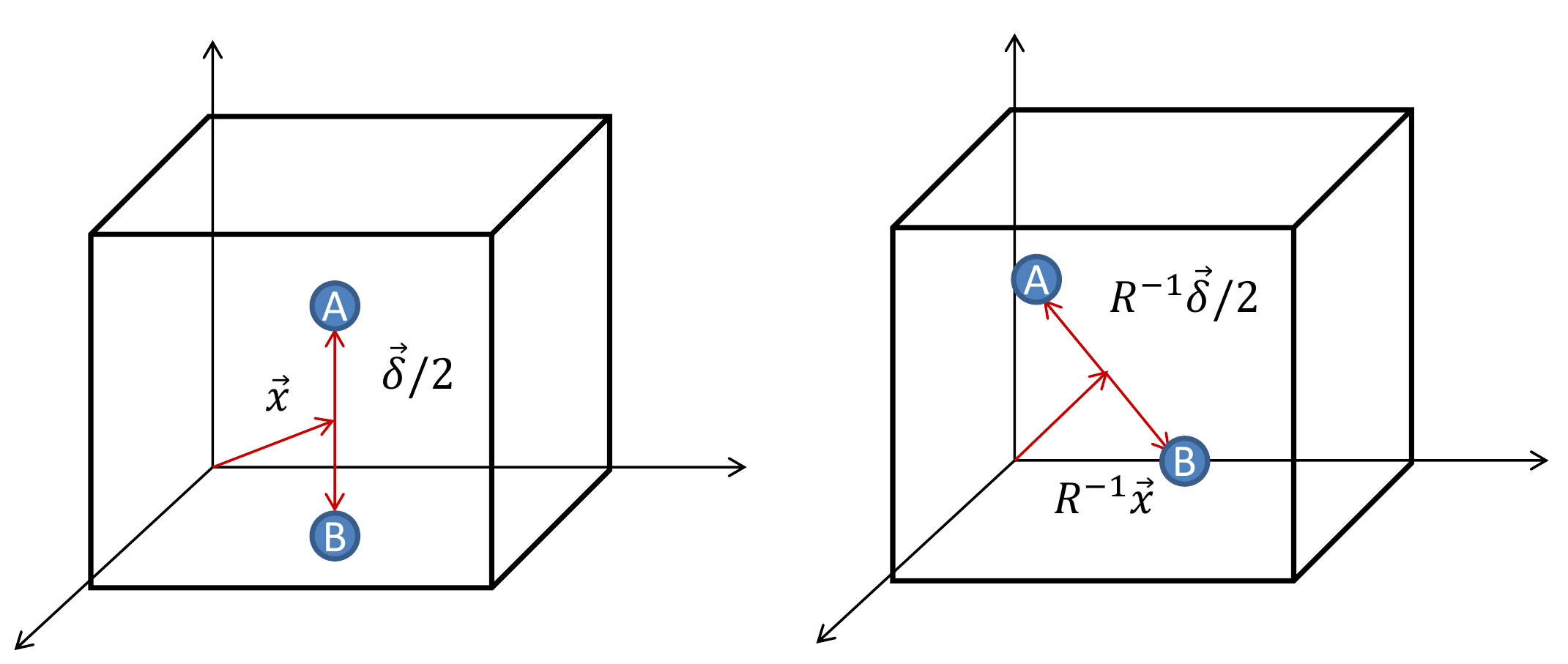}
\caption{Illustration of the 2-hadron ``dumbbell'' interpolating operator. A fixed
  source location is indicated in the left image, and the sink on the
  right indicates that the separation of the pointwise interpolating
  fields are summed over all lattice rotations, $R$.
\label{fig:dumbbell}}
\end{center}
\end{figure}

Our goal is to construct extended interpolating operators which
project onto states of both definite momenta and irreps
of the lattice rotation group. To minimise the
numerical cost associated with inversion of Dirac matrices, we seek a
construction which allows our correlation functions to be constructed
from just two conventional local sources. The projection onto definite
Fourier momenta and rotational irreps are to be performed at the sink,
as depicted in Fig.~\ref{fig:dumbbell}.
To construct the appropriate projections we start from
a composite operator of two hadrons with a
separation $\boldsymbol{\delta}$ between them:
\begin{eqnarray}
\Phi(\vec{x},\boldsymbol{\delta})
\equiv
\phi(\vec{x}+\boldsymbol{\delta}/2) \phi'(\vec{x}-\boldsymbol{\delta}/2)\,,
\end{eqnarray}
where time-dependence has been suppressed and the operator $\phi$ (or $\phi'$)
denotes a conventional, local single-hadron operator.
For example, in the following calculation, we consider the standard $\pi^-$ operator given by
\begin{eqnarray}
\phi(\vec{x})=\phi'(\vec{x})\equiv\sum_a  \bar{u}^{a}(\vec{x})\gamma_5 d^a(\vec{x})\,,\label{eq:pionoperator1}
\end{eqnarray}
with a sum over the color index $a$.

We consider the set of operators, $\{\Phi_{\hat{R}}\}$, which 
are related by a lattice rotation, $\hat{R} \in O_h$.
Under such a rotation, the transformed operators take the form:
\begin{eqnarray}
\Phi_{\hat{R}}(\vec{x},\boldsymbol{\delta})
&\equiv&\hat{P}_{\hat{R}}\,\Phi(\vec{x},\boldsymbol{\delta})\hat{P}_{\hat{R}^{-1}}
=\Phi(\hat{R}^{-1}\vec{x},\hat{R}^{-1}\boldsymbol{\delta})
\nonumber\\
&=&\phi(\hat{R}^{-1}(\vec{x}+\boldsymbol{\delta}/2)) \phi'(\hat{R}^{-1}(\vec{x}-\boldsymbol{\delta}/2)),  \label{eq:prpsipr}
\end{eqnarray}
as being represented in the right panel of Fig.~\ref{fig:dumbbell}.
To maximally span the space of lattice irreps, we choose to work with separation vectors satisfying
$0<\delta_x<\delta_y<\delta_z$ such that
$\hat{R}\,\boldsymbol{\delta} \neq \boldsymbol{\delta}$ (for
$\hat{R}\ne I$).
We then have 24 different operators that are related by a lattice
rotation --- in the case of non-identical particles, there are 48 operators.
While the single-hadron operators must lie on lattice sites, the
center of the composite operator, $\vec{x}$, need not be on a lattice site.
In the numerical results presented here, we work with the choice $\boldsymbol{\delta} = (1,\,3,\,5)$ and
$\vec{x}=(1/2,\,1/2,\,1/2)$.
We note that choosing all even values for $\boldsymbol{\delta}$ would place the origin of
the extended operator on a lattice site and maintain the same discrete
rotational symmetries.
In principle, combinations of even and odd displacements by
$\boldsymbol{\delta}$ would be possible, but it would lead to a (short
distance) modification of the rotational symmetries discussed here.

A Fourier transform with respect to the coordinate $\vec{x}$ project
onto states of definite momenta:
\begin{equation}
\langle \Phi(\vec{P},\vec\delta)|  = \langle\Omega|  \sum_{\vec{x}}e^{-i\vec{P}\cdot\vec{x}}\Phi(\vec{x},\boldsymbol{\delta}).
\end{equation}
In just the same way that the Fourier transform projects onto states
of definite momenta, particular linear combinations of operators
related by lattice rotations, Eq.~(\ref{eq:prpsipr}), will project
onto particular irreducible representations.
In particular, for $(|\vec{p}|L/2\pi)^2=0$, $1$, $2$, and $3$, operators
are constructed to project onto the irreps of the groups commonly
denoted $O_h$, $C_{4\nu}$, $C_{2\nu}$ and $C_{3\nu}$, respectively
(see for instance, Ref.~\cite{weissbluth2012atoms}).
The projection onto these irreps has been discussed in a number of
previous works
\cite{Luu:2011ep,Gockeler:2012yj,Leskovec:2012gb,Doring:2012eu,Dudek:2012xn}.
For completeness and to set our notation, we briefly summarise the
relevant features here.

The states $|\Phi_{\hat{R}}(t;\vec{x},\boldsymbol{\delta}) \rangle$ where
$\hat{R}$ belong to the corresponding group will transform as vectors
of the regular representation as follows:
\begin{eqnarray}
  \hat{P}_{\hat{R}}\,|\Phi_{\hat{R}'}(\vec{x},\boldsymbol{\delta}) \rangle
  &=&
  \sum_{\hat{R}'' \in G^{\vec{p}}}
  |\Phi_{\hat{R}''}(\vec{x},\boldsymbol{\delta}) \rangle
  \left(\bar{B}(\hat{R})\,\right)_{\hat{R}'', \hat{R}'}, \label{eq:PrB}
\end{eqnarray}
where $ \left(\bar{B}(\hat{R})\,\right)_{\hat{R}'', \hat{R}'} =
\delta_{\hat{R}\hat{R}',\, \hat{R}''}$, and $G^{\vec{p}}$ denotes the rotation group for specific total
momentum $\vec{p}$ (see \cref{tab:momentum}).
The regular representation of any non-trivial group is reducible. 
Thus, $\bar{B}$ can be made block diagonal according to the irreps
of the symmetry group via a unitary transformation matrix
$\bar{S}$. For example, in the $O_h$ group,
\begin{eqnarray}
\bar{S}^{-1} \bar{B}(\hat{R})\,\bar{S} =1 \bar{A}^\pm_1(\hat{R}) \oplus
1 \bar{A}^\pm_2(\hat{R}) \oplus 2 \bar{E}^\pm(\hat{R})  \oplus 3
\bar{T}^\pm_1(\hat{R})  \oplus 3 \bar{T}^\pm_2(\hat{R})  \equiv
\bar{A}(\hat{R})\,,
\end{eqnarray}
where $\bar{\Gamma}(\hat{R})$ denotes the representation matrix of
$\hat{R}$ in the irrep $\Gamma$, and the number before $\bar{\Gamma}$
indicates the number of occurrences of $\Gamma$ in the regular
representation.
Using $(i, \Gamma, n)$ to label the $n$-th vector of the $i$-th
occurrence of an irrep $\Gamma$, the matrix $\bar{A}$ takes the block
diagonal form:
\begin{eqnarray}
  \bar{A}_{i\Gamma\,n, i'\Gamma'\,n'}(\hat{R}) =
  \delta_{i'i}\,\delta_{\Gamma'\Gamma}\,\bar{\Gamma}_{n,n'}(\hat{R})\,.
  \label{eq:A1}
\end{eqnarray}

With the unitary transformation matrix $\bar{S}$, one can construct
the states $|\Psi^\dag_{i,\Gamma,n}\rangle$ as
\begin{align}
  |\Phi^\dag_{i, \Gamma, n}\rangle =
  \sum_{\hat{R}} |\Phi^\dag_{\hat{R}}\rangle \bar{S}_{R, i\Gamma n}\,,
  \label{eq:newPhi}
\end{align}
which will satisfy
\begin{align}
  \hat{P}_{R}|\Phi^\dag_{i,\Gamma,n}\rangle =
  \sum_{i',\Gamma', n'} |\Phi^\dag_{i',\Gamma',n'}\rangle
  \left(\bar{A}(\hat{R})\,\right)_{i'\Gamma'\,n', i\,\Gamma\,n}\,.
  \label{eq:PrA}
\end{align}
Correspondingly, a new type of two-hadron operator can be defined as in Eq.(\ref{eq:newPhi}) as
\begin{eqnarray}
  \Phi^\dag_{i, \Gamma, n} =
  \sum_{\hat{R}} \Phi^\dag_{\hat{R}} \bar{S}_{R, i\Gamma n}\,.
  \label{eq:newPhioper}
\end{eqnarray}


\subsection{Correlation function}

The elementary two-point correlation function is constructed from
$\Phi_{\hat{R}}(t;\vec{x},\boldsymbol{\delta})$ at the source and sink as
follows:
\begin{align}
G_{\hat{R},\hat{R}'}(t;\vec{p};\vec{x},\boldsymbol{\delta})
=
\sum_{(\vec{y}-\vec{x})\in \mathbb{Z}^3} 
e^{-i\vec{p}\cdot(\vec{y}-\vec{x})}
\;\left\langle \mathrm{T}\left( \Phi_{\hat{R}}(t;\vec{y},\boldsymbol{\delta}),\, \Phi^{\dag}_{\hat{R}'}(0;\vec{x},\boldsymbol{\delta})  \right) \right\rangle
\label{eq:psical2}
\nonumber
\end{align}
where the angle brackets denote the ensemble average across gauge ensembles
and $\mathrm{T}$ the time-ordered product of field operators.
While this generally involves the full set of rotations at source and
sink, we can exploit the translational and rotational symmetry of this
correlator to obtain:
\begin{equation}
  G_{\hat{R},\hat{R}'}(t;\vec{p};\vec{x},\boldsymbol{\delta}) =
  G_{\hat{R}\hat{R}'^{-1},\hat{I}}(t;\vec{p};\vec{x},\boldsymbol{\delta})\quad
  \forall \hat{R},\hat{R}'\in G^{\vec{p}}.
\end{equation}

The projection of the correlation function onto definite irreps of the
lattice rotation group is then given by
\begin{align}
  \tilde{G}_{\Gamma}(t;\vec{p};\vec{x},\boldsymbol{\delta})
&=
\sum_{(\vec{y}-\vec{x})\in \mathbb{Z}^3} 
e^{-i\vec{p}\cdot(\vec{y}-\vec{x})}
\;\langle \mathrm{T}\left( \Phi_{i,\Gamma,n}(t;\vec{y},\boldsymbol{\delta}),\, \Phi^{\dag}_{{i,\Gamma,n}}(0;\vec{x},\boldsymbol{\delta})  \right) \rangle
  \label{eq:Psicorle1}\\
  &=
  \frac{1}{l_{\Gamma}}\sum_{\hat{R}}\chi^{\Gamma}_{\hat{R}}
  G_{\hat{R},\hat{I}}(t;\vec{p};\vec{x},\boldsymbol{\delta})\,,\label{eq:Psicorle2}
\end{align}
where $\chi^{\Gamma}_{\hat{R}}$ is the character number of element
$\hat{R}$ of the group in the irrep $\Gamma$, and $l_\Gamma$ is the
dimension of the irrep $\Gamma$.
See \cref{app:group} for an in depth discussion.

\begin{figure}[thbp]
\begin{center}
\includegraphics[width=0.7\columnwidth]{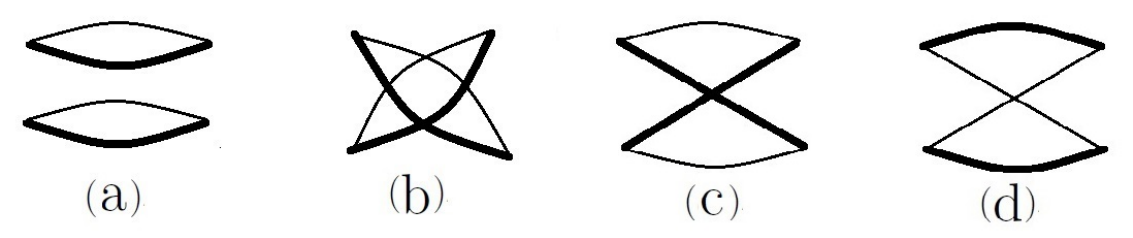}
\caption{Diagrams for Wick contractions. Thick and thin lines are to distinguish 
$d$ and $u$ propagators, respectively.}
\label{fg:diagram}
\end{center}
\end{figure}

A demonstration of the technique introduced here is performed in the
$\pi^-\pi^-$ system.
The individual contributions, $G_{\hat{R}^{-1},\hat{I}}$, to the
target correlation functions are given in terms of the Wick
contractions shown in Fig.~\ref{fg:diagram}, given explicitly by:
\begin{align}
&G_{\hat{R}^{-1},\hat{I}}(t;\vec{p};\vec{x},\boldsymbol{\delta})=
\sum_{(\vec{y}-\vec{x})\in \mathbb{Z}^3} 
e^{i\vec{p}\cdot(\vec{y}-\vec{x})}\;
\nonumber\\
&\times \left\langle\left\{\Tr\left[
S_d( \vec{y}_R^-, t;\vec{x}^-)
S^{\dag}_u(\vec{y}_R^-, t;\,\vec{x}^-,0) \right]
\Tr\left[
S_d( \vec{y}_R^+, t;\, \vec{x}^+,0)
S^{\dag}_u(\vec{y}_R^+, t;\,\vec{x}^+,0) \right] \right.\right.\nonumber\\
&\quad\quad\left. + \Tr\left[
S_d(\vec{y}_R^+, t;\, \vec{x}^-,0)
S^{\dag}_u(\vec{y}_R^+, t;\,\vec{x}^-,0)\right]
\Tr\left[
S_d(\vec{y}_R^-, t;\, \vec{x}^+,0)
S^{\dag}_u(\vec{y}_R^-, t;\,\vec{x}^+,0)\right]\right. \nonumber\\
&\quad\quad\left. - \Tr\left[
S_d( \vec{y}_R^-, t;\,\vec{x}^-,0)
S^{\dag}_u(\vec{y}_R^+, t;\,\vec{x}^-,0)
S_d(\vec{y}_R^+, t;\,\vec{x}^+,0 )
S^{\dag}_u( \vec{y}_R^-, t;\,\vec{x}^+,0) \right]\right. \nonumber\\
& \quad\quad\left.\left. - \Tr\left[
S_d(\vec{y}_R^+, t;\,\vec{x}^-,0)
S^{\dag}_u(\vec{y}_R^-, t;\,\vec{x}^-)
S_d(\vec{y}_R^-, t;\,\vec{x}^+ ,0)
S^{\dag}_u(\vec{y}_R^+, t;\,\vec{x}^+,0) \right] 
\right\}\right\rangle.
\label{eq:detailcal}
\end{align}
Here we have made use of the notation
\begin{align}
\vec{x}^{\pm}=\frac{2\vec{x}\pm\boldsymbol{\delta}}{2},\quad
\vec{y}_R^{\pm}=\frac{2\vec{y}\pm\hat{R}\boldsymbol{\delta}}{2},    
\end{align}
and $S_{q}(\vec{y},t;\vec{x},0)$ denotes a conventional
point-to-all propagator.
The quark flavours, $q=u\textrm{ or }d$, are shown explicitly, however
in the following numerical calculation we assume isospin symmetry, $S_u=S_d$.

Given the form of the correlator construction, we note that the
correlation function can be efficiently calculated by only performing
Dirac matrix inversions from two distinct sites $\vec{x}^\pm$.
Furthermore, in the moving frame, the same spectra are extracted from
the set of $\tilde{G}_{\Gamma}(t;\vec{p};\vec{x},\boldsymbol{\delta})$ with the same
$|\vec{p}|$.
One can therefore sum over each direction to reduce the statistical variance, {\it i.e.},
\begin{align}
\tilde{G}_{\Gamma}(t;P;\vec{x},\boldsymbol{\delta}) 
&=
\sum_{\vec{p},\,|\vec{p}|=P}\sum_{\hat{R} \in G^{\vec{p}}}\chi^{\Gamma}_{\hat{R}} G_{\hat{R},\hat{I}}(t;\vec{p};\vec{x},\boldsymbol{\delta}),
\label{eq:correlationmom}
\end{align}
where, as above, $G^{\vec{p}}$ denotes the
rotation group for specific total momentum $\vec{p}$.

In the following Section we present numerical results for the
determination of the ground states in each of the considered irreps up
to $P^2=3P_0^2$, where $P_0$ defines the basic momentum unit in the
box $P_0\equiv 2\pi/L$.



\section{Numerical Results}
\label{sec:III}

\subsection{Lattice setup}
Following the prescription given by
\cref{eq:detailcal,eq:correlationmom,tab:charterp,tab:momentum}, the
correlation functions of the $\pi^-\pi^-$ system are analysed for various
total momenta and irreps of the lattice rotation group.
The present calculation is performed on an ensemble with 2 flavours of
dynamical ${\cal O}(a)$-improved Wilson fermions with $\beta=5.29$, $\kappa=0.13550$
on a $24^3\times 48$ volume,
corresponding to $a=0.71\,{\rm fm}$ and $m_\pi\simeq 900\,{\rm MeV}$,
from the QCDSF Collaboration~\cite{Bali:2012qs}.

Results are collected from 376 configurations using 16 different
randomised source locations, totalling $\mathcal{O}$(6,000)
measurements.
With two distinct propagators required for each source, the
comparative computational cost of the present calculation is
$\mathcal{O}$(12,000) measurements.
For comparison, other calculations of the isospin-2 $\pi\pi$ system in
lattice QCD have used various amounts of computation, depending the
complexity of the observable: $\mathcal{O}$(4,500) by NPLQCD to
extract the $S$-wave scattering length \cite{Beane:2007xs};
$\mathcal{O}$(290,000) by NPLQCD \cite{Beane:2009kya} for the
energy-dependent $S$-wave phase shifts \cite{Beane:2011sc}; and
$\mathcal{O}$(270,000) by Dudek et al. to isolate $S$- and $D$-wave
phase shifts \cite{Dudek:2012gj}.


\subsection{Spectra}

In this study, we consider correlation functions with total momentum
up to three lattice units $|\vec{p}|\le \sqrt{3}P_0$.
The correlation functions for each irrep are fit with a
parameterisation taking the form:
\begin{equation}
G(t)
= A\left(e^{-Et}+e^{-E(T-t)}\right)
+B\left(e^{-\Delta E t}+e^{-\Delta E (T-t)}\right),
\label{eq:fitting}
\end{equation}
where the fit parameters $A$ and $E$ correspond to the amplitude and
2-point energy of interest.
The term involving $B$ is provided to isolate the leading contribution
arising from thermal states, as is familiar in studies of multi-hadron correlators
\cite{Feng:2009ij,Detmold:2010unv,Detmold:2011kw,Dudek:2012gj,Detmold:2012wc,Culver:2019qtx,Culver:2019vvu}.
For the present study, this corresponds to one pion propagating forwards
and the other backwards in Euclidean time.
The value of the exponent in the thermal contribution is held fixed to $\Delta E=E_\pi(\vec{p}-\vec{k})-E_\pi(\vec{k})$, for
single-pion energies $E_\pi$, and $\vec{k}$ chosen to correspond to the
lightest single pion state contributing to the given correlator.
At large temporal extent, the coefficient $B$ should scale according
to $e^{-E_\pi(\vec{k})T}$.
While we don't have numerical results at different $T$, we see that
the fitted values of $B$ are always suppressed by this order of
magnitude compared to $A$.

\begin{figure}[htbp] \vspace{-0.cm}
\begin{center}
\includegraphics[width=0.9\columnwidth]{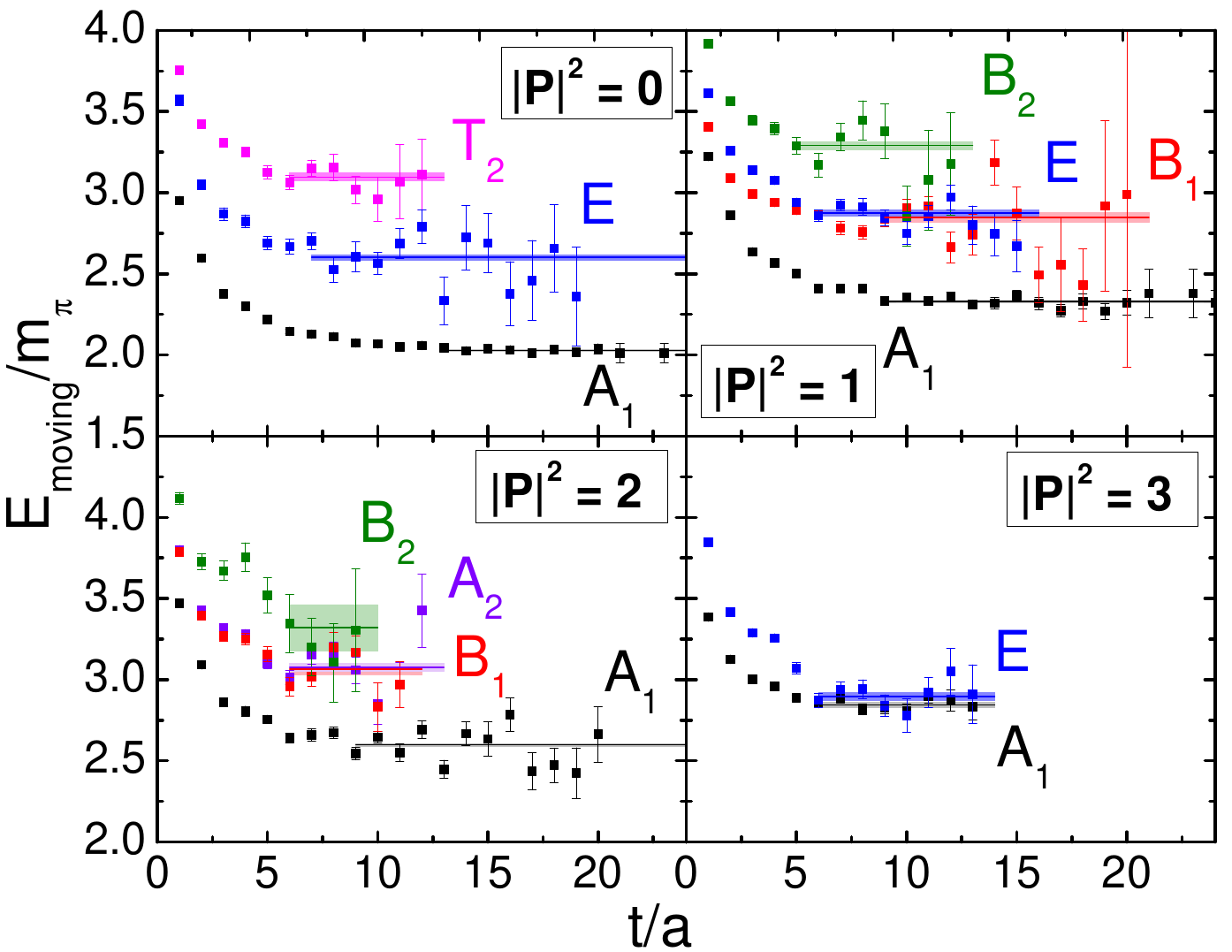}
\caption{The effective energies
for different total momenta $P$ (in multiples of $2\pi/L$), and
different irreps. The bands display the two-pion energies fitted to
Eq.~(\ref{eq:fitting}). The
horizontal width of the bands indicates the corresponding fit window.
\label{fg:latticedata}}
\end{center}
\end{figure}

After subtracting the contibutions from thermal states,
Figure~\ref{fg:latticedata} displays the effective mass for different
total momenta and irreducible representation.
We see a clear separation of the energy levels in distinct irreps.
As expected, the low-lying $A_1$ irreps are generally cleaner
statistically, whereas the signal quality degrades for the irreps 
corresponding to the resolution of higher-spin partial waves.

\begin{figure}[htbp] \vspace{-0.cm}
\begin{center}
\includegraphics[width=0.7\columnwidth]{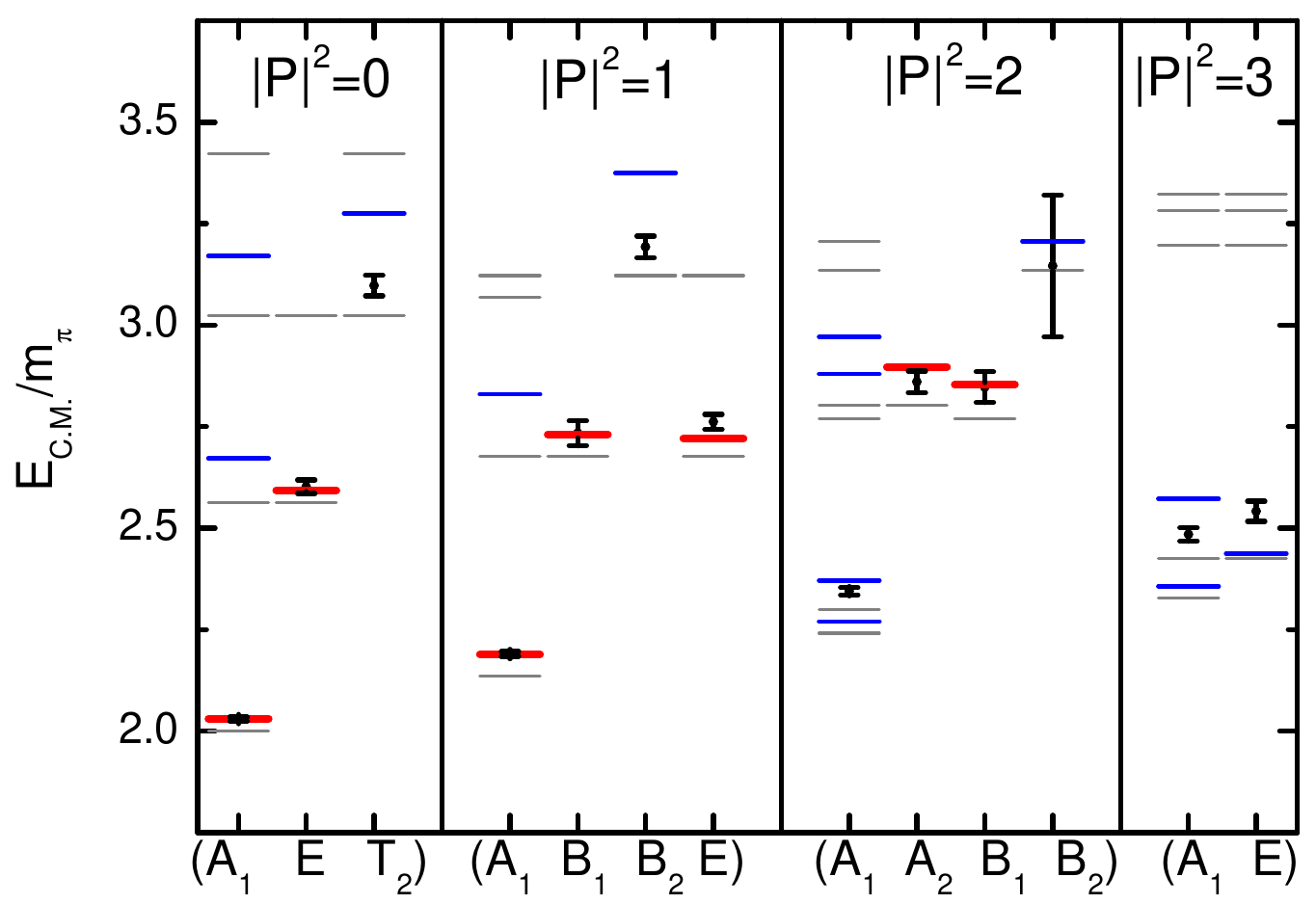}
\caption{The energy levels of the various systems with different total
  momenta and irreps. The black points show the extracted
  centre-of-momentum (CM) energies.
  The grey lines display the locations of the corresponding 
  non-interacting energies. The red lines show the fitted energies
  according to fit (iii), as described in Table~\ref{tab:fits}, and the blue lines display further predicted
  eigenvalues based upon this fit.
\label{fg:spectra}}
\end{center}
\end{figure}

The results for the extracted energy levels are shown by the black
circles in \cref{fg:spectra}.
For comparison, the low-lying non-interacting energy levels in each
system are displayed by the grey lines.
Each of the energy levels isolated are consistent with some degree of
weak repulsion, as expected for the $I=2$ state.
For most of the states considered, the first excited state is expected
to be clearly separated, and hence the ground-state isolation should
be reliable (to within the statistical uncertainties of this work).
However, there are three particular channels where multiple
low-lying states are anticipated, arising from the clustering of
non-interacting two-particle energy levels. These include $A_1$ at $P^2=2P_0^2,3P_0^2$ and
$B_2$ at $P^2=2P_0^2$.
In these cases, we do not expect that our correlation functions are
dominated by a single ground-state energy and hence the fitted
parameters are not representative of eigenenergies of the system.

Within the operator construction presented, only the total momentum is
specified, whereas the momentum of each pion is not.
In contrast to Refs.~\cite{Beane:2011sc,Berkowitz:2015eaa}, which
involve momentum-projected hadrons at the sink, we use the same operators
at both the source and sink.
While staying within the paradigm of local sources, this method then
lends itself to a variational analysis~\cite{Blossier:2009kd,Mahbub:2013ala,Stokes:2013fgw}, where
the operator basis can be extended by varying $|\boldsymbol{\delta}|$.
%



\subsection{Phase shifts}

We use the following L\"uscher formula \cite{Dudek:2010ew,
  Gockeler:2012yj, Luscher:1990ux}, assuming that
exponentially-suppressed corrections can be neglected, to extract
phase shifts from finite-volume spectra:
\begin{align}
  \det[M^{\Gamma, \vec{p}}_{\ell n,\ell'n'}(q(\Gamma))-\delta_{\ell,\ell'}\delta_{n,n'}\cot\delta_\ell(q(\Gamma))] = 0.
  \label{eq:Luscher}
\end{align}
The matrix $M^{\Gamma, \vec{p}}_{\ell n,\ell'n'}$ has been discussed
extensively in the literature, see
e.g.~\cite{Rummukainen:1995vs,Gockeler:2012yj,Leskovec:2012gb}. For completeness, we
provide detail relevant to the present investigation in
Appendix~\ref{app:luscher}.

As encoded by \cref{eq:Luscher}, each energy level determined on the
lattice is constrained by multiple partial waves --- see
\cref{tab:angirr}.
This necessitates the use of a parameterisation of the energy
dependence of the phase shifts in order to isolate the individual
partial waves.
For the purpose of this investigation, we consider the
parameterisation of the $\ell$-wave phase shifts by the effective range expansion:
\begin{eqnarray}
  q^{2\ell+1} \cot\delta_\ell = \frac{1}{ a_\ell } + \frac{1}{2}r_\ell q^2, 
   \label{eq:relationphpar}
\end{eqnarray}
for parameters $a_\ell$ and $r_\ell$ --- for $\ell=0$ these are
familiarly recognised as the scattering length and effective range,
respectively.
Such a parameterisation should be reasonable for the weakly-repulsive
interactions anticipated in $I=2$ scattering.

\begin{table}[htbp]
\begin{center}\caption{The relationship between angular momentum and
    irrep in the various momentum. The total angular momentum quantum
    number for
    exact spherical symmetry are only quoted up to $\ell=4$.}
\begin{tabular}{cccccc}\hline\hline
    Group & $|\vec{p}L/2\pi|^2$  &   $\Gamma$    &     $\ell$        \\  
    \hline
     $O_h$ & $0$                 &       $A^+_1$        &      0, 4      \\
           &                                                         &        $A^+_2$        &      $>$4           \\ 
            &                                                        &        $E^+$            &      2, 4            \\ 
             &                                                       &        $T^+_1$        &      4            \\ 
              &                                                      &        $T^+_2$        &      2, 4            \\ 
     $C_{4\nu}$ & $1$                   &         $A_1$            &      0, 2,4              \\
            &                                                        &        $A_2$             &      $>$4            \\ 
             &                                                       &        $B_1$             &      2, 4            \\ 
              &                                                      &        $B_2$             &     2, 4            \\ 
               &                                                     &        $E$                 &      2, 4            \\ 
     $C_{2\nu}$ & $2$                   &         $A_1$            &      0, 2, 4              \\
            &                                                        &        $A_2$             &      2, 4            \\ 
             &                                                       &        $B_1$             &      2, 4            \\ 
              &                                                      &        $B_2$             &      2, 4            \\ 
     $C_{3\nu}$ & $3$                   &         $A_1$            &      0, 2,  4              \\
            &                                                        &        $A_2$             &      $>$ 4            \\ 
             &                                                       &        $E$                 &      2, 4            \\ 
      \hline\hline
\end{tabular}  \label{tab:angirr}
\end{center}
\end{table}

As described above, we do not expect that our extracted energy levels in $A_1$ at $P^2=2P_0^2,3P_0^2$ or
$B_2$ at $P^2=2P_0^2$ are meaningful representations of an energy eigenstate, and hence these are excluded from any fits.
This leaves up to 10 data points for constraining the phase shift
parameterisation.
We summarise a selection of fit prescriptions in Table~\ref{tab:fits}.
In fit (i), we attempt to describe all 10 data points with just a
leading order $a_\ell$ in each partial wave.
We find that the $E$ representation at $P^2=3P_0^2$ is incompatible
with the fit form, as shown in Table~\ref{tab:params}, and hence we drop this point also from subsequent
fits.
Just dropping this one point improves the reduced $\chi^2$
($\chi^2_r$) significantly
in fit (ii), yet still suggests some tension with the data.
As further modifications, we try removing the large centre-of-mass ($E^*/m>3$)
points in fit (iii).
In fits (iv) and (v) we introduce parameters $r_0$ or
$r_2$ to capture any additional curvature in the energy
dependence of the phase shifts.
However, we find
that these additional parameters are poorly constrained and lead to
weaker $\chi^2_r$ values.
Hence we conclude that the best description of our lattice results is that
of fit (iii), with a restricition to the lower-energy spectra and fitting just a 
single parameter in each partial wave, $\ell=0,2,4$.

\begin{table}[htbp]
     \setlength{\tabcolsep}{0.15cm}
\begin{center}
    \caption{Fitted parameters. \label{tab:fits}}
\begin{tabular}{l|ccc|ccc}
\hline\hline
Fit & $\{3_E\}$  & $\{E^*/m>3\}$ & $N_{\rm data}$ & $a_{0,2,4}$ & $r_0$ & $r_2$ \\
\hline 
i   & \checkmark & \checkmark   & 10 & \checkmark & $\times$ & $\times$  \\
ii  & $\times$   & \checkmark   & 9  & \checkmark & $\times$ & $\times$  \\
iii & $\times$   & $\times$     & 7  & \checkmark & $\times$ & $\times$  \\
iv  & $\times$   & \checkmark   & 9  & \checkmark & \checkmark & $\times$  \\
v   & $\times$   & \checkmark   & 9  & \checkmark & $\times$ & \checkmark  \\
\hline\hline
\end{tabular}
\end{center}
\end{table}

\begin{table}[htbp]
\begin{center}
    \caption{Fitted parameters. \label{tab:params}}
\begin{tabular}{lccccccc}
\hline\hline
Fit & $a_0$        & $r_0$ & $a_2$         & $r_2\times 10^6$ & $a_4$ & $\chi^2$ & $\chi^2_r$ \\
\hline
i   & $-0.690(53)$ & ---   & $-0.0111(84)$ & ---   & $-0.0200(41)$ & 36.2 & 5.2 \\
ii   & $-0.691(65)$ & ---   & $-0.0092(64)$ & ---   & $-0.0208(43)$ & 15.5 & 2.6 \\
iii  & $-0.683(65)$ & ---   & $-0.0602(58)$ & ---   & $-0.0118(48)$ & 7.2 & 1.8 \\
iv  & $-0.65(11)$  & 0.7(19)& $-0.0091(90)$ & --- & $-0.0208(67)$ & $15.4$ & 3.1\\
v   & $-0.691(53)$ & ---    & $-0.0092(96)$ & 0.5(51) & $-0.0208(67)$ & $15.5$ & 3.9\\
\hline\hline
\end{tabular}
\end{center}
\end{table}

\begin{figure}[htbp] \vspace{-0.0cm}
\begin{center}
\includegraphics[width=0.7\columnwidth]{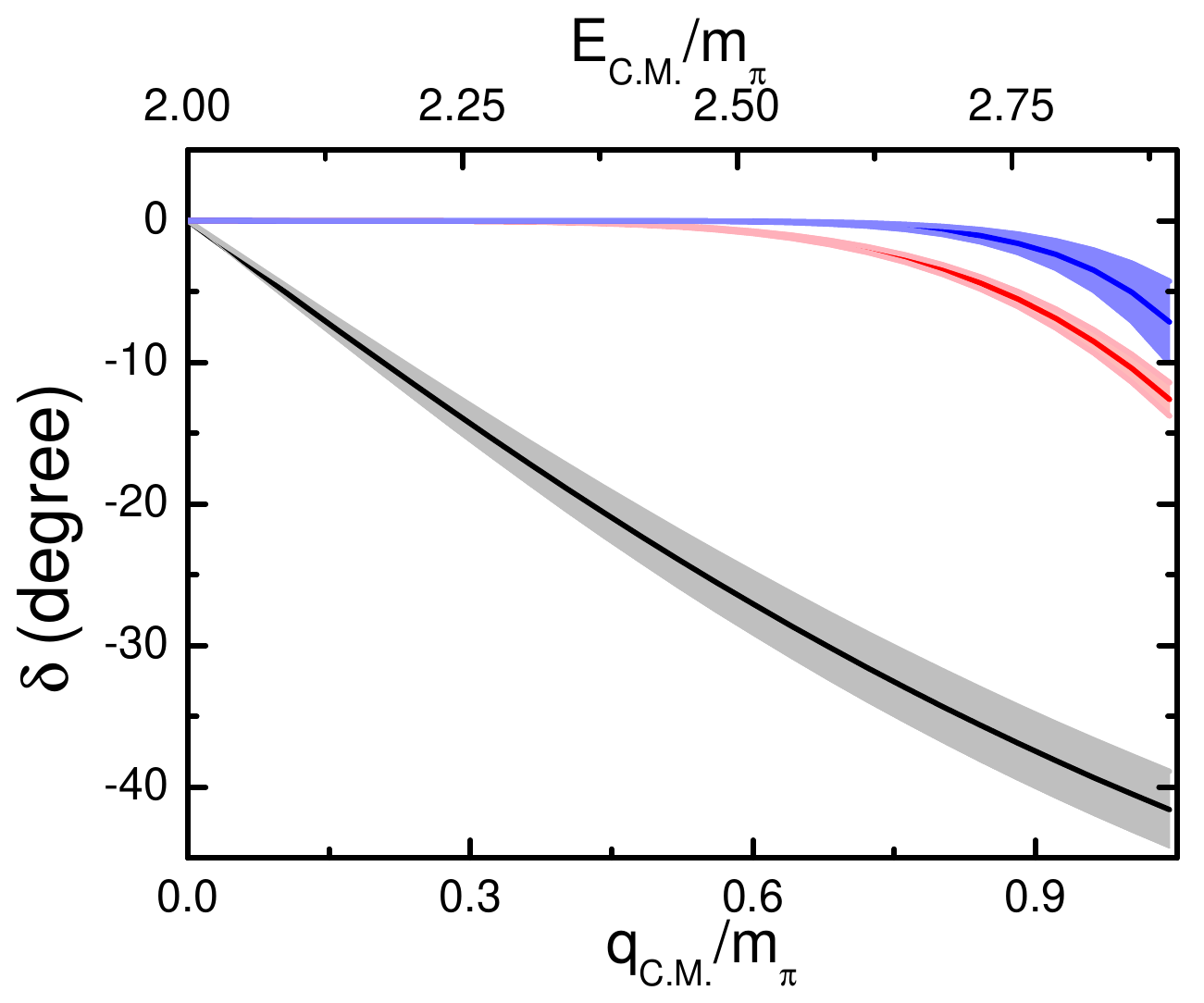}
\caption{The phase shifts from the fit parameterisation (iii), as
  described in the text. The black, red
  and blue curves are for the $S$-, $D$- and $G$-wave phase shifts.
  Note the top and bottom horizontal
  axes provide the total energy and the corresponding on-shell
  momentum, respectively, in the CM system.
\label{fg:phaseshift}}
\end{center}
\end{figure}

The fitted spectra for fit (iii) are displayed by the red lines in
Fig.~\ref{fg:spectra}.
For this fit, we also show additional predicted finite-volume spectra for
this parameterisation by the blue lines.
From the fit parameters in Table~\ref{tab:params}, we
see a clear signal for a $G$-wave interaction.
This gets somewhat stronger for the other fit forms considered,
however at the expense of a reduced fit quality.
For the preferred fit, we show the corresponding phase shifts in
Fig.~\ref{fg:phaseshift}. 

\section{Summary and Outlook}
\label{sec:sum}

In this paper, we introduce a new extended operator to extract the
spectra of irreducible representations at rest and in moving systems.
In coordinate space, the two-particle operator projects onto an
irrep by summing appropriately over a spherical
shell.
The method is straightforward to implement as a generalisation of
conventional point sources, and hence offers an alternative for cases
where stochastic momentum sources are impractical.

For the numerical investigation in this work, we studied the isospin-2
$\pi\pi$ system at a range of total momenta, on a $24^3\times 48$
volume  with a
lattice spacing of $a=0.071$~fm and $m_\pi\approx 900$~MeV.
The correlation functions of various irreps with a total
momentum-squared ranging from
$0$ to $3$ have been studied, with 13 plateaus---10 of which were
considered as viable ground-state candidates.
These discrete finite volume spectra have then been analysed with
the L\"uscher quantisation condition.
Using a simple effective range expansion of the phase shifts, we
identify $S$-, $D$- and $G$-wave interactions.

In the future, this method can also be readily extended to particles with spin, particularly for the two baryon system.
Including a basis of operators at different hadronic separations
would allow for a variational analysis to be performed, and thereby
allow for a determination of the excited energy levels on the lattice.
This would correspond to an analogue of mapping out the quantum
mechanical coordinate-space wave function.


\begin{acknowledgments}
The calculations presented in this manuscript made use of 
the Chroma software library~\cite{Edwards:2004sx}.
This research was supported with supercomputing resources provided by
the Phoenix HPC service at the University of Adelaide and the National
Computational Infrastructure (NCI).
NCI resources were
provided through the National Computational Merit Allocation Scheme,
supported by the Australian Government through Grants Nos.~LE160100051
and LE190100021 (DBL $\&$ JMZ) and the University of Adelaide Partner Share. 
This investigation has been supported by the Australian Research
Council under grants DP140103067 (RDY, JMZ $\&$ DBL), DP190100297 (JMZ
$\&$ RDY) and DP190102215 (DBL) and DP210103706 (DBL).
GS was supported by DFG grant SCHI 179/8-1.
JJW was supported by the Fundamental Research Funds for the Central Universities and National Key R$\&$D Program of China under Contract No. 2020YFA0406400. 
\end{acknowledgments}


\appendix

\section{Octahedral Group ($O_h$) and Its Little Groups}
\label{app:group}

\subsection{The 48 elements of $O_h$ group}

The cubic group ($O$) has 24 elements indicated as $R_i$ ($i=1$--$24$)
which correspond to 24 rotations, $\hat{R}_i$ as listed in Table
\ref{tab:delta}.
Starting with one vector $\boldsymbol{\delta}_1 \equiv (q_1, q_2,
q_3)$, one can then construct 23 vectors via $\hat{R}_i$ ($i=2$--$24$)
($\hat{R}_1$ is the identity operator) as follows:
%
\begin{eqnarray}
\boldsymbol{\delta}_i&=&\hat{R}_i \boldsymbol{\delta}_1,
\end{eqnarray}
In Table \ref{tab:delta}, the 24 vectors $\boldsymbol{\delta}_i$ are all listed. 
%
The $O_h$ group can be recognized as the product of the $O$ group and
the $C_2=\{e,\hat\sigma\}$ group, {\it i.e.}, $O_h = O \otimes C_2$. 
Then the other 24 operators belonging to $O_h$ group rather than $O$
group will be $\hat{R}_{i+24} = \hat{\sigma}\hat{R}_i$, and
correspondingly, $\boldsymbol{\delta}_{i+24}=-\boldsymbol{\delta}_i$.

\begin{table}[htbp]
\renewcommand\arraystretch{1.5}
\begin{center}\caption{For the $O$ group, 24 vectors $\boldsymbol{\delta}_i$ and
    operators $R_i$ ($i=1$--$24$) are listed. $\boldsymbol{\delta}_1^\top \equiv (q_1, q_2,
    q_3)$ and $R_1 \equiv E$ which is the identity. $R_i$ includes the rotation axis and angle.
}
\begin{tabular}{ccccc}\hline\hline
  Class      &  $R_i$         & Axis-Angle                            &  Euler    Angle                  &   $\boldsymbol{\delta}_i^\top$                                            \\  
    \hline
$E$         &   $R_{1 }$   &  Any                   $   0^o$    &$  (     0^o,     0^o,    0^o)$  &  $( q_1,\, q_2,\, q_3)$   \\ 
$8C'_3$   &   $R_{2 }$   &  $( 1,\, 1,\,1)$  $-120^o$  &$  (   90^o,   90^o,180^o)$  &  $( q_2,\, q_3,\, q_1)$   \\       
               &   $R_{3 }$   &  $( 1,\, 1,\,1)$  $+120^o$ &$  (     0^o,   90^o,  90^o)$  &  $( q_3,\, q_1,\, q_2)$   \\        
               &   $R_{4 }$   &  $(-1,\, 1,\,1)$  $-120^o$  &$  ( 180^o,  90^o,   90^o)$  &  $(-q_3,\,-q_1,\,q_2)$   \\        
               &   $R_{5 }$   &  $(-1,\, 1,\,1)$  $+120^o$ &$  (   90^o,  90^o,    0^o)$   &  $(-q_2,\,q_3,\,-q_1)$   \\        
               &   $R_{6 }$   &  $(-1,\,-1,\,1)$  $-120^o$  &$  (  -90^o,  90^o,    0^o)$   &  $(q_2,\,-q_3,\,-q_1)$   \\        
               &  $R_{7 }$    &  $(-1,\,-1,\,1)$  $+120^o$ &$  ( 180^o,  90^o, -90^o)$   &  $(-q_3,\,q_1,\,-q_2)$   \\        
               &  $R_{8 }$    &  $( 1,\,-1,\,1)$  $-120^o$  &$  (      0^o,  90^o, -90^o)$   &  $(q_3,\,-q_1,\,-q_2)$   \\        
               &  $R_{9 }$    &  $( 1,\,-1,\,1)$  $+120^o $&$  (  -90^o,  90^o, 180^o)$   &  $(-q_2,\,-q_3,\,q_1)$   \\        
$6C_4$   &  $R_{10}$   &  $( 1,\, 0,\,0)$  $-90^o $   &$  (    90^o,  90^o, -90^o)$   &  $( q_1,\, q_3,\,-q_2)$   \\         
               &  $R_{11}$   &  $( 1,\, 0,\,0)$  $+90^o $  &$  (  -90^o,  90^o,   90^o)$   &  $( q_1,\,-q_3,\, q_2)$   \\         
               &  $R_{12}$   &  $( 0,\, 1,\,0)$  $-90^o $   &$  (  180^o,  90^o, 180^o)$   &  $(-q_3,\, q_2,\, q_1)$   \\         
               &  $R_{13}$   &  $( 0,\, 1,\,0)$  $+90^o $  &$  (     0^o,   90^o,    0^o)$     &  $( q_3,\, q_2,\,-q_1)$   \\         
               &  $R_{14}$   &  $( 0,\, 0,\,1)$  $-90^o $   &$  (  -90^o,    0^o,     0^o)$    &  $( q_2,\,-q_1,\, q_3)$   \\         
               &  $R_{15}$   &  $( 0,\, 0,\,1)$  $+90^o$   &$  (   90^o,    0^o,     0^o)$   &  $(-q_2,\, q_1,\, q_3)$   \\         
$6C''_2$  &  $R_{16}$   &  $( 0,\, 1,\,1)$  $-180^o$  &$  (   90^o,   90^o,  90^o)$   &  $(-q_1,\, q_3,\, q_2)$   \\         
               &  $R_{17}$   &  $( 0,\,-1,\,1)$  $-180^o$  &$  ( -90^o,   90^o, -90^o)$   &  $(-q_1,\,-q_3,\,-q_2)$   \\         
               &  $R_{18}$   &  $( 1,\, 1,\,0)$  $-180^o$  &$  ( -90^o, 180^o,     0^o)$   &  $( q_2,\, q_1,\,-q_3)$   \\         
               &  $R_{19}$   &  $( 1,\,-1,\,0)$  $-180^o$  &$  (  90^o, 180^o,     0^o)$   &  $(-q_2,\,-q_1,\,-q_3)$   \\         
               &  $R_{20}$   &  $( 1,\, 0,\,1)$  $-180^o$  &$  (    0^o,    90^o,180^o)$    &  $( q_3,\,-q_2,\, q_1)$   \\         
               &  $R_{21}$   &  $(-1,\, 0,\,1)$  $-180^o$  &$  (180^o,   90^o,     0^o)$    &  $(-q_3,\,-q_2,\,-q_1)$   \\         
$3C^2_4$   &  $R_{22}$   &  $( 1,\, 0,\,0)$  $-180^o$  &$  (180^o, 180^o,     0^o)$    &  $( q_1,\,-q_2,\,-q_3)$   \\         
               &  $R_{23}$   &  $( 0,\, 1,\,0)$  $-180^o$  &$  (    0^o, 180^o,     0^o)$    & $(-q_1,\, q_2,\,-q_3)$   \\         
               &  $R_{24}$   &  $( 0,\, 0,\,1)$  $-180^o$  &$  (180^o,     0^o,     0^o)$    &  $(-q_1,\,-q_2,\, q_3)$   \\            \hline\hline
\end{tabular}  \label{tab:delta}
\end{center}
\end{table}

\subsection{The Classes and Irreps of $O_h$ group }

There are 48 elements in the $O_h$ group and they can be partitioned
into 10 different classes. There are ten irreps, $A^\pm_1 (1)$,
$A^\pm_2 (1)$,  $E^\pm (2)$,  $T^\pm_1 (3)$,  and $T^\pm_2 (3)$,
where the numbers in the parentheses are the dimensions of these irreps.
The character table of the cubic group is shown in Table \ref{tab:charterp}.

\subsection{Regular Representation}

Using the $O_h$ group, a scalar function $\phi(\boldsymbol{\delta})$ can be extended to 48 functions as follows
\begin{eqnarray}
  \phi_R(\boldsymbol{\delta}) &=& \hat{P}_{R}\,\phi(\boldsymbol{\delta}) =\phi (\hat{R}^{-1} \boldsymbol{\delta})\,. \label{eq:Pr1}
\end{eqnarray}
Under the group action, they should transform as
\begin{eqnarray}
  \hat{P}_{R}\phi_{R'}(\boldsymbol{\delta}) &=& \sum_{R''}
  \phi_{R''}(\boldsymbol{\delta}) \left(\bar{B}(\hat{R})\right)_{R'',
    R}
  =\hat{P}_{R}\,\hat{P}_{R'}\,\phi(\boldsymbol{\delta})
  = \hat{P}_{RR'}\,\phi(\boldsymbol{\delta})
  = \phi(\hat{R'}^{-1}\hat{R}^{-1}\boldsymbol{\delta})     
  = \phi_{RR'}(\boldsymbol{\delta})\,.\nonumber \label{eq:PrB}
\end{eqnarray}
Here $\bar{B}(\hat{R})$ is the representation matrix of $\hat{R}$ for
the regular representation.
The dimension of the regular representation is the same as the order of the group. 

\subsection{From Regular Representation to Irreps}

The regular representation of any non-trivial group is reducible. So $\bar{B}$ can be made block diagonal according to the irreps of $O_h$ via a unitary transformation matrix $\bar{S}$ as follows
%
 %
\begin{eqnarray}
\bar{S}^{-1} \bar{B}(\hat{R}) \bar{S} =1 \bar{A}^\pm_1(\hat{R}) \oplus 1 \bar{A}^\pm_2(\hat{R}) \oplus 2 \bar{E}^\pm(\hat{R})  \oplus 3 \bar{T}^\pm_1(\hat{R})  \oplus 3 \bar{T}^\pm_2(\hat{R})  \equiv \bar{A}(\hat{R})
\end{eqnarray}
The number before the irrep indicates the occurrence of that irrep. At
last you will find $48 =  2(1^2 +1^2 +2^2+3^2+3^2)$. And the matrix
$\bar{A}$ can be written as:
\begin{eqnarray}
\bar{A}_{i\Gamma\,n, i'\Gamma'\,n'}(\hat{R}) = \delta_{i'i}\,\delta_{\Gamma'\Gamma}\, \bar{\Gamma}_{n,n'}(\hat{R}),
\label{eq:A1}
\end{eqnarray}
where $\Gamma$ is the name of the irrep, and $i$ shows how many times
it appears, for example $i = 1, 2, 3$ for $T^\pm_1$ and $T^\pm_2$, and
$i = 1, 2$ for $E$, and  $i = 1$ for $A^\pm_1$ and $A^\pm_2$; and $n$
indicates the order of the irrep $\Gamma$. The matrix
$\bar\Gamma(\hat{R})$ is the matrix representation of element
$\hat{R}$ in the irrep $\Gamma$.
Because $O_h$ is a finite group, the matrices $\bar \Gamma$ can be
chosen to be unitary.  

As shown in Eq.(\ref{eq:PrB}), the matrices $\bar{B}(\hat{R})$ show
the rotations of 48 scalar functions $\phi$.
Then matrices $\bar{A}$ also have 48 scalar functions satisfied:
\begin{eqnarray}
\hat{P}_{R}\Phi_{i,\Gamma,n}(\boldsymbol{\delta}) &=& \sum_{i',\Gamma', n'} \Phi_{i',\Gamma',n'}(\boldsymbol{\delta}) \left(\bar{A}(\hat{R})\right)_{i'\Gamma'\,n', i\,\Gamma\,n}.\label{eq:PrA}
\end{eqnarray} 
The transformation matrix $\bar{S}$ can connect $\Phi_R$ and $\Phi_{i,
  \Gamma, n}$ as follows:

\begin{eqnarray}
\Phi_{i, \Gamma, n} = \sum_ R \phi_{R} \bar{S}_{R, i \Gamma n}.\label{eq:newPhi}
\end{eqnarray}
The row index of $\bar{S}$ is the name of the elements of the cubic
group, and the column index is the same as the indices of $\Phi$, $ (i, \Gamma, n)$.

On the other hand, from Eq.(\ref{eq:PrB},\ref{eq:PrA},\ref{eq:newPhi}), we have:
\begin{eqnarray}
\hat{P_{R}}\Phi_{i,\,\Gamma,\,n}&=&\sum_{n'}\Phi_{i,\,\Gamma,\,n'}\bar{\Gamma}_{n'n}(\hat{R})
=\sum_{R'}\sum_{n'}\phi_{R'}\bar{S}_{R'\, i\Gamma\,n'}\bar{\Gamma}_{n'n}(\hat{R})\nonumber\\
&=&\hat{P_{R}}\sum_{R'}\phi_{R'}\bar{S}_{R',\,i\Gamma\,n}=\sum_{R'}\phi_{RR'}\bar{S}_{R',\,i\Gamma\,n},
\end{eqnarray}
Then we have:
\begin{eqnarray}
\sum_{R'}\sum_{n'}\phi_{R'}\bar{S}_{R'\, i\Gamma\,n'}\bar{\Gamma}_{n'n}(\hat{R})
&=&\sum_{R'}\phi_{RR'}\bar{S}_{R',\,i\Gamma\,n}, \\
\bar{S}_{R, i \Gamma n} & =& \sum_ m C_{i \Gamma\,m} \bar{\Gamma}_{m, n}(R^{-1}).\label{eq:S1}\\
C_{i \Gamma\,m} &=& \bar{S}_{I,\,i \Gamma\,m}
\end{eqnarray}
%
The $ C_{i \Gamma\,m}$ satisfy the orthogonality relations:
\begin{eqnarray}
\frac{l_\Gamma}{G}\delta_{i,i'} = \sum_ m C_{i \Gamma\,m}C^*_{i' \Gamma\,m} ,\label{eq:C1}\\
\frac{l_\Gamma}{G}\delta_{m,m'} = \sum_ i C_{i \Gamma\,m}C^*_{i \Gamma\,m'} .\label{eq:C2}
\end{eqnarray}
where $G$ and $l_{\Gamma}$ are the orders of $O_h$ group and irrep $\Gamma$, respectively.
%

\subsection{The inner product of $\Phi_R$ and $\Phi_{(i, \Gamma, n)}$ }
We use the Dirac symbol for the inner product of  $\Phi_R$ and
$\Phi_{(i, \Gamma, n)}$. The normalization is given by:
\begin{eqnarray}
\delta_{R,R'} &=& \langle \Phi_R | \Phi_{R'} \rangle, \label{eq:phi1}\\
\delta_{i,i'}\delta_{\Gamma,\Gamma'}\delta_{n,n'} &=& \langle \Phi_{i,\Gamma,n} | \Phi_{i',\Gamma',n'}  \rangle, \label{eq:Phi1}
\end{eqnarray}

If we have some operator, $\hat{H}$, which is invariant under the
rotation, such as the Hamiltonian operator, through
Eq.(\ref{eq:newPhi},\ref{eq:S1},\ref{eq:C2}), we have:
\begin{eqnarray}
&&\sum_{i}\langle \Phi_{i,\Gamma,n} |\hat{H}| \Phi_{i,\Gamma',n'}  \rangle
=\sum_i\sum_{R,R'} \bar{S}^*_{R, i\Gamma\,n}\langle \phi_{R} |\hat{H}| \phi_{R'}  \rangle\bar{S}_{R',i\Gamma'\,n'} \nonumber\\
&=&\sum_i\sum_{R,R'} \sum_{m,m'} 
C^*_{i \Gamma\,m} \bar{\Gamma}^*_{m, n}(R^{-1})
\langle \phi_{R'^{-1}R} |\hat{H}| \phi_{I}  \rangle
 C_{i \Gamma'\,m'} \bar{\Gamma}_{m', n'}(R'^{-1}) \nonumber\\
 &=&\sum_i\sum_{\tilde{R},R'} \sum_{m,m'} C^*_{i \Gamma\,m} \bar{\Gamma}^*_{m, n}(\tilde{R}^{-1}R'^{-1})
\langle \phi_{\tilde{R}} |\hat{H}| \phi_{I}  \rangle
 C_{i \Gamma'\,m'} \bar{\Gamma'}_{m', n'}(R'^{-1}) \nonumber\\
   &=&\sum_i\sum_{\tilde{R}} \langle \phi_{\tilde{R}} |\hat{H}| \phi_{I}  \rangle
  \sum_{m,m'} C^*_{i \Gamma\,m}  C_{i \Gamma'\,m'}
  \sum_{l}
 \delta_{n,n'} \delta_{\Gamma,\Gamma'}\delta_{l,m'}\bar{\Gamma}^*_{m, l}(\tilde{R}^{-1}) \nonumber\\
&=&\sum_i\sum_{\tilde{R}} \langle \phi_{\tilde{R}} |\hat{H}| \phi_{I}  \rangle
  \sum_{m} C^*_{i \Gamma\,m}  C_{i \Gamma\,m}
  \frac{G}{l_\Gamma}\delta_{\Gamma,\Gamma'}\delta_{n,n'}\bar{\Gamma}_{m', m}(\tilde{R}) \nonumber\\
&=&\delta_{\Gamma,\Gamma'} \delta_{n,n'}\sum_{\tilde{R}}
 \left(\frac{G}{l_\Gamma}\sum_{m,m'}  \sum_{i}C_{i \Gamma\,m'} \bar{\Gamma}_{m', m}(\tilde{R})C^*_{i \Gamma\,m}\right)
\langle \phi_{\tilde{R}} |\hat{H}| \phi_{I}  \rangle\label{eq:Phical21}
\nonumber\\
&=&\delta_{\Gamma,\Gamma'} \delta_{n,n'}\sum_{\tilde{R}}
 \left(\frac{G}{l_\Gamma}\sum_{m,m'} \frac{l_\Gamma}{G} \delta_{m,m'} \bar{\Gamma}_{m', m}(\tilde{R})\right)
\langle \phi_{\tilde{R}} |\hat{H}| \phi_{I}  \rangle \label{eq:Phical22}
\nonumber\\
&=&\delta_{\Gamma,\Gamma'} \delta_{n,n'}\sum_{\tilde{R}}
 \left(\chi^{\Gamma}(\tilde{R})\right)
\langle \phi_{\tilde{R}} |\hat{H}| \phi_{I}  \rangle \label{eq:Phical23}
\end{eqnarray}

At last, we find 
%
\begin{eqnarray}
\sum_i\langle \Phi_{i,\Gamma,n} |\hat{H}| \Phi_{i,\Gamma',n'}  \rangle 
&=& \delta_{\Gamma,\Gamma'}\delta_{n,n'}\sum_{R} \chi^{\Gamma}(R)\langle \phi_{R} |\hat{H}| \phi_{I}  \rangle, \label{eq:Phical2}
\end{eqnarray}
where $\chi^{\Gamma}_{\hat{R}}$ is the charter of element $\hat{R}$ in the $\Gamma$ irrep.
The character tables for $O_h$ group and the little group are listed in Table \ref{tab:charterp}.

\begin{table}[htbp]
     \setlength{\tabcolsep}{0.15cm}
\begin{center}\caption{Character table of $O_h$, $C_{4\nu}$, $C_{2\nu}$ and $C_{3\nu}$ for $|\vec{p}L/2\pi|^2 = 0$, $1$, $2$, and $3$, respectively.}
\begin{tabular}{cccccccccccccccccc}
\hline\hline
 $O_h$  &  $\Gamma$/Class      &     $I$\,\,                                    &   $8C'_3$\,\,    
                                    & $6C_4$\,\,                                 &  $6C'_4$\,\,   
                                    & $3C^2_4$\,\,                             &      $\hat{\pi}$\,\,       
                                    &   $8C'_3\times\hat{\pi}$\,\,        & $6C_4\times\hat{\pi}$\,\,   
                                    &  $6C'_4\times\hat{\pi}$\,\,         & $3C^2_4\times\hat{\pi}$\,\,  \\     
    \hline
 &    $A^\pm_1$  & $1$ & $1$ & $1$ & $1$ & $1$ & $\pm1$ & $\pm1$ & $\pm1$ & $\pm1$ & $\pm1$  \\
 &    $A^\pm_2$  & $1$ & $1$ & $-1$& $-1$& $1$ & $\pm1$ & $\pm1$ & $\mp1$ & $\mp1$ & $\pm1$  \\ 
 &    $E^\pm$    & $2$ & $-1$& $0$ & $0$ & $2$ & $\pm2$ & $\mp1$ & $0$    & $0$    & $\pm2$  \\
 &    $T^\pm_1$  & $3$ & $0$ & $-1$& $1$ & $-1$& $\pm3$ & $0$    & $\mp1$ & $\pm1$ & $\mp1$  \\
 &    $T^\pm_2$  & $3$ & $0$ & $1$ & $-1$& $-1$& $\pm3$ & $0$    & $\pm1$ & $\mp1$ & $\mp1$  \\        
\hline\hline
 $C_{4\nu}$  &   $\Gamma$/Class      &     $I$       &   $2C_4$   & $2C'_2\times\hat{\pi}$   &  $2C_2\times\hat{\pi}$  & $C_2$  \\  
    \hline
 &    $A_1$      &       1   &        1  &      1   &       1   &       1   \\
 &    $A_2$      &       1   &        1  &     -1   &      -1   &       1       \\ 
 &    $B_1$      &       1   &       -1  &     -1   &       1   &       1       \\
 &    $B_2$      &       1   &       -1  &      1   &      -1   &      1       \\
 &    $E$        &       2   &        0  &      0   &       0   &      -2          
\\\hline\hline
  $C_{2\nu}$  &   $\Gamma$/Class      &     $I$       &   $C'_2$   & $C'_2\times\hat{\pi}$   &  $C_2\times\hat{\pi}$   \\  
    \hline
 &    $A_1$     &       1         &        1         &      1        &       1          \\
 &    $A_2$     &       1         &        1         &     -1        &      -1         \\ 
 &    $B_1$     &       1         &       -1         &      1        &      -1          \\
 &    $B_2$     &       1         &        -1        &     -1        &       1           
     \\  \hline\hline
 $C_{3\nu}$  &   $\Gamma$/Class      &     $I$       &   $2C_3$   & $3C'_2\times\hat{\pi}$    \\  
    \hline
 &    $A_1$        &       1         &        1         &      1         \\
 &    $A_2$        &       1         &        1         &     -1             \\ 
 &    $E$          &       2         &        -1         &      0           
\\  \hline\hline

\end{tabular}  \label{tab:charterp}
\end{center}
\end{table}

\subsection{The rotation operator in the little group}

The $O_h$ group is discussed in detail in the above sections, and it is the symmetry group in the rest frame, i.e., $\vec{p} = 0$.
In the nonzero momentum system, the symmetry group becomes the
subgroup of $O_h$, named as the little group. 
In each little group, the rotations satisfying $\hat{R}\vec{p} = \vec{p}$ will survive. 
Therefore, for the moving system, one just needs to keep the surviving
rotations and do the same procedure as that in the rest frame. 
All the rotations for different momentum with $|\vec{p}| = 1,2,3$ are listed in Table \ref{tab:momentum}.

\begin{table}[htbp]
\renewcommand\arraystretch{1.5}
\begin{center}\caption{The rotation for the each class in the $O_h$ group, $C_{4\nu}$, $C_{2\nu}$ and $C_{3\nu}$ for $|\vec{p}| = 0$, $1$, $2$ and $3$, respectively.
}
\begin{tabular}{cccccccccccc}\hline\hline
 $O_h$  & $\vec{p}$/Class      &     $I$\,\,                                    &   $8C'_3$\,\,    
                                    & $6C_4$\,\,                                 &  $6C'_4$\,\,   
                                    & $3C^2_4$\,\,                             &      $\hat{\pi}$\,\,       
                                    &   $8C'_3\times\hat{\pi}$\,\,        & $6C_4\times\hat{\pi}$\,\,   
                                    &  $6C'_4\times\hat{\pi}$\,\,         & $3C^2_4\times\hat{\pi}$\,\,  \\  
    \hline
 &    $(0,0,0)$                     &      $ \hat{R}_1          $           &       $\hat{R}_{2-9}       $ 
                                         &       $\hat{R}_{10-15}$          &       $\hat{R}_{16-21}   $       
                                         &       $\hat{R}_{22-24}$          &       $\hat{R}_{25}                            $ 
                                         &       $\hat{R}_{26-33}$          &       $\hat{R}_{33-38}   $       
                                         &       $\hat{R}_{39-44}$          &       $ \hat{R}_{45-48}  $   
\\            \hline\hline
$C_{4\nu}$  &    $\vec{p}$/Class\,\,\,\,\,\,
                                       &  $I$ \,\,      &   $2C_4$\,\,  
                                        & $2C'_2\times\hat{\pi}$ \,\,  &  $2C_2\times\hat{\pi}$\,\, 
                                         & $C_2$\,\, \\  
    \hline
 &    $(0,0,\pm 1)$             &      $ \hat{R}_1            $           &       $\hat{R}_{14,\,15}       $ 
                                         &       $\hat{R}_{42,\,43}$          &       $\hat{R}_{46,\,47}   $       
                                         &       $\hat{R}_{24}$        \\
  &   $(0,\pm 1, 0)$             &      $ \hat{R}_1            $           &       $\hat{R}_{12,\,13}       $ 
                                         &       $\hat{R}_{44,\,45}$          &       $\hat{R}_{46,\,48}   $       
                                         &       $\hat{R}_{23}$        \\
   &  $(\pm 1,0,0)$             &      $ \hat{R}_1            $           &       $\hat{R}_{10,\,11}       $ 
                                         &       $\hat{R}_{40,\,41}$          &       $\hat{R}_{47,\,48}   $       
                                         &       $\hat{R}_{22}$        \\
\hline\hline
 $C_{2\nu}$ &  $\vec{p}$/Class\,\,\,\,\,\,
                                       &  $I$ \,\,      &   $C'_2$\,\,  
                                        & $C'_2\times\hat{\pi}$ \,\,  &  $C_2\times\hat{\pi}$\,\, \\  
    \hline
  &   $(\pm 1,\pm 1, 0)$      &      $ \hat{R}_1            $           &       $\hat{R}_{18}       $ 
                                         &       $\hat{R}_{43}$                  &       $\hat{R}_{48}   $          \\
  &   $(\pm 1, 0, \pm 1)$     &      $ \hat{R}_1            $           &       $\hat{R}_{20}       $ 
                                         &       $\hat{R}_{45}$                  &       $\hat{R}_{47}   $         \\
  &   $(0,\pm 1,\pm 1)$       &      $ \hat{R}_1            $           &       $\hat{R}_{16}       $ 
                                         &       $\hat{R}_{41}$                  &       $\hat{R}_{46}   $        \\
  &   $(0,\pm 1,\mp 1)$       &      $ \hat{R}_1            $           &       $\hat{R}_{17}       $ 
                                         &       $\hat{R}_{40}$                  &       $\hat{R}_{46}   $        \\
  &   $(\pm 1,\mp 1, 0)$      &      $ \hat{R}_1            $           &       $\hat{R}_{21}       $ 
                                         &       $\hat{R}_{44}$                  &       $\hat{R}_{47}   $          \\
  &   $(\pm 1, 0, \mp 1)$     &      $ \hat{R}_1            $           &       $\hat{R}_{19}       $ 
                                         &       $\hat{R}_{42}$                  &       $\hat{R}_{48}   $         \\
  \hline\hline
   $C_{3\nu}$ &  $\vec{p}$/Class\,\,\,\,\,\,
                                       &  $I$ \,\,      &   $2C_3$\,\,  
                                        & $3C'_2\times\hat{\pi}$ \,\, \\  
    \hline
   &  $(\pm 1,\pm 1, \pm 1)$      &      $ \hat{R}_1            $           &       $\hat{R}_{2,\,3}       $ 
                                                &       $\hat{R}_{41,\,43,\,45}$                     \\
   &  $(\pm 1, \pm 1, \mp 1)$     &      $ \hat{R}_1            $           &       $\hat{R}_{6,\,7}       $ 
                                                &       $\hat{R}_{40,\,43,\,44}$                   \\
   &  $(\pm 1,\mp 1,\pm 1)$       &      $ \hat{R}_1            $           &       $\hat{R}_{8,\,9}       $ 
                                                &       $\hat{R}_{40,\,42,\,45}$                \\
   &  $(\mp 1,\pm 1,\pm 1)$       &      $ \hat{R}_1            $           &       $\hat{R}_{4,\,5}       $ 
                                                &       $\hat{R}_{41,\,42,\,44}$                 \\
 \hline\hline
\end{tabular}  \label{tab:momentum}
\end{center}
\end{table}



\section{L\"uscher's quantisation condition}
\label{app:luscher}

The L\"uscher formalism provides a model-independent relationship
between the phase shifts and the energy levels, assuming
exponentially-suppressed finite-volume effects can be safely neglected.
In this section, we give the relationship between the spectra of
irreps considered in this work and the phase shifts up to $\ell=4$. We
have confirmed that the partial waves $\ell=0$ and $\ell=2$ agree with
previous results reported in Ref.~\cite{Gockeler:2012yj}.
The general quantisation condition equation is summarised by:
\begin{eqnarray}
\det[M^{\Gamma, \vec{p}}_{ln,l'n'}(q(\Gamma))-\delta_{l,l'}\delta_{n,n'}\cot\delta_l(q(\Gamma))] = 0\,,
 \label{eq:momn0rel1}
\end{eqnarray}
where $\Gamma$, $\vec{p}$, $l$($l'$) and $n$($n'$) indicate the irrep,
total momentum, angular momentum and the $n$-th $\Gamma$ appearing in
the representation of this angular momentum, respectively.
$q(\Gamma)$ is the on-shell momentum of the energy level of irrep
$\Gamma$ in the C.M. system.

The matrix $M$ is calculated from:
\begin{align}
M^{\Gamma, \vec{p}}_{ln,l'n'}(q(\Gamma)) &= \sum_{m,m'}
C^{\Gamma,\,\alpha,n\, *}_{l,m} C^{\Gamma,\,\alpha,n'}_{l',m'}
M^{\vec{p}}_{lm,\,l'm'}(q(\Gamma))\,,
\label{eq:MtoM1}\\
 M^{\vec{p}}_{lm,\,l'm'}(q(\Gamma)) &= (-1)^{l}\sum^{l+l'}_{j=|l-l'|}\sum_{s=-j}^{j}
 i^{j}\sqrt{2j+1}\omega^{\vec{d}=\vec{p}L/2\pi}_{js}(\tilde{q}=q(\Gamma)L/2\pi)
 C_{lm,\,js,\,l'm'}, \label{eq:omegatoM1}\\
 \omega^{\vec{d}}_{js}(\tilde{q})&=
  \frac{1}{\pi^{3/2}\sqrt{2j+1}}
  \frac{Z^{\vec{d}}_{js}(1,\tilde{q})}{\gamma\,\tilde{q}^{j+1}}^{-1}\,,
  \label{eq:Mequation1}
\end{align}
where the index $\alpha$ is an index denoting the dimension of the irrep $\Gamma$. $\gamma$ is the Lorentz
factor
\begin{align}
\gamma &= \frac{W}{E_{C.M.}}=\frac{\sqrt{\vec{p}^2+E^2_{C.M.}}}{E_{C.M.}}\,, \label{eq:lorentzgamma}
\end{align}
where $E_{C.M.}=2\sqrt{q^2+m^2_\pi}$ is the energy level in the C.M. system.

The factor $C_{lm,\,js,\,l'm'}$ is related to the Wigner 3-j symbols as follows,
\begin{eqnarray}
 C_{lm,\,js,\,l'm'} &=&(-1)^{m'} i^{l-j+l'}\sqrt{(2l+1)(2j+1)(2l'+1)} 
 \left( \begin{array}{ccc}
  l & j & l' \\ 
 m& s &-m' \\
\end{array} \right)
 \left( \begin{array}{ccc}
  l & j & l' \\ 
 0& 0 &0 \\
\end{array} \right).  \label{eq:Ccoef}
\end{eqnarray}

Now we only need to know the coefficients
$C^{\Gamma,\,\alpha,n}_{l,m}$ in Eq.~(\ref{eq:MtoM1}). We give these
values in Table~\ref{tab:ccoefinM} for the moving system, while for the
rest frame, the matrices $M^{\Gamma, \vec{p}}_{ln,l'n'}(q(\Gamma))$
can be read from Ref.~\cite{Luscher:1990ux}.
It is worth mentioning that in our calculation we average over
all momenta with fixed $|\vec{p}|$. Since the spectra of them are the
same, we choose one case to list each
$C^{\Gamma,\,\alpha,n}_{l,m}$.
Finally, Eq.~(\ref{eq:momn0rel1}) for each case are listed in the following.

For the $A_1$ irrep in the rest frame, $\vec{d}=\vec{p}L/2\pi =
\vec{0}$,
\begin{eqnarray}
 0&=&\det
 \left( \begin{array}{ccc}
  -\cot\delta_0 + \omega^{\vec{d}}_{00} & \frac{6\sqrt{21}}{7}\omega^{\vec{d}}_{40}  \\ 
 \frac{6\sqrt{21}}{7}\omega^{\vec{d}}_{40}
 & -\cot\delta_4+\omega^{\vec{d}}_{00}+\frac{324}{143}\omega^{\vec{d}}_{40}
 +\frac{80}{11}\omega^{\vec{d}}_{60}+\frac{560}{143}\omega^{\vec{d}}_{80} \\
\end{array} \right).
\label{eq:p0A1}
\end{eqnarray}
For the $E$ irrep in the rest frame, $\vec{d}=\vec{0}$, 
\begin{eqnarray}
 0&=&\det
 \left( \begin{array}{ccc}
  -\cot\delta_2 + \omega^{\vec{d}}_{00}+\frac{18}{7}\omega^{\vec{d}}_{40} 
  & -\frac{120\sqrt{3}}{77}\omega^{\vec{d}}_{40}-\frac{30\sqrt{3}}{11}\omega^{\vec{d}}_{60}  \\ 
  -\frac{120\sqrt{3}}{77}\omega^{\vec{d}}_{40}-\frac{30\sqrt{3}}{11}\omega^{\vec{d}}_{60} 
 & -\cot\delta_4+\omega^{\vec{d}}_{00}+\frac{324}{1001}\omega^{\vec{d}}_{40}
 -\frac{64}{11}\omega^{\vec{d}}_{60}+\frac{392}{143}\omega^{\vec{d}}_{80} \\
\end{array} \right).
\label{eq:p0E}
\end{eqnarray}
For the $T_2$ irrep in the rest frame, $\vec{d}=\vec{0}$, 
\begin{eqnarray}
 0&=&\det
 \left( \begin{array}{ccc}
  -\cot\delta_2 + \omega^{\vec{d}}_{00}-\frac{12}{7}\omega^{\vec{d}}_{40} 
  & -\frac{60\sqrt{3}}{77}\omega^{\vec{d}}_{40}-\frac{40\sqrt{3}}{11}\omega^{\vec{d}}_{60}  \\ 
  -\frac{60\sqrt{3}}{77}\omega^{\vec{d}}_{40}-\frac{40\sqrt{3}}{11}\omega^{\vec{d}}_{60} 
 & -\cot\delta_4+\omega^{\vec{d}}_{00}-\frac{162}{77}\omega^{\vec{d}}_{40}
 +\frac{20}{11}\omega^{\vec{d}}_{60} \\
\end{array} \right).
\label{eq:p0T2}
\end{eqnarray}
%
For the $A_1$ irrep in the moving frame with $|\vec{p}L/2\pi|=1$, we
choose $\vec{d}=\vec{p}L/2\pi=(0,0,1)$,
\begin{eqnarray}
 0&=&\det
 \left( \begin{array}{cccc}
  -\cot\delta_0 + \omega^{\vec{d}}_{00} 
  & -\sqrt{5}\omega^{\vec{d}}_{20}  
  &\frac{3}{\sqrt{2}}\omega^{\vec{d}}_{40}+3\omega^{\vec{d}}_{44}
  &-\frac{3}{\sqrt{2}}\omega^{\vec{d}}_{40}+3\omega^{\vec{d}}_{44}  
  \\ 
  -\sqrt{5}\omega^{\vec{d}}_{20}
  &-\cot\delta_2+M^{A_1, \vec{p}}_{21,21}
 & M^{A_1, \vec{p}}_{21,41} & M^{A_1, \vec{p}}_{21,42}
 \\
   \frac{3}{\sqrt{2}}\omega^{\vec{d}}_{40}+3\omega^{\vec{d}}_{44}
  &M^{A_1, \vec{p}}_{21,41}
 & -\cot\delta_4+M^{A_1, \vec{p}}_{41,41}
 &M^{A_1, \vec{p}}_{41,42}
   \\ 
 -\frac{3}{\sqrt{2}}\omega^{\vec{d}}_{40}+3\omega^{\vec{d}}_{44} 
  &M^{A_1, \vec{p}}_{21,41}
 & M^{A_1, \vec{p}}_{41,42} & -\cot\delta_4+M^{A_1, \vec{p}}_{42,42}
\end{array} \right).
\label{eq:p1A1}
\end{eqnarray}
where
\begin{align}
 M^{A_1, \vec{p}}_{21,21}&=\omega^{\vec{d}}_{00}+\frac{10}{7}\omega^{\vec{d}}_{20}
 +\frac{18}{7}\omega^{\vec{d}}_{40}\,,\nonumber\\
M^{A_1, \vec{p}}_{21,41}&= -\frac{3\sqrt{10}}{7}\omega^{\vec{d}}_{20}
           +\frac{12\sqrt{5}}{11}\omega^{\vec{d}}_{44}
           -\frac{15}{11}\omega^{\vec{d}}_{64}
           -\frac{30\sqrt{10}}{77}\omega^{\vec{d}}_{40}
           -\frac{15\sqrt{10}}{22}\omega^{\vec{d}}_{60}\,,
           \nonumber\\
M^{A_1, \vec{p}}_{21,42}&= +\frac{3\sqrt{10}}{7}\omega^{\vec{d}}_{20}
           +\frac{12\sqrt{5}}{11}\omega^{\vec{d}}_{44}
           -\frac{15}{11}\omega^{\vec{d}}_{64}
           +\frac{30\sqrt{10}}{77}\omega^{\vec{d}}_{40}
           +\frac{15\sqrt{10}}{22}\omega^{\vec{d}}_{60}\,,
 \nonumber\\
  M^{A_1, \vec{p}}_{41,41}&= +\omega^{\vec{d}}_{00}
           -\frac{20}{77}\omega^{\vec{d}}_{20}
           +\frac{1296}{1001}\omega^{\vec{d}}_{40}
           +\frac{8}{11}\omega^{\vec{d}}_{60}
           +\frac{497}{286}\omega^{\vec{d}}_{80}
           +\frac{162\sqrt{2}}{143}\omega^{\vec{d}}_{44}
           \nonumber\\&\quad
           -\frac{12\sqrt{10}}{11} \omega^{\vec{d}}_{64}
           +\frac{42}{13}\sqrt{\frac{5}{22}}\omega^{\vec{d}}_{84}
           +21\sqrt{\frac{5}{286}}\omega^{\vec{d}}_{88}\,,
 \nonumber\\
M^{A_1, \vec{p}}_{41,42}&= -\frac{120}{77}\omega^{\vec{d}}_{20}
           -\frac{162}{1001}\omega^{\vec{d}}_{40}
           -\frac{12}{11}\omega^{\vec{d}}_{60}
           -\frac{483}{286}\omega^{\vec{d}}_{80}
           +21\sqrt{\frac{5}{286}}\omega^{\vec{d}}_{88}\,,
 \nonumber\\
 M^{A_1, \vec{p}}_{42,42}&= +\omega^{\vec{d}}_{00}
           -\frac{20}{77}\omega^{\vec{d}}_{20}
           +\frac{1296}{1001}\omega^{\vec{d}}_{40}
           +\frac{8}{11}\omega^{\vec{d}}_{60}
           +\frac{497}{286}\omega^{\vec{d}}_{80}
           -\frac{162\sqrt{2}}{143}\omega^{\vec{d}}_{44}
           \nonumber\\&\quad
           +\frac{12\sqrt{10}}{11} \omega^{\vec{d}}_{64}
           -\frac{42}{13}\sqrt{\frac{5}{22}}\omega^{\vec{d}}_{84}
           +21\sqrt{\frac{5}{286}}\omega^{\vec{d}}_{88}\,.
 \nonumber
\end{align}
For the $B_1$ irrep in the moving frame with $|\vec{p}L/2\pi|=1$, we
choose $\vec{d}=\vec{p}L/2\pi=(0,0,1)$,
\begin{eqnarray}
 0&=&\det
 \left( \begin{array}{cc}
  -\cot\delta_2 + \omega^{\vec{d}}_{00}
  -\frac{10}{7}\omega^{\vec{d}}_{00}
  +\frac{3}{7}\omega^{\vec{d}}_{40}
  +6\sqrt{\frac{5}{14}}\omega^{\vec{d}}_{44}
  & M^{B_1, \vec{p}}_{21,41}  
  \\ 
 M^{B_1, \vec{p}}_{21,41}  & -\cot\delta_4 +M^{B_1, \vec{p}}_{41,41} 
 \\
\end{array} \right).
\label{eq:p1B1}
\end{eqnarray}
where
\begin{align}
M^{B_1, \vec{p}}_{21,41}
       &= -\frac{5\sqrt{3}}{7}\omega^{\vec{d}}_{20}
           +\frac{90\sqrt{3}}{77}\omega^{\vec{d}}_{40}
           -\frac{5\sqrt{3}}{11}\omega^{\vec{d}}_{60}
           +\frac{6\sqrt{210}}{77}\omega^{\vec{d}}_{44}
           -\frac{5\sqrt{42}}{11}\omega^{\vec{d}}_{64}\,,
\nonumber\\
 M^{B_1, \vec{p}}_{41,41}
       &= +\omega^{\vec{d}}_{00}
           +\frac{40}{77}\omega^{\vec{d}}_{20}
           -\frac{81}{91}\omega^{\vec{d}}_{40}
           -2\omega^{\vec{d}}_{60}
           +\frac{196}{143}\omega^{\vec{d}}_{80}
           +\frac{243}{143}\sqrt{\frac{10}{7}}\omega^{\vec{d}}_{44}
           \nonumber\\&\quad
           +\frac{6\sqrt{14}}{11} \omega^{\vec{d}}_{64}
           +\frac{42}{13}\sqrt{\frac{14}{11}}\omega^{\vec{d}}_{84}\,.
 \nonumber
\end{align}
%
For the $B_2$ irrep in the moving frame with $|\vec{p}L/2\pi|=1$, we
choose $\vec{d}=\vec{p}L/2\pi=(0,0,1)$,
\begin{align}
 0&=\det
 \left( \begin{array}{cc}
  -\cot\delta_2 + \omega^{\vec{d}}_{00}
  -\frac{10}{7}\omega^{\vec{d}}_{00}
  +\frac{3}{7}\omega^{\vec{d}}_{40}
  -\frac{3\sqrt{70}}{7}\omega^{\vec{d}}_{44}
  & M^{B_2, \vec{p}}_{21,41}  
  \\ 
 M^{B_2, \vec{p}}_{21,41}  & -\cot\delta_4 +M^{B_2, \vec{p}}_{41,41} 
 \\
\end{array} \right).
\label{eq:p1B2}
\end{align}
where
\begin{align}
M^{B_2, \vec{p}}_{21,41}
       &= +\frac{5\sqrt{3}}{7}\omega^{\vec{d}}_{20}
           -\frac{90\sqrt{3}}{77}\omega^{\vec{d}}_{40}
           +\frac{5\sqrt{3}}{11}\omega^{\vec{d}}_{60}
           +\frac{6\sqrt{210}}{77}\omega^{\vec{d}}_{44}
           -\frac{5\sqrt{42}}{11}\omega^{\vec{d}}_{64}\,,
\nonumber\\
 M^{B_2, \vec{p}}_{41,41}
       &= +\omega^{\vec{d}}_{00}
           +\frac{40}{77}\omega^{\vec{d}}_{20}
           -\frac{81}{91}\omega^{\vec{d}}_{40}
           -2\omega^{\vec{d}}_{60}
           +\frac{196}{143}\omega^{\vec{d}}_{80}
           -\frac{243}{143}\sqrt{\frac{10}{7}}\omega^{\vec{d}}_{44}
           \nonumber\\&\quad
           -\frac{6\sqrt{14}}{11} \omega^{\vec{d}}_{64}
           -\frac{42}{13}\sqrt{\frac{14}{11}}\omega^{\vec{d}}_{84}\,.
 \nonumber
\end{align}
%
For the $E$ irrep in the moving frame with $|\vec{p}L/2\pi|=1$, we
choose $\vec{d}=\vec{p}L/2\pi=(0,0,1)$,
\begin{align}
 0&=\det
 \left( \begin{array}{ccc}
  -\cot\delta_2 + \omega^{\vec{d}}_{00}
  +\frac{5}{7}\omega^{\vec{d}}_{20}
  -\frac{12}{7}\omega^{\vec{d}}_{40}
  & M^{E, \vec{p}}_{21,41}  
  & M^{E, \vec{p}\,*}_{21,41}  
  \\ 
 M^{E, \vec{p}\,*}_{21,41}    
 & -\cot\delta_4 +M^{E, \vec{p}}_{41,41} 
 & M^{E, \vec{p}}_{41,42} 
   \\ 
 M^{E, \vec{p}}_{21,41} 
 & M^{E, \vec{p}\,*}_{41,42}    
 & -\cot\delta_4 +M^{E, \vec{p}}_{41,41} 
\end{array} \right).
\label{eq:p1E}
\end{align}
where
\begin{eqnarray}
M^{E, \vec{p}}_{21,41}
       &=& -\frac{5\sqrt{3}}{7}\omega^{\vec{d}}_{20}
           -\frac{15\sqrt{3}}{77}\omega^{\vec{d}}_{40}
           +\frac{10\sqrt{3}}{11}\omega^{\vec{d}}_{60}
           -i\frac{3\sqrt{30}}{11}\omega^{\vec{d}}_{44}
           -i\frac{10\sqrt{6}}{11}\omega^{\vec{d}}_{64}\,,
\nonumber\\
 M^{E, \vec{p}}_{41,41}
       &=& +\omega^{\vec{d}}_{00}
           +\frac{25}{77}\omega^{\vec{d}}_{20}
           -\frac{486}{1001}\omega^{\vec{d}}_{40}
           +\frac{8}{11}\omega^{\vec{d}}_{60}
           -\frac{224}{143}\omega^{\vec{d}}_{80}\,,
 \nonumber\\
M^{E, \vec{p}}_{41,42}
       &=& +\frac{60}{77}\omega^{\vec{d}}_{20}
           +\frac{1215}{1001}\omega^{\vec{d}}_{40}
           -\frac{9}{11}\omega^{\vec{d}}_{60}
           -\frac{168}{143}\omega^{\vec{d}}_{80}
           +i\frac{81\sqrt{10}}{143}\omega^{\vec{d}}_{44}
           +i\frac{3\sqrt{2}}{11}\omega^{\vec{d}}_{64}
           -i\frac{84}{13}\sqrt{\frac{2}{11}}\omega^{\vec{d}}_{84}\,.
 \nonumber
\end{eqnarray}
%
For the $A_1$ irrep in the moving frame with
$|\vec{p}L/2\pi|=\sqrt{2}$, 
we choose $\vec{d}=\vec{p}L/2\pi=(1,1,0)$,
\begin{align}
 0=\det
 \left( 
 \setlength{\arraycolsep}{0.02pt}
 \begin{array}{cccccc}
  -\cot\delta_0 + \omega^{\vec{d}}_{00}   
  & -\sqrt{5}\omega^{\vec{d}}_{20}
  &  \sqrt{10}\omega^{\vec{d}}_{22}
 &3(\omega^{\vec{d}}_{44}+ \omega^{\vec{d}}_{42})
 &3( \omega^{\vec{d}}_{44}-\omega^{\vec{d}}_{42})
 &3\omega^{\vec{d}}_{40}
 \\
   -\sqrt{5}\omega^{\vec{d}}_{20}
  &-\cot\delta_2 + M^{A_1, \vec{p}}_{21,21}
 &M^{A_1, \vec{p}}_{21,22}
 &M^{A_1, \vec{p}}_{21,41}
 &M^{A_1, \vec{p}\,*}_{21,41}
 &M^{A_1, \vec{p}}_{21,43}
   \\ 
 -\sqrt{10}\omega^{\vec{d}}_{22}
  &M^{A_1, \vec{p}\,*}_{21,22}
 & M^{A_1, \vec{p}}_{22,22}-\cot\delta_2 
 & M^{A_1, \vec{p}}_{22,41}
 & -M^{A_1, \vec{p}\,*}_{22,41}
 & M^{A_1, \vec{p}}_{22,43}
    \\ 
 3( \omega^{\vec{d}}_{44}-\omega^{\vec{d}}_{42})
  &M^{A_1, \vec{p}\,*}_{21,41}
 & M^{A_1, \vec{p}\,*}_{22,41}
 & M^{A_1, \vec{p}}_{41,41}-\cot\delta_4 
 & M^{A_1, \vec{p}}_{41,42}
 & M^{A_1, \vec{p}}_{41,43}
    \\ 
 3( \omega^{\vec{d}}_{44}+\omega^{\vec{d}}_{42})
  &M^{A_1, \vec{p}}_{21,41}
 & -M^{A_1, \vec{p}}_{22,41}
 & M^{A_1, \vec{p}\,*}_{41,42}
 & M^{A_1, \vec{p}}_{41,41}-\cot\delta_4 
 & M^{A_1, \vec{p}\,*}_{41,43}
     \\ 
 3\omega^{\vec{d}}_{40}
  &M^{A_1, \vec{p}\,*}_{21,43}
 & M^{A_1, \vec{p}\,*}_{22,43}
 & M^{A_1, \vec{p}\,*}_{41,43}
 & M^{A_1, \vec{p}}_{41,43}
 & M^{A_1, \vec{p}}_{43,43}-\cot\delta_4 
\end{array} \right),
\nonumber\\
\label{eq:p2A1}
\end{align}
where
\begin{align}
 M^{A_1, \vec{p}}_{21,21}&=\omega^{\vec{d}}_{00}+\frac{10}{7}\omega^{\vec{d}}_{20}
 +\frac{18}{7}\omega^{\vec{d}}_{40}\,,
 \nonumber\\
M^{A_1, \vec{p}}_{21,22}&= \frac{10\sqrt{2}}{7}\omega^{\vec{d}}_{22}
           -\frac{3\sqrt{30}}{7}\omega^{\vec{d}}_{42}\,,
\nonumber\\
M^{A_1, \vec{p}}_{21,41}&= 
           -\frac{5\sqrt{10}}{7}\omega^{\vec{d}}_{22}
           -\frac{24\sqrt{5}}{77}\omega^{\vec{d}}_{42}
           -\frac{2\sqrt{210}}{11}\omega^{\vec{d}}_{62}
           +\frac{12\sqrt{5}}{11}\omega^{\vec{d}}_{44}
           -\frac{15}{11}\omega^{\vec{d}}_{64}\,,
 \nonumber\\
M^{A_1, \vec{p}}_{21,43}&= 
           -\frac{6\sqrt{5}}{77}\omega^{\vec{d}}_{20}
           -\frac{60\sqrt{5}}{77}\omega^{\vec{d}}_{40}
           -\frac{15}{11}\omega^{\vec{d}}_{60}\,,
 \nonumber\\
 M^{A_1, \vec{p}}_{22,22}&=\omega^{\vec{d}}_{00}-\frac{10}{7}\omega^{\vec{d}}_{20}
 +\frac{3}{7}\omega^{\vec{d}}_{40}-6\sqrt{\frac{5}{14}}\omega^{\vec{d}}_{44}\,,
 \nonumber\\
M^{A_1, \vec{p}}_{22,41}&= 
           +\frac{5\sqrt{6}}{14}\omega^{\vec{d}}_{20}
           -\frac{45\sqrt{6}}{77}\omega^{\vec{d}}_{40}
           +\frac{5\sqrt{6}}{22}\omega^{\vec{d}}_{60}
           +\frac{5\sqrt{7}}{7}\omega^{\vec{d}}_{22}
           -\frac{6\sqrt{105}}{77}\omega^{\vec{d}}_{42}
           +\frac{\sqrt{10}}{22}\omega^{\vec{d}}_{62}
           \nonumber\\&\quad
           +\frac{6\sqrt{105}}{77}\omega^{\vec{d}}_{44}
           -\frac{5\sqrt{21}}{11}\omega^{\vec{d}}_{64}
           -\frac{15}{\sqrt{22}}\omega^{\vec{d}}_{66}\,,
\nonumber\\
M^{A_1, \vec{p}}_{22,43}&= 
           -\frac{\sqrt{10}}{7}\omega^{\vec{d}}_{22}
           +\frac{90\sqrt{6}}{77}\omega^{\vec{d}}_{42}
           -\frac{10\sqrt{7}}{11}\omega^{\vec{d}}_{62}\,,
 \nonumber\\
 M^{A_1, \vec{p}}_{41,41}&=
                      \omega^{\vec{d}}_{00}
         -\frac{50}{77}\omega^{\vec{d}}_{20}
         +\frac{243}{2002}\omega^{\vec{d}}_{40}
         -\frac{13}{11}\omega^{\vec{d}}_{60}
         +\frac{203}{286}\omega^{\vec{d}}_{80}
         -\frac{243}{143}\sqrt{\frac{5}{14}}\omega^{\vec{d}}_{44}
           \nonumber\\&\quad
         -\frac{3\sqrt{14}}{11}\omega^{\vec{d}}_{64}
         -\frac{42}{13}\sqrt{\frac{7}{22}}\omega^{\vec{d}}_{84}
         +21\sqrt{\frac{5}{286}}\omega^{\vec{d}}_{88} \,,        
 \nonumber\\
M^{A_1, \vec{p}}_{41,42}&= 
           -\frac{90}{77}\omega^{\vec{d}}_{20}
           +\frac{2025}{2002}\omega^{\vec{d}}_{40}
           +\frac{9}{11}\omega^{\vec{d}}_{60}
           -\frac{189}{286}\omega^{\vec{d}}_{80}
           -\frac{10}{11}\sqrt{\frac{6}{7}}\omega^{\vec{d}}_{22}
           \nonumber\\&\quad
           +\frac{243}{143}\sqrt{\frac{10}{7}}\omega^{\vec{d}}_{42}
           -\frac{4\sqrt{15}}{11}\omega^{\vec{d}}_{62}
           +\frac{21\sqrt{5}}{143}\omega^{\vec{d}}_{82}
           +\frac{243}{143}\sqrt{\frac{5}{14}}\omega^{\vec{d}}_{44}
           +\frac{3\sqrt{14}}{11}\omega^{\vec{d}}_{64},
           \nonumber\\&\quad
           +\frac{42}{13}\sqrt{\frac{7}{22}}\omega^{\vec{d}}_{84}
           +4\sqrt{\frac{3}{11}}\omega^{\vec{d}}_{66}
           -7\sqrt{\frac{21}{143}}\omega^{\vec{d}}_{86}
           +21\sqrt{\frac{5}{286}}\omega^{\vec{d}}_{88}\,,
 \nonumber\\
 M^{A_1, \vec{p}}_{41,43}&=
           +\frac{30\sqrt{15}}{77}\omega^{\vec{d}}_{22}
           +\frac{81}{91}\omega^{\vec{d}}_{42}
           -\frac{105\sqrt{14}}{143}\omega^{\vec{d}}_{82}
           +\frac{162}{143}\omega^{\vec{d}}_{44}
           -\frac{12\sqrt{5}}{11}\omega^{\vec{d}}_{64}
           +\frac{21}{13}\sqrt{\frac{5}{11}}\omega^{\vec{d}}_{84}\,,
\nonumber\\           
M^{A_1, \vec{p}}_{43,43}&=
                      \omega^{\vec{d}}_{00}
         +\frac{100}{77}\omega^{\vec{d}}_{20}
         +\frac{1458}{1001}\omega^{\vec{d}}_{40}
         +\frac{20}{11}\omega^{\vec{d}}_{60}
         +\frac{490}{143}\omega^{\vec{d}}_{80}\,.
 \nonumber
\end{align}
%
%
For the $A_2$ irrep in moving frame with $|\vec{p}L/2\pi|=\sqrt{2}$, we choose $\vec{d}=\vec{p}L/2\pi=(1,1,0)$, 
\begin{eqnarray}
 0&=&\det
 \left( \begin{array}{ccc}
  -\cot\delta_0 + M^{A_2, \vec{p}}_{21,21}
  &
   M^{A_2, \vec{p}}_{21,41}
  &
   -M^{A_2, \vec{p}\,*}_{21,41}
 \\
    M^{A_2, \vec{p}\,*}_{21,41}
  &-\cot\delta_4 + M^{A_2, \vec{p}}_{41,41}
 & M^{A_2, \vec{p}}_{41,42}
  \\ 
  -M^{A_2, \vec{p}}_{21,41}
  &M^{A_2, \vec{p}\,*}_{41,42}
 & -\cot\delta_4+ M^{A_2, \vec{p}}_{41,41}
\end{array} \right),
\label{eq:p2A2}
\end{eqnarray}
where
\begin{eqnarray}
 M^{A_2, \vec{p}}_{21,21}&=&
            \omega^{\vec{d}}_{00}+\frac{5}{7}\omega^{\vec{d}}_{20}
 -\frac{12}{7}\omega^{\vec{d}}_{40}
 +i\frac{\sqrt{30}}{7}\left(\sqrt{5}\omega^{\vec{d}}_{22}
                     +2\sqrt{3}\omega^{\vec{d}}_{42}\right),
 \nonumber\\
 M^{A_2, \vec{p}}_{21,41}&=& 
           -\frac{5\sqrt{3}}{7}\omega^{\vec{d}}_{20}
           -\frac{15\sqrt{3}}{77}\omega^{\vec{d}}_{40}
           +\frac{10\sqrt{3}}{11}\omega^{\vec{d}}_{60}
           +\frac{5\sqrt{2}}{14}\left(i+\sqrt{7}\right)\omega^{\vec{d}}_{22}
           \nonumber\\&&
           +\frac{3}{11}\sqrt{\frac{15}{14}}\left(5-i\frac{9}{\sqrt{7}}\right)\omega^{\vec{d}}_{42}
           -\frac{4\sqrt{5}}{11}\left(1+i\sqrt{7}\right)\omega^{\vec{d}}_{62}
           +i\frac{3\sqrt{30}}{11}\omega^{\vec{d}}_{44}
           +i\frac{10\sqrt{6}}{11}\omega^{\vec{d}}_{64},
 \nonumber\\
 M^{A_2, \vec{p}}_{41,41}&=& +\omega^{\vec{d}}_{00}
           +\frac{25}{77}\omega^{\vec{d}}_{20}
           -\frac{486}{1001}\omega^{\vec{d}}_{40}
           +\frac{8}{11}\omega^{\vec{d}}_{60}
           -\frac{224}{143}\omega^{\vec{d}}_{80}
           +i\frac{25\sqrt{6}}{77}\omega^{\vec{d}}_{22}
           \nonumber\\&&
           +i\frac{243\sqrt{10}}{1001} \omega^{\vec{d}}_{42}
           +i\frac{\sqrt{105}}{11}\omega^{\vec{d}}_{62}
           +i\frac{42\sqrt{35}}{143}\omega^{\vec{d}}_{82}
           -i\sqrt{\frac{21}{11}}\omega^{\vec{d}}_{66}
           -i14\sqrt{\frac{3}{143}}\omega^{\vec{d}}_{86},
 \nonumber\\
 M^{A_2, \vec{p}}_{41,42}&=& 
           -\frac{60}{77}\omega^{\vec{d}}_{20}
           -\frac{1215}{1001}\omega^{\vec{d}}_{40}
           +\frac{9}{11}\omega^{\vec{d}}_{60}
           +\frac{168}{143}\omega^{\vec{d}}_{80}
           +\frac{5\sqrt{6}}{77}\left(3\sqrt{7}-i5\right)\omega^{\vec{d}}_{22}
           \nonumber\\&&
           +\frac{81\sqrt{10}}{1001}\left(-\sqrt{7}-i3\right)\omega^{\vec{d}}_{42}
           +\frac{\sqrt{15}}{11}\left(-6-i\sqrt{7}\right)\omega^{\vec{d}}_{62}
           +\frac{42\sqrt{5}}{143}\left(2-i\sqrt{7}\right)\omega^{\vec{d}}_{82}
           \nonumber\\&&
           +i\frac{81\sqrt{10}}{143}\omega^{\vec{d}}_{44}
           +i\frac{3\sqrt{2}}{11}\omega^{\vec{d}}_{64}
           -i\frac{84}{13}\sqrt{\frac{2}{11}}\omega^{\vec{d}}_{84}
           -i\sqrt{\frac{21}{11}}\omega^{\vec{d}}_{66}
           -i14\sqrt{\frac{3}{143}}\omega^{\vec{d}}_{86}.
 \nonumber
\end{eqnarray}
%
%
For the $B_1$ irrep in the moving frame with
$|\vec{p}L/2\pi|=\sqrt{2}$, we choose $\vec{d}=\vec{p}L/2\pi=(1,1,0)$,
\begin{eqnarray}
 0&=&\det
 \left( \begin{array}{ccc}
  -\cot\delta_0 + M^{B_1, \vec{p}}_{21,21}
  &
   M^{B_1, \vec{p}}_{21,41}
  &
  -M^{B_1, \vec{p}\,*}_{21,41}
 \\
    M^{B_1, \vec{p}\,*}_{21,41}
  &-\cot\delta_4 + M^{B_1, \vec{p}}_{41,41}
 & M^{B_1, \vec{p}}_{41,42}
  \\ 
  -M^{B_1, \vec{p}}_{21,41}
  &M^{B_1, \vec{p}\,*}_{41,42}
 & -\cot\delta_4+ M^{B_1, \vec{p}}_{41,41}
\end{array} \right),
\label{eq:p2B1}
\end{eqnarray}
where
\begin{eqnarray}
 M^{B_1, \vec{p}}_{21,21}&=&
            \omega^{\vec{d}}_{00}+\frac{5}{7}\omega^{\vec{d}}_{20}
 -\frac{12}{7}\omega^{\vec{d}}_{40}
 -i\frac{\sqrt{30}}{7}\left(\sqrt{5}\omega^{\vec{d}}_{22}
                     +2\sqrt{3}\omega^{\vec{d}}_{42}\right)
 \nonumber\\
 M^{B_1, \vec{p}}_{21,41}&=& 
           -\frac{5\sqrt{3}}{7}\omega^{\vec{d}}_{20}
           -\frac{15\sqrt{3}}{77}\omega^{\vec{d}}_{40}
           +\frac{10\sqrt{3}}{11}\omega^{\vec{d}}_{60}
           +\frac{5\sqrt{2}}{14}\left(-i+\sqrt{7}\right)\omega^{\vec{d}}_{22}
           \nonumber\\&&
           +\frac{3}{11}\sqrt{\frac{15}{14}}\left(5+i\frac{9}{\sqrt{7}}\right)\omega^{\vec{d}}_{42}
           -\frac{4\sqrt{5}}{11}\left(1-i\sqrt{7}\right)\omega^{\vec{d}}_{62}
           -i\frac{3\sqrt{30}}{11}\omega^{\vec{d}}_{44}
           -i\frac{10\sqrt{6}}{11}\omega^{\vec{d}}_{64}\,,
\nonumber\\
 M^{B_1, \vec{p}}_{41,41}&=& +\omega^{\vec{d}}_{00}
           +\frac{25}{77}\omega^{\vec{d}}_{20}
           -\frac{486}{1001}\omega^{\vec{d}}_{40}
           +\frac{8}{11}\omega^{\vec{d}}_{60}
           -\frac{224}{143}\omega^{\vec{d}}_{80}
           -i\frac{25\sqrt{6}}{77}\omega^{\vec{d}}_{22}
           \nonumber\\&&
           -i\frac{243\sqrt{10}}{1001} \omega^{\vec{d}}_{42}
           -i\frac{\sqrt{105}}{11}\omega^{\vec{d}}_{62}
           -i\frac{42\sqrt{35}}{143}\omega^{\vec{d}}_{82}
           +i\sqrt{\frac{21}{11}}\omega^{\vec{d}}_{66}
           +i14\sqrt{\frac{3}{143}}\omega^{\vec{d}}_{86}\,,
 \nonumber\\
 M^{B_1, \vec{p}}_{41,42}&=& 
           -\frac{60}{77}\omega^{\vec{d}}_{20}
           -\frac{1215}{1001}\omega^{\vec{d}}_{40}
           +\frac{9}{11}\omega^{\vec{d}}_{60}
           +\frac{168}{143}\omega^{\vec{d}}_{80}
           +\frac{5\sqrt{6}}{77}\left(3\sqrt{7}+i5\right)\omega^{\vec{d}}_{22}
           \nonumber\\&&
           +\frac{81\sqrt{10}}{1001}\left(-\sqrt{7}+i3\right)\omega^{\vec{d}}_{42}
           +\frac{\sqrt{15}}{11}\left(-6+i\sqrt{7}\right)\omega^{\vec{d}}_{62}
           +\frac{42\sqrt{5}}{143}\left(2+i\sqrt{7}\right)\omega^{\vec{d}}_{82}
           \nonumber\\&&
           -i\frac{81\sqrt{10}}{143}\omega^{\vec{d}}_{44}
           -i\frac{3\sqrt{2}}{11}\omega^{\vec{d}}_{64}
           +i\frac{84}{13}\sqrt{\frac{2}{11}}\omega^{\vec{d}}_{84}
           +i\sqrt{\frac{21}{11}}\omega^{\vec{d}}_{66}
           +i14\sqrt{\frac{3}{143}}\omega^{\vec{d}}_{86}\,.
 \nonumber
\end{eqnarray}
%
%
For the $B_2$ irrep in the moving frame with
$|\vec{p}L/2\pi|=\sqrt{2}$, we choose $\vec{d}=\vec{p}L/2\pi=(1,1,0)$,
\begin{eqnarray}
 0&=&\det
 \left( \begin{array}{ccc}
  -\cot\delta_0 + M^{B_2, \vec{p}}_{21,21}
  &
   M^{B_2, \vec{p}}_{21,41}
  &
   M^{B_2, \vec{p}\,*}_{21,41}
 \\
    M^{B_2, \vec{p}\,*}_{21,41}
  &-\cot\delta_4 + M^{B_2, \vec{p}}_{41,41}
 & M^{B_2, \vec{p}}_{41,42}
  \\ 
  M^{B_2, \vec{p}}_{21,41}
  &M^{B_2, \vec{p}\,*}_{41,42}
 & -\cot\delta_4+ M^{B_2, \vec{p}}_{41,41}
\end{array} \right),
\label{eq:p2B1}
\end{eqnarray}
where
\begin{eqnarray}
 M^{B_2, \vec{p}}_{21,21}&=&
            \omega^{\vec{d}}_{00}
            -\frac{10}{7}\omega^{\vec{d}}_{20}
            +\frac{3}{7}\omega^{\vec{d}}_{40}
            +6\sqrt{\frac{5}{14}}\omega^{\vec{d}}_{44}\,,
 \nonumber\\
 M^{B_2, \vec{p}}_{21,41}&=& 
           -\frac{5\sqrt{6}}{14}\omega^{\vec{d}}_{20}
           +\frac{45\sqrt{6}}{77}\omega^{\vec{d}}_{40}
           -\frac{5\sqrt{6}}{22}\omega^{\vec{d}}_{60}
           -\frac{5}{\sqrt{7}}\omega^{\vec{d}}_{22}
           \nonumber\\&&
           +\frac{6\sqrt{105}}{77}\omega^{\vec{d}}_{42}
           -\frac{\sqrt{10}}{22}\omega^{\vec{d}}_{62}
           +\frac{6\sqrt{105}}{77}\omega^{\vec{d}}_{44}
           -\frac{5\sqrt{21}}{11}\omega^{\vec{d}}_{64}
           -\frac{15}{\sqrt{22}}\omega^{\vec{d}}_{66}\,,
\nonumber\\
 M^{B_2, \vec{p}}_{41,41}&=& +\omega^{\vec{d}}_{00}
           -\frac{50}{77}\omega^{\vec{d}}_{20}
           +\frac{243}{2002}\omega^{\vec{d}}_{40}
           -\frac{13}{11}\omega^{\vec{d}}_{60}
           +\frac{203}{286}\omega^{\vec{d}}_{80}
           \nonumber\\&&
           +\frac{243}{143}\sqrt{\frac{5}{14}}\omega^{\vec{d}}_{44}
           +\frac{3\sqrt{14}}{11}\omega^{\vec{d}}_{64}
           +\frac{42}{13}\sqrt{\frac{7}{22}}\omega^{\vec{d}}_{84}
           -21\sqrt{\frac{5}{286}}\omega^{\vec{d}}_{88}\,,
 \nonumber\\
 M^{B_2, \vec{p}}_{41,42}&=& 
           +\frac{90}{77}\omega^{\vec{d}}_{20}
           -\frac{2025}{2002}\omega^{\vec{d}}_{40}
           -\frac{9}{11}\omega^{\vec{d}}_{60}
           +\frac{189}{286}\omega^{\vec{d}}_{80}
           +\frac{10}{11}\sqrt{\frac{6}{7}}\omega^{\vec{d}}_{22}
           -\frac{243}{143}\sqrt{\frac{10}{7}}\omega^{\vec{d}}_{42}
           \nonumber\\&&
           +\frac{4\sqrt{5}}{11}\omega^{\vec{d}}_{62}
           -\frac{21\sqrt{5}}{143}\omega^{\vec{d}}_{82}
           +\frac{243}{143}\sqrt{\frac{5}{14}}\omega^{\vec{d}}_{44}
           +\frac{3\sqrt{14}}{11}\omega^{\vec{d}}_{64}
           \nonumber\\&&
           +\frac{42}{13}\sqrt{\frac{7}{22}}\omega^{\vec{d}}_{84}
           +4\sqrt{\frac{3}{11}}\omega^{\vec{d}}_{66}
           -7\sqrt{\frac{21}{143}}\omega^{\vec{d}}_{86}
           +21\sqrt{\frac{5}{286}}\omega^{\vec{d}}_{88}\,.
 \nonumber
\end{eqnarray}
%
%
For the $A_1$ irrep in the moving frame with $|\vec{p}L/2\pi|=\sqrt{3}$, we
choose $\vec{d}=\vec{p}L/2\pi=(1,1,1)$,
\begin{eqnarray}
 0&=&\det
 \left( \begin{array}{cccc}
  -\cot\delta_0 + \omega^{\vec{d}}_{00}   
  & \sqrt{30}\omega^{\vec{d}}_{22}
  & \frac{6\sqrt{21}}{7}\omega^{\vec{d}}_{40}
 &3\sqrt{3}(1-i)\omega^{\vec{d}}_{42}
  \\
   -\sqrt{30}\omega^{\vec{d}}_{22}
  &-\cot\delta_2 + M^{A_1, \vec{p}}_{21,21}
 & M^{A_1, \vec{p}}_{21,41}
 &M^{A_1, \vec{p}}_{21,42}
   \\ 
 \frac{6\sqrt{21}}{7}\omega^{\vec{d}}_{40}
  &M^{A_1, \vec{p}\,*}_{21,41}
 & M^{A_1, \vec{p}}_{41,41}-\cot\delta_2 
 & M^{A_1, \vec{p}}_{41,42}
    \\ 
 3\sqrt{3}(-1-i)\omega^{\vec{d}}_{42}
  &M^{A_1, \vec{p}\,*}_{21,42}
 & M^{A_1, \vec{p}\,*}_{41,42}
 & M^{A_1, \vec{p}}_{42,42}-\cot\delta_4 
\end{array} \right),
\nonumber\\
\label{eq:p3A1}
\end{eqnarray}
where
\begin{eqnarray}
 M^{A_1, \vec{p}}_{21,21}&=&\omega^{\vec{d}}_{00}
 -\frac{12}{7}\omega^{\vec{d}}_{40}
 -i\frac{10\sqrt{6}}{7}\omega^{\vec{d}}_{22}
 -i\frac{12\sqrt{10}}{7}\omega^{\vec{d}}_{42}\,,
 \nonumber\\
M^{A_1, \vec{p}}_{21,41}&=& 
           +2\sqrt{\frac{10}{7}}\omega^{\vec{d}}_{22}
           +\frac{30}{11}\sqrt{\frac{6}{7}}\omega^{\vec{d}}_{42}
           -\frac{260}{99}(1-i)\omega^{\vec{d}}_{63}
           -\frac{100}{9}\sqrt{\frac{5}{11}}\omega^{\vec{d}}_{66}\,,
 \nonumber\\
 M^{A_1, \vec{p}}_{21,42}&=& 
           -\frac{20}{77}(1+i)\omega^{\vec{d}}_{22}
           -\frac{30\sqrt{6}}{77}(1-i)\omega^{\vec{d}}_{40}
           -\frac{39\sqrt{15}}{77}(1+i)\omega^{\vec{d}}_{42}
           +\frac{20\sqrt{6}}{11}(1-i)\omega^{\vec{d}}_{60}
           \nonumber\\&&
           +\frac{64\sqrt{70}}{99}\omega^{\vec{d}}_{63}
           +\frac{20}{9}\sqrt{\frac{14}{11}}(1+i)\omega^{\vec{d}}_{66}\,,
 \nonumber\\
 M^{A_1, \vec{p}}_{41,41}&=&
         \omega^{\vec{d}}_{00}
         +\frac{324}{143}\omega^{\vec{d}}_{40}
         +\frac{80}{11}\omega^{\vec{d}}_{60}
         +\frac{560}{143}\omega^{\vec{d}}_{80}\,,
 \nonumber\\
M^{A_1, \vec{p}}_{41,42}&=& 
           -\frac{10}{11}\sqrt{\frac{15}{7}}(1-i)\omega^{\vec{d}}_{22}
           -\frac{81}{11}\sqrt{\frac{1}{7}}(1-i)\omega^{\vec{d}}_{42}
           -i\frac{80}{33}\sqrt{\frac{2}{3}}\omega^{\vec{d}}_{63}
           -\frac{4}{3}\sqrt{\frac{10}{33}}(1-i)\omega^{\vec{d}}_{66}
           \nonumber\\&&
           +\frac{8}{11}\sqrt{\frac{210}{143}}(1-i)\omega^{\vec{d}}_{86}
           +i\frac{240}{11}\sqrt{\frac{7}{143}}\omega^{\vec{d}}_{87}\,,
\nonumber\\           
M^{A_1, \vec{p}}_{42,42}&=&
          \omega^{\vec{d}}_{00}
         -\frac{162}{77}\omega^{\vec{d}}_{40}
         +\frac{20}{11}\omega^{\vec{d}}_{60}
         -i\frac{65\sqrt{6}}{77}\omega^{\vec{d}}_{22}
         +i\frac{162\sqrt{10}}{1001}\omega^{\vec{d}}_{42}
         +\frac{20}{33}\sqrt{\frac{35}{3}}(1+i)\omega^{\vec{d}}_{63}
           \nonumber\\&&
         +i\frac{4}{3}\sqrt{\frac{77}{3}}\omega^{\vec{d}}_{66}
         +i\frac{7952}{143}\sqrt{\frac{3}{143}}\omega^{\vec{d}}_{86}
         +\frac{1008}{143}\sqrt{\frac{10}{143}}(1+i)\omega^{\vec{d}}_{87}\,.
 \nonumber
\end{eqnarray}
%
%
For the $E$ irrep in the moving frame with $|\vec{p}L/2\pi|=\sqrt{3}$, we
choose $\vec{d}=\vec{p}L/2\pi=(1,1,1)$,
\begin{eqnarray}
 0&=&\det
 \left( 
 \setlength{\arraycolsep}{0.02pt}
 \begin{array}{ccccc}
  -\cot\delta_2 +M^{E, \vec{p}}_{21,21} 
  & M^{E, \vec{p}}_{21,22}  
  & M^{E, \vec{p}}_{21,41}  
  & M^{E, \vec{p}}_{21,42}  
  & M^{E, \vec{p}}_{21,43}  
  \\ 
 M^{E, \vec{p}\,*}_{21,22}    
  &  -\cot\delta_2+ M^{E, \vec{p}}_{22,22}
  & M^{E, \vec{p}}_{22,41}  
  & M^{E, \vec{p}}_{22,42}  
  & M^{E, \vec{p}}_{22,43}  
   \\ 
 M^{E, \vec{p}\,*}_{21,41}    
  & M^{E, \vec{p}\,*}_{22,41}    
  &  -\cot\delta_4+ M^{E, \vec{p}}_{41,41}  
  & M^{E, \vec{p}}_{41,42}  
  & M^{E, \vec{p}}_{41,43}  
   \\ 
 M^{E, \vec{p}\,*}_{21,42}    
  & M^{E, \vec{p}\,*}_{22,42}    
  & M^{E, \vec{p}\,*}_{41,42}  
  &  -\cot\delta_4+ M^{E, \vec{p}}_{42,42}  
  & M^{E, \vec{p}}_{42,43}  
   \\ 
 M^{E, \vec{p}\,*}_{21,43}    
  & M^{E, \vec{p}\,*}_{22,43}    
  & M^{E, \vec{p}\,*}_{41,43}  
  & M^{E, \vec{p}\,*}_{42,43}  
  &  -\cot\delta_4+ M^{E, \vec{p}}_{43,43}  
\end{array} \right), \nonumber
\label{eq:p1E}
\end{eqnarray}
where
\begin{eqnarray}
M^{E, \vec{p}}_{21,21} 
       &=& \omega^{\vec{d}}_{00}
  +\frac{18}{7}\omega^{\vec{d}}_{40}\,,
\nonumber\\
M^{E, \vec{p}}_{21,22}
       &=&\frac{\sqrt{10}}{14}(1-i)(\sqrt{60}\omega^{\vec{d}}_{22}-9\omega^{\vec{d}}_{42})\,,
\nonumber\\
M^{E, \vec{p}}_{21,41}
       &=& +\frac{15\sqrt{2}}{14}(1-i)\omega^{\vec{d}}_{22}
           +\frac{12\sqrt{30}}{77}(1-i)\omega^{\vec{d}}_{42}
           -i\frac{32\sqrt{35}}{33}\omega^{\vec{d}}_{63}
           +\frac{10}{3}\sqrt{\frac{7}{11}}(1-i)\omega^{\vec{d}}_{66}\,,
\nonumber\\
 M^{E, \vec{p}}_{21,42}
       &=& \frac{30\sqrt{3}}{77}(-4\omega^{\vec{d}}_{40}-7\omega^{\vec{d}}_{60})\,,
 \nonumber\\
 M^{E, \vec{p}}_{21,43}
       &=& -\frac{5\sqrt{21}}{7}\omega^{\vec{d}}_{22}
           +\frac{18\sqrt{35}}{77}\omega^{\vec{d}}_{42}
           -\frac{4}{33}\sqrt{\frac{10}{3}}(1-i)\omega^{\vec{d}}_{63}
           -\frac{70}{3}\sqrt{\frac{2}{33}}\omega^{\vec{d}}_{66}\,,
 \nonumber\\
 M^{E, \vec{p}}_{22,22}
       &=& \omega^{\vec{d}}_{00}
          +i\frac{5\sqrt{6}}{7}\omega^{\vec{d}}_{22}
          -\frac{12}{7}\omega^{\vec{d}}_{40}
          +i\frac{6\sqrt{10}}{7}\omega^{\vec{d}}_{42}\,,
 \nonumber\\
M^{E, \vec{p}}_{22,41}
       &=& -i\frac{10\sqrt{2}}{7}\omega^{\vec{d}}_{22}
           +\frac{60\sqrt{3}}{77}\omega^{\vec{d}}_{40}
           -i\frac{39\sqrt{30}}{154}\omega^{\vec{d}}_{42}
           -\frac{40\sqrt{3}}{11}\omega^{\vec{d}}_{60}
           +i\frac{32\sqrt{35}}{99}(1+i)\omega^{\vec{d}}_{63}
           +i\frac{20\sqrt{77}}{99}\omega^{\vec{d}}_{66}\,,
 \nonumber\\
M^{E, \vec{p}}_{22,42}
       &=& -\frac{10\sqrt{2}}{7}(1+i)\omega^{\vec{d}}_{22}
           +\frac{3\sqrt{30}}{7}(1+i)\omega^{\vec{d}}_{42}
           -\frac{4\sqrt{35}}{9}\omega^{\vec{d}}_{63}
           +i\frac{20}{9}\sqrt{\frac{7}{11}}(1+i)\omega^{\vec{d}}_{66}\,,
 \nonumber\\
M^{E, \vec{p}}_{22,43}
       &=& -\frac{9\sqrt{35}}{22}(1-i)\omega^{\vec{d}}_{42}
           +i\frac{32}{11}\sqrt{\frac{10}{3}}\omega^{\vec{d}}_{63}
           -10\sqrt{\frac{2}{33}}(1-i)\omega^{\vec{d}}_{66}\,,
 \nonumber\\
 M^{E, \vec{p}}_{41,41}
       &=& \omega^{\vec{d}}_{00}
          -\frac{162}{77}\omega^{\vec{d}}_{40}
          +\frac{20}{11}\omega^{\vec{d}}_{60}
          +i\frac{65\sqrt{6}}{154}\omega^{\vec{d}}_{22}
          -i\frac{81\sqrt{10}}{1001}\omega^{\vec{d}}_{42}
          -\frac{10}{33}\sqrt{\frac{35}{3}}(1+i)\omega^{\vec{d}}_{63}
          -i\frac{2}{3}\sqrt{\frac{77}{3}}\omega^{\vec{d}}_{66}
           \nonumber\\&&
          -i\frac{3976}{143}\sqrt{\frac{3}{143}}\omega^{\vec{d}}_{86}
          -\frac{504}{143}\sqrt{\frac{10}{143}}(1+i)\omega^{\vec{d}}_{87}\,,
 \nonumber\\
 M^{E, \vec{p}}_{41,42}
       &=& -\frac{5\sqrt{6}}{7}(1+i)\omega^{\vec{d}}_{22}
           +\frac{405\sqrt{10}}{2002}(1+i)\omega^{\vec{d}}_{42}
           -\frac{16}{33}\sqrt{\frac{35}{3}}\omega^{\vec{d}}_{63}
           +\frac{4}{3}\sqrt{\frac{7}{33}}(1+i)\omega^{\vec{d}}_{66}
           \nonumber\\&&
           -\frac{238}{13}\sqrt{\frac{3}{143}}(1+i)\omega^{\vec{d}}_{86}
           -\frac{168}{13}\sqrt{\frac{10}{143}}\omega^{\vec{d}}_{87}\,,
 \nonumber\\
 M^{E, \vec{p}}_{41,43}
       &=& -\frac{15\sqrt{7}}{22}(1-i)\omega^{\vec{d}}_{22}
           -i\frac{2\sqrt{10}}{3}\omega^{\vec{d}}_{63}
           -\frac{\sqrt{22}}{33}(1-i)\omega^{\vec{d}}_{66}
           -\frac{24}{11}\sqrt{\frac{14}{143}}(1-i)\omega^{\vec{d}}_{86}
           -i\frac{48}{11}\sqrt{\frac{105}{143}}\omega^{\vec{d}}_{87}\,,
 \nonumber\\
 M^{E, \vec{p}}_{42,42}
       &=& \omega^{\vec{d}}_{00}
          +\frac{324}{1001}\omega^{\vec{d}}_{40}
          -\frac{64}{11}\omega^{\vec{d}}_{60}
          +\frac{392}{143}\omega^{\vec{d}}_{80}\,,
 \nonumber\\
 M^{E, \vec{p}}_{42,43}
       &=& +\frac{30\sqrt{7}}{77}\omega^{\vec{d}}_{22}
           -\frac{243}{143}\sqrt{\frac{15}{7}}\omega^{\vec{d}}_{42}
           +\frac{16\sqrt{10}}{33}(1-i)\omega^{\vec{d}}_{63}
           +\frac{28}{3}\sqrt{\frac{2}{11}}\omega^{\vec{d}}_{66}
           \nonumber\\&&
           -\frac{1434}{143}\sqrt{\frac{14}{143}}\omega^{\vec{d}}_{86}
           +\frac{48}{143}\sqrt{\frac{105}{143}}(1-i)\omega^{\vec{d}}_{87}\,,
 \nonumber\\
 M^{E, \vec{p}}_{43,43}
       &=& \omega^{\vec{d}}_{00}
          +\frac{162}{143}\omega^{\vec{d}}_{40}
          -\frac{4}{11}\omega^{\vec{d}}_{60}
          -\frac{448}{143}\omega^{\vec{d}}_{80}
          -i\frac{5\sqrt{6}}{22}\omega^{\vec{d}}_{22}
          -i\frac{81\sqrt{10}}{143}\omega^{\vec{d}}_{42}
          -\frac{26}{33}\sqrt{\frac{35}{3}}(1+i)\omega^{\vec{d}}_{63}
           \nonumber\\&&
          -i\frac{14}{3}\sqrt{\frac{7}{33}}\omega^{\vec{d}}_{66}
          +i\frac{1288}{143}\sqrt{\frac{3}{143}}\omega^{\vec{d}}_{86}
          +\frac{840}{143}\sqrt{\frac{10}{143}}(1+i)\omega^{\vec{d}}_{87}\,.
 \nonumber
\end{eqnarray}

\begin{table}[htbp]
\begin{center}\caption{The relationship between angular momentum and
    irrep at momenta up to $P^2=3P_0^2$.}
\resizebox{480pt}{310pt}{
\begin{tabular}{cccccc}\hline\hline
   $\vec{p}$       & $l$    &       $n$        &      $\Gamma$  & $\alpha$  & $C^{\Gamma,\,\alpha,n}_{l,m}  |l,\,m\rangle $  \\
    \hline
   $(0,\,0,\,1)$   & $0$  &       $1$        &      $A_1$        & $1$          & $ |0,\,0\rangle$   \\     
                          & $2$   &       $1$        &      $A_1$        & $1$          & $ |2,\,0\rangle$   \\     
                           &         &       $1$        &      $B_1$        & $1$          & $ \frac{1}{\sqrt{2}}(|2,\,-2\rangle+|2,\,2\rangle)  $ \\     
                           &         &       $1$        &      $B_2$        & $1$          & $  \frac{1}{\sqrt{2}}(|2,\,-2\rangle-|2,\,2\rangle)  $ \\     
                           &         &       $1$        &      $E$            & $1$          & $ \frac{1}{\sqrt{2}}(|2,\,-1\rangle-i|2,\,1\rangle)   $ \\     
                           &         &       $  $        &      $  $            & $2$          & $ \frac{1-i}{2}(|2,\,-1\rangle+i|2,\,1\rangle)$   \\     
                           & $4$  &       $1$        &      $A_1$        & $1$          
                           & $ \frac{1}{2}(|4,\,-4\rangle+\sqrt{2}|4,\,0\rangle+|4,\,4\rangle) $  \\     
                           &         &       $2$        &      $A_1$        & $1$          
                           & $ \frac{1}{2}(|4,\,-4\rangle-\sqrt{2}|4,\,0\rangle+|4,\,4\rangle) $  \\     
                           &         &       $1$        &      $A_2$        & $1$          & $\frac{1}{\sqrt{2}}(-|4,\,-4\rangle+|4,\,4\rangle) $  \\     
                           &         &       $1$        &      $B_1$        & $1$          & $\frac{1}{\sqrt{2}}(  |4,\,-2\rangle+|4,\,2\rangle) $  \\     
                           &         &       $1$        &      $B_2$        & $1$          & $\frac{1}{\sqrt{2}}(-|4,\,-2\rangle+|4,\,2\rangle) $  \\     
                           &         &       $1$        &      $E$            & $1$          
                           & $\frac{1+i}{2\sqrt{2}}(-|4,\,-3\rangle+|4,\,-1\rangle-i|4,\,1\rangle-i|4,\,3\rangle) $  \\     
                           &         &       $  $        &      $  $            & $2$          
                           & $\frac{1}{2}(|4,\,-3\rangle+|4,\,-1\rangle+i|4,\,1\rangle-i|4,\,3\rangle) $  \\     
                           &         &       $2$        &      $E$            & $1$          
                           & $\frac{1+i}{2\sqrt{2}}(|4,\,-3\rangle+|4,\,-1\rangle-i|4,\,1\rangle+i|4,\,3\rangle) $  \\     
                           &         &       $  $        &      $  $            & $2$          
                           & $\frac{1}{2}(-|4,\,-3\rangle+|4,\,-1\rangle+i|4,\,1\rangle+i|4,\,3\rangle) $ \\     
                           \hline 
   $(1,\,1,\,0)$   & $0$  &       $1$        &      $A_1$        & $1$          & $ |0,\,0\rangle$   \\     
                          & $2$   &       $1$        &      $A_1$        & $1$          & $ |2,\,0\rangle$   \\     
                           &         &       $2$        &      $A_1$        & $1$          & $ \frac{1}{\sqrt{2}}(|2,\,-2\rangle-|2,\,2\rangle)  $ \\     
                           &         &       $1$        &      $A_2$        & $1$          & $  \frac{1}{\sqrt{2}}(|2,\,-1\rangle-i|2,\,1\rangle)  $ \\     
                           &         &       $1$        &      $B_1$        & $1$          & $ \frac{1}{\sqrt{2}}(|2,\,-1\rangle+i|2,\,1\rangle)   $ \\     
                           &         &       $1$        &      $B_2$        & $1$          & $ \frac{1}{\sqrt{2}}(|2,\,-2\rangle+|2,\,2\rangle)$   \\     
                           & $4$  &       $1$        &      $A_1$        & $1$          
                           & $ \frac{1}{2}(|4,\,-4\rangle-|4,\,-2\rangle+|4,\,2\rangle+|4,\,4\rangle) $  \\     
                           &         &       $2$        &      $A_1$        & $1$          
                           & $ \frac{1}{2}(|4,\,-4\rangle+|4,\,-2\rangle-|4,\,2\rangle+|4,\,4\rangle) $  \\     
                           &         &       $3$        &      $A_1$        & $1$          & $|4,\,0\rangle $  \\     
                           &         &       $1$        &      $A_2$        & $1$          
                           & $ \frac{1}{2}(|4,\,-3\rangle+|4,\,-1\rangle-i|4,\,1\rangle+i|4,\,3\rangle) $  \\     
                           &         &       $2$        &      $A_2$        & $1$              
                           & $ \frac{1}{2}(|4,\,-3\rangle-|4,\,-1\rangle+i|4,\,1\rangle+i|4,\,3\rangle) $  \\     
                           &         &       $1$        &      $B_1$            & $1$          
                           & $ \frac{1}{2}(|4,\,-3\rangle+|4,\,-1\rangle+i|4,\,1\rangle-i|4,\,3\rangle) $  \\     
                           &         &       $2$        &      $B_1$            & $1$          
                           & $ \frac{1}{2}(|4,\,-3\rangle-|4,\,-1\rangle-i|4,\,1\rangle-i|4,\,3\rangle) $  \\     
                           &         &       $1$        &      $B_2$            & $1$          
                           & $ \frac{1}{2}(-|4,\,-4\rangle+|4,\,-2\rangle+|4,\,2\rangle+|4,\,4\rangle) $  \\     
                           &         &       $2$        &      $B_2$            & $1$          
                           & $ \frac{1}{2}(|4,\,-4\rangle+|4,\,-2\rangle+|4,\,2\rangle-|4,\,4\rangle) $  \\  
                           \hline 
   $(1,\,1,\,1)$   & $0$  &       $1$        &      $A_1$        & $1$          & $ |0,\,0\rangle$   \\     
                          & $2$   &       $1$        &      $A_1$        & $1$         
                           & $ \frac{1}{\sqrt{6}}(|2,\,-2\rangle+(1-i)|2,\,-1\rangle+(1+i)|2,\,1\rangle-|2,\,2\rangle) $  \\  
                           &         &       $1$        &      $E$            & $1$          & $ \frac{1}{\sqrt{2}}(|2,\,-2\rangle+|2,\,2\rangle)  $ \\     
                           &         &       $  $        &      $  $            & $2$          & $  -|2,\,0\rangle  $ \\     
                           &         &       $2$        &      $E$            & $1$          & $ \frac{1}{\sqrt{2}}(|2,\,-1\rangle-i|2,\,1\rangle)   $ \\     
                           &         &       $  $        &      $  $            & $2$             
                           & $ \frac{1}{\sqrt{6}}(-(1-i)|2,\,-2\rangle-i|2,\,-1\rangle+|2,\,1\rangle+(1-i)|2,\,2\rangle) $  \\  
                           & $4$  &       $1$        &      $A_1$        & $1$          
                           & $ \frac{1}{2}\sqrt{\frac{5}{6}}(|4,\,-4\rangle+\sqrt{\frac{14}{5}}|4,\,0\rangle+|4,\,4\rangle) $  \\     
                           &         &       $2$        &      $A_1$        & $1$          
                           & $ \frac{1}{2}\left(-\sqrt{\frac{7}{6}}|4,\,-3\rangle+\frac{-1+i}{\sqrt{3}}|4,\,-2\rangle
                                +\frac{-i}{\sqrt{6}}|4,\,-1\rangle +\frac{1}{\sqrt{6}}|4,\,1\rangle 
                                 +\frac{1-i}{\sqrt{3}}|4,\,2\rangle +i\sqrt{\frac{7}{6}}|4,\,3\rangle\right)$  \\     
                           &         &       $3$        &      $A_2$        & $1$          
                           & $ \frac{1}{2}\left(\frac{-1+i}{\sqrt{3}}|4,\,-4\rangle-\frac{i}{\sqrt{6}}|4,\,-3\rangle
                                +\sqrt{\frac{7}{6}}|4,\,-1\rangle -i\sqrt{\frac{7}{6}}|4,\,1\rangle
                                 +\frac{1}{\sqrt{6}}|4,\,3\rangle +\frac{1-i}{\sqrt{3}}|4,\,4\rangle\right) $ \\     
                           &         &       $1$        &      $E$            & $1$          
                           & $ \frac{1}{4}(-i\sqrt{7}|4,\,-3\rangle-|4,\,-1\rangle+i|4,\,1\rangle+\sqrt{7}|4,\,3\rangle) $  \\     
                           &         &       $  $        &      $E$            & $2$              
                           & $ \frac{1}{4}\left(\sqrt{\frac{7}{3}}|4,\,-3\rangle+\frac{4(-1+i)}{\sqrt{6}}|4,\,-2\rangle
                                +\frac{i}{\sqrt{3}}|4,\,-1\rangle -\frac{1}{\sqrt{3}}|4,\,1\rangle
                                 +\frac{4(1-i)}{\sqrt{6}}|4,\,2\rangle -i\sqrt{\frac{7}{3}}|4,\,4\rangle\right)$  \\     
                           &         &       $2$        &      $E$            & $1$          
                           & $ -\frac{1}{\sqrt{2}}(|4,\,-2\rangle+|4,\,2\rangle) $  \\     
                           &         &       $ $        &       $  $            & $2$          
                           & $ \frac{1}{2}(\sqrt{\frac{7}{6}}|4,\,-4\rangle-\sqrt{\frac{5}{3}}|4,\,0\rangle+\sqrt{\frac{7}{6}}|4,\,4\rangle)  $  \\     
                           &         &       $3$        &      $E$            & $1$          
                           & $ \frac{1}{\sqrt{3}}\left(-|4,\,-4\rangle-\frac{1-i}{4\sqrt{2}}|4,\,-3\rangle
                                -\frac{\sqrt{7}(1+i)}{4\sqrt{2}}|4,\,-1\rangle +\frac{\sqrt{7}(-1+i)}{4\sqrt{2}}|4,\,1\rangle
                                 -\frac{1+i}{4\sqrt{2}}|4,\,3\rangle +|4,\,4\rangle\right)$  \\     
                           &         &       $ $        &       $ $            & $2$          
                           & $ \frac{1+i}{4\sqrt{2}}(|4,\,-3\rangle-i\sqrt{7}|4,\,-1\rangle+\sqrt{7}|4,\,1\rangle-i|4,\,3\rangle) $  \\     
\hline\hline
\end{tabular}
}\label{tab:ccoefinM}
\end{center}
\end{table}


\bibliography{ref}

\begin{thebibliography}{55}%
\makeatletter
\providecommand \@ifxundefined [1]{%
 \@ifx{#1\undefined}
}%
\providecommand \@ifnum [1]{%
 \ifnum #1\expandafter \@firstoftwo
 \else \expandafter \@secondoftwo
 \fi
}%
\providecommand \@ifx [1]{%
 \ifx #1\expandafter \@firstoftwo
 \else \expandafter \@secondoftwo
 \fi
}%
\providecommand \natexlab [1]{#1}%
\providecommand \enquote  [1]{``#1''}%
\providecommand \bibnamefont  [1]{#1}%
\providecommand \bibfnamefont [1]{#1}%
\providecommand \citenamefont [1]{#1}%
\providecommand \href@noop [0]{\@secondoftwo}%
\providecommand \href [0]{\begingroup \@sanitize@url \@href}%
\providecommand \@href[1]{\@@startlink{#1}\@@href}%
\providecommand \@@href[1]{\endgroup#1\@@endlink}%
\providecommand \@sanitize@url [0]{\catcode `\\12\catcode `\$12\catcode
  `\&12\catcode `\#12\catcode `\^12\catcode `\_12\catcode `\%12\relax}%
\providecommand \@@startlink[1]{}%
\providecommand \@@endlink[0]{}%
\providecommand \url  [0]{\begingroup\@sanitize@url \@url }%
\providecommand \@url [1]{\endgroup\@href {#1}{\urlprefix }}%
\providecommand \urlprefix  [0]{URL }%
\providecommand \Eprint [0]{\href }%
\providecommand \doibase [0]{http://dx.doi.org/}%
\providecommand \selectlanguage [0]{\@gobble}%
\providecommand \bibinfo  [0]{\@secondoftwo}%
\providecommand \bibfield  [0]{\@secondoftwo}%
\providecommand \translation [1]{[#1]}%
\providecommand \BibitemOpen [0]{}%
\providecommand \bibitemStop [0]{}%
\providecommand \bibitemNoStop [0]{.\EOS\space}%
\providecommand \EOS [0]{\spacefactor3000\relax}%
\providecommand \BibitemShut  [1]{\csname bibitem#1\endcsname}%
\let\auto@bib@innerbib\@empty
\bibitem [{\citenamefont {Liu}(2017)}]{Liu:2016kbb}%
  \BibitemOpen
  \bibfield  {author} {\bibinfo {author} {\bibfnamefont {C.}~\bibnamefont
  {Liu}},\ }\bibfield  {booktitle} {\emph {\bibinfo {booktitle} {{Proceedings,
  34th International Symposium on Lattice Field Theory (Lattice 2016):
  Southampton, UK, July 24-30, 2016}}},\ }\href@noop {} {\bibfield  {journal}
  {\bibinfo  {journal} {PoS}\ }\textbf {\bibinfo {volume} {LATTICE2016}},\
  \bibinfo {pages} {006} (\bibinfo {year} {2017})},\ \Eprint
  {http://arxiv.org/abs/1612.00103} {arXiv:1612.00103 [hep-lat]} \BibitemShut
  {NoStop}%
\bibitem [{\citenamefont {Brice\~no}\ \emph {et~al.}(2018)\citenamefont
  {Brice\~no}, \citenamefont {Dudek},\ and\ \citenamefont
  {Young}}]{Briceno:2017max}%
  \BibitemOpen
  \bibfield  {author} {\bibinfo {author} {\bibfnamefont {R.~A.}\ \bibnamefont
  {Brice\~no}}, \bibinfo {author} {\bibfnamefont {J.~J.}\ \bibnamefont
  {Dudek}}, \ and\ \bibinfo {author} {\bibfnamefont {R.~D.}\ \bibnamefont
  {Young}},\ }\href {\doibase 10.1103/RevModPhys.90.025001} {\bibfield
  {journal} {\bibinfo  {journal} {Rev. Mod. Phys.}\ }\textbf {\bibinfo {volume}
  {90}},\ \bibinfo {pages} {025001} (\bibinfo {year} {2018})},\ \Eprint
  {http://arxiv.org/abs/1706.06223} {arXiv:1706.06223 [hep-lat]} \BibitemShut
  {NoStop}%
\bibitem [{\citenamefont {Padmanath}(2018)}]{Padmanath:2019wid}%
  \BibitemOpen
  \bibfield  {author} {\bibinfo {author} {\bibfnamefont {M.}~\bibnamefont
  {Padmanath}},\ }\href {\doibase 10.22323/1.334.0013} {\bibfield  {journal}
  {\bibinfo  {journal} {PoS}\ }\textbf {\bibinfo {volume} {LATTICE2018}},\
  \bibinfo {pages} {013} (\bibinfo {year} {2018})},\ \Eprint
  {http://arxiv.org/abs/1905.09651} {arXiv:1905.09651 [hep-lat]} \BibitemShut
  {NoStop}%
\bibitem [{\citenamefont {Dudek}\ \emph {et~al.}(2013)\citenamefont {Dudek},
  \citenamefont {Edwards},\ and\ \citenamefont {Thomas}}]{Dudek:2012xn}%
  \BibitemOpen
  \bibfield  {author} {\bibinfo {author} {\bibfnamefont {J.~J.}\ \bibnamefont
  {Dudek}}, \bibinfo {author} {\bibfnamefont {R.~G.}\ \bibnamefont {Edwards}},
  \ and\ \bibinfo {author} {\bibfnamefont {C.~E.}\ \bibnamefont {Thomas}}
  (\bibinfo {collaboration} {Hadron Spectrum}),\ }\href {\doibase
  10.1103/PhysRevD.87.034505} {\bibfield  {journal} {\bibinfo  {journal} {Phys.
  Rev. D}\ }\textbf {\bibinfo {volume} {87}},\ \bibinfo {pages} {034505}
  (\bibinfo {year} {2013})},\ \bibinfo {note} {[Erratum: Phys.Rev.D 90, 099902
  (2014)]},\ \Eprint {http://arxiv.org/abs/1212.0830} {arXiv:1212.0830
  [hep-ph]} \BibitemShut {NoStop}%
\bibitem [{\citenamefont {Dudek}\ \emph {et~al.}(2011)\citenamefont {Dudek},
  \citenamefont {Edwards}, \citenamefont {Peardon}, \citenamefont {Richards},\
  and\ \citenamefont {Thomas}}]{Dudek:2010ew}%
  \BibitemOpen
  \bibfield  {author} {\bibinfo {author} {\bibfnamefont {J.~J.}\ \bibnamefont
  {Dudek}}, \bibinfo {author} {\bibfnamefont {R.~G.}\ \bibnamefont {Edwards}},
  \bibinfo {author} {\bibfnamefont {M.~J.}\ \bibnamefont {Peardon}}, \bibinfo
  {author} {\bibfnamefont {D.~G.}\ \bibnamefont {Richards}}, \ and\ \bibinfo
  {author} {\bibfnamefont {C.~E.}\ \bibnamefont {Thomas}},\ }\href {\doibase
  10.1103/PhysRevD.83.071504} {\bibfield  {journal} {\bibinfo  {journal} {Phys.
  Rev.}\ }\textbf {\bibinfo {volume} {D83}},\ \bibinfo {pages} {071504}
  (\bibinfo {year} {2011})},\ \Eprint {http://arxiv.org/abs/1011.6352}
  {arXiv:1011.6352 [hep-ph]} \BibitemShut {NoStop}%
\bibitem [{\citenamefont {Dudek}\ \emph {et~al.}(2012)\citenamefont {Dudek},
  \citenamefont {Edwards},\ and\ \citenamefont {Thomas}}]{Dudek:2012gj}%
  \BibitemOpen
  \bibfield  {author} {\bibinfo {author} {\bibfnamefont {J.~J.}\ \bibnamefont
  {Dudek}}, \bibinfo {author} {\bibfnamefont {R.~G.}\ \bibnamefont {Edwards}},
  \ and\ \bibinfo {author} {\bibfnamefont {C.~E.}\ \bibnamefont {Thomas}},\
  }\href {\doibase 10.1103/PhysRevD.86.034031} {\bibfield  {journal} {\bibinfo
  {journal} {Phys. Rev.}\ }\textbf {\bibinfo {volume} {D86}},\ \bibinfo {pages}
  {034031} (\bibinfo {year} {2012})},\ \Eprint {http://arxiv.org/abs/1203.6041}
  {arXiv:1203.6041 [hep-ph]} \BibitemShut {NoStop}%
\bibitem [{\citenamefont {Beane}\ \emph {et~al.}(2012)\citenamefont {Beane},
  \citenamefont {Chang}, \citenamefont {Detmold}, \citenamefont {Lin},
  \citenamefont {Luu}, \citenamefont {Orginos}, \citenamefont {Parre\~no},
  \citenamefont {Savage}, \citenamefont {Torok},\ and\ \citenamefont
  {Walker-Loud}}]{Beane:2011sc}%
  \BibitemOpen
  \bibfield  {author} {\bibinfo {author} {\bibfnamefont {S.~R.}\ \bibnamefont
  {Beane}}, \bibinfo {author} {\bibfnamefont {E.}~\bibnamefont {Chang}},
  \bibinfo {author} {\bibfnamefont {W.}~\bibnamefont {Detmold}}, \bibinfo
  {author} {\bibfnamefont {H.~W.}\ \bibnamefont {Lin}}, \bibinfo {author}
  {\bibfnamefont {T.~C.}\ \bibnamefont {Luu}}, \bibinfo {author} {\bibfnamefont
  {K.}~\bibnamefont {Orginos}}, \bibinfo {author} {\bibfnamefont
  {A.}~\bibnamefont {Parre\~no}}, \bibinfo {author} {\bibfnamefont {M.~J.}\
  \bibnamefont {Savage}}, \bibinfo {author} {\bibfnamefont {A.}~\bibnamefont
  {Torok}}, \ and\ \bibinfo {author} {\bibfnamefont {A.}~\bibnamefont
  {Walker-Loud}} (\bibinfo {collaboration} {NPLQCD}),\ }\href {\doibase
  10.1103/PhysRevD.85.034505} {\bibfield  {journal} {\bibinfo  {journal} {Phys.
  Rev.}\ }\textbf {\bibinfo {volume} {D85}},\ \bibinfo {pages} {034505}
  (\bibinfo {year} {2012})},\ \Eprint {http://arxiv.org/abs/1107.5023}
  {arXiv:1107.5023 [hep-lat]} \BibitemShut {NoStop}%
\bibitem [{\citenamefont {Bulava}\ \emph {et~al.}(2016)\citenamefont {Bulava},
  \citenamefont {Fahy}, \citenamefont {H\"orz}, \citenamefont {Juge},
  \citenamefont {Morningstar},\ and\ \citenamefont {Wong}}]{Bulava:2016mks}%
  \BibitemOpen
  \bibfield  {author} {\bibinfo {author} {\bibfnamefont {J.}~\bibnamefont
  {Bulava}}, \bibinfo {author} {\bibfnamefont {B.}~\bibnamefont {Fahy}},
  \bibinfo {author} {\bibfnamefont {B.}~\bibnamefont {H\"orz}}, \bibinfo
  {author} {\bibfnamefont {K.~J.}\ \bibnamefont {Juge}}, \bibinfo {author}
  {\bibfnamefont {C.}~\bibnamefont {Morningstar}}, \ and\ \bibinfo {author}
  {\bibfnamefont {C.~H.}\ \bibnamefont {Wong}},\ }\href {\doibase
  10.1016/j.nuclphysb.2016.07.024} {\bibfield  {journal} {\bibinfo  {journal}
  {Nucl. Phys. B}\ }\textbf {\bibinfo {volume} {910}},\ \bibinfo {pages} {842}
  (\bibinfo {year} {2016})},\ \Eprint {http://arxiv.org/abs/1604.05593}
  {arXiv:1604.05593 [hep-lat]} \BibitemShut {NoStop}%
\bibitem [{\citenamefont {Michael}(1985)}]{Michael:1985ne}%
  \BibitemOpen
  \bibfield  {author} {\bibinfo {author} {\bibfnamefont {C.}~\bibnamefont
  {Michael}},\ }\href {\doibase 10.1016/0550-3213(85)90297-4} {\bibfield
  {journal} {\bibinfo  {journal} {Nucl. Phys. B}\ }\textbf {\bibinfo {volume}
  {259}},\ \bibinfo {pages} {58} (\bibinfo {year} {1985})}\BibitemShut
  {NoStop}%
\bibitem [{\citenamefont {{L\"uscher}}\ and\ \citenamefont
  {Wolff}(1990)}]{Luscher:1990ck}%
  \BibitemOpen
  \bibfield  {author} {\bibinfo {author} {\bibfnamefont {M.}~\bibnamefont
  {{L\"uscher}}}\ and\ \bibinfo {author} {\bibfnamefont {U.}~\bibnamefont
  {Wolff}},\ }\href {\doibase 10.1016/0550-3213(90)90540-T} {\bibfield
  {journal} {\bibinfo  {journal} {Nucl. Phys. B}\ }\textbf {\bibinfo {volume}
  {339}},\ \bibinfo {pages} {222} (\bibinfo {year} {1990})}\BibitemShut
  {NoStop}%
\bibitem [{\citenamefont {Blossier}\ \emph {et~al.}(2009)\citenamefont
  {Blossier}, \citenamefont {Della~Morte}, \citenamefont {von Hippel},
  \citenamefont {Mendes},\ and\ \citenamefont {Sommer}}]{Blossier:2009kd}%
  \BibitemOpen
  \bibfield  {author} {\bibinfo {author} {\bibfnamefont {B.}~\bibnamefont
  {Blossier}}, \bibinfo {author} {\bibfnamefont {M.}~\bibnamefont
  {Della~Morte}}, \bibinfo {author} {\bibfnamefont {G.}~\bibnamefont {von
  Hippel}}, \bibinfo {author} {\bibfnamefont {T.}~\bibnamefont {Mendes}}, \
  and\ \bibinfo {author} {\bibfnamefont {R.}~\bibnamefont {Sommer}},\ }\href
  {\doibase 10.1088/1126-6708/2009/04/094} {\bibfield  {journal} {\bibinfo
  {journal} {JHEP}\ }\textbf {\bibinfo {volume} {04}},\ \bibinfo {pages} {094}
  (\bibinfo {year} {2009})},\ \Eprint {http://arxiv.org/abs/0902.1265}
  {arXiv:0902.1265 [hep-lat]} \BibitemShut {NoStop}%
\bibitem [{\citenamefont {Mahbub}\ \emph {et~al.}(2009)\citenamefont {Mahbub},
  \citenamefont {{\'O Cais}}, \citenamefont {Kamleh}, \citenamefont {Lasscock},
  \citenamefont {Leinweber},\ and\ \citenamefont {Williams}}]{Mahbub:2009nr}%
  \BibitemOpen
  \bibfield  {author} {\bibinfo {author} {\bibfnamefont {M.~S.}\ \bibnamefont
  {Mahbub}}, \bibinfo {author} {\bibfnamefont {A.}~\bibnamefont {{\'O Cais}}},
  \bibinfo {author} {\bibfnamefont {W.}~\bibnamefont {Kamleh}}, \bibinfo
  {author} {\bibfnamefont {B.~G.}\ \bibnamefont {Lasscock}}, \bibinfo {author}
  {\bibfnamefont {D.~B.}\ \bibnamefont {Leinweber}}, \ and\ \bibinfo {author}
  {\bibfnamefont {A.~G.}\ \bibnamefont {Williams}},\ }\href {\doibase
  10.1103/PhysRevD.80.054507} {\bibfield  {journal} {\bibinfo  {journal} {Phys.
  Rev. D}\ }\textbf {\bibinfo {volume} {80}},\ \bibinfo {pages} {054507}
  (\bibinfo {year} {2009})},\ \Eprint {http://arxiv.org/abs/0905.3616}
  {arXiv:0905.3616 [hep-lat]} \BibitemShut {NoStop}%
\bibitem [{\citenamefont {Hudspith}\ \emph {et~al.}(2020)\citenamefont
  {Hudspith}, \citenamefont {Colquhoun}, \citenamefont {Francis}, \citenamefont
  {Lewis},\ and\ \citenamefont {Maltman}}]{Hudspith:2020tdf}%
  \BibitemOpen
  \bibfield  {author} {\bibinfo {author} {\bibfnamefont {R.~J.}\ \bibnamefont
  {Hudspith}}, \bibinfo {author} {\bibfnamefont {B.}~\bibnamefont {Colquhoun}},
  \bibinfo {author} {\bibfnamefont {A.}~\bibnamefont {Francis}}, \bibinfo
  {author} {\bibfnamefont {R.}~\bibnamefont {Lewis}}, \ and\ \bibinfo {author}
  {\bibfnamefont {K.}~\bibnamefont {Maltman}},\ }\href {\doibase
  10.1103/PhysRevD.102.114506} {\bibfield  {journal} {\bibinfo  {journal}
  {Phys. Rev. D}\ }\textbf {\bibinfo {volume} {102}},\ \bibinfo {pages}
  {114506} (\bibinfo {year} {2020})},\ \Eprint
  {http://arxiv.org/abs/2006.14294} {arXiv:2006.14294 [hep-lat]} \BibitemShut
  {NoStop}%
\bibitem [{\citenamefont {Chen}\ \emph {et~al.}(2014)\citenamefont {Chen} \emph
  {et~al.}}]{Chen:2014afa}%
  \BibitemOpen
  \bibfield  {author} {\bibinfo {author} {\bibfnamefont {Y.}~\bibnamefont
  {Chen}} \emph {et~al.},\ }\href {\doibase 10.1103/PhysRevD.89.094506}
  {\bibfield  {journal} {\bibinfo  {journal} {Phys. Rev. D}\ }\textbf {\bibinfo
  {volume} {89}},\ \bibinfo {pages} {094506} (\bibinfo {year} {2014})},\
  \Eprint {http://arxiv.org/abs/1403.1318} {arXiv:1403.1318 [hep-lat]}
  \BibitemShut {NoStop}%
\bibitem [{\citenamefont {Adlarson}\ \emph {et~al.}(2011)\citenamefont
  {Adlarson} \emph {et~al.}}]{Adlarson:2011bh}%
  \BibitemOpen
  \bibfield  {author} {\bibinfo {author} {\bibfnamefont {P.}~\bibnamefont
  {Adlarson}} \emph {et~al.} (\bibinfo {collaboration} {WASA-at-COSY}),\ }\href
  {\doibase 10.1103/PhysRevLett.106.242302} {\bibfield  {journal} {\bibinfo
  {journal} {Phys. Rev. Lett.}\ }\textbf {\bibinfo {volume} {106}},\ \bibinfo
  {pages} {242302} (\bibinfo {year} {2011})},\ \Eprint
  {http://arxiv.org/abs/1104.0123} {arXiv:1104.0123 [nucl-ex]} \BibitemShut
  {NoStop}%
\bibitem [{\citenamefont {Adlarson}\ \emph {et~al.}(2014)\citenamefont
  {Adlarson} \emph {et~al.}}]{Adlarson:2014pxj}%
  \BibitemOpen
  \bibfield  {author} {\bibinfo {author} {\bibfnamefont {P.}~\bibnamefont
  {Adlarson}} \emph {et~al.} (\bibinfo {collaboration} {WASA-at-COSY}),\ }\href
  {\doibase 10.1103/PhysRevLett.112.202301} {\bibfield  {journal} {\bibinfo
  {journal} {Phys. Rev. Lett.}\ }\textbf {\bibinfo {volume} {112}},\ \bibinfo
  {pages} {202301} (\bibinfo {year} {2014})},\ \Eprint
  {http://arxiv.org/abs/1402.6844} {arXiv:1402.6844 [nucl-ex]} \BibitemShut
  {NoStop}%
\bibitem [{\citenamefont {Wita\l{}a}\ and\ \citenamefont
  {Gl\"ockle}(1991)}]{Witala:1991emr}%
  \BibitemOpen
  \bibfield  {author} {\bibinfo {author} {\bibfnamefont {H.}~\bibnamefont
  {Wita\l{}a}}\ and\ \bibinfo {author} {\bibfnamefont {W.}~\bibnamefont
  {Gl\"ockle}},\ }\href {\doibase 10.1016/0375-9474(91)90419-7} {\bibfield
  {journal} {\bibinfo  {journal} {Nucl. Phys. A}\ }\textbf {\bibinfo {volume}
  {528}},\ \bibinfo {pages} {48} (\bibinfo {year} {1991})}\BibitemShut
  {NoStop}%
\bibitem [{\citenamefont {Tornow}\ \emph {et~al.}(1998)\citenamefont {Tornow},
  \citenamefont {Wita\l{}a},\ and\ \citenamefont {Kievsky}}]{Tornow:1998zz}%
  \BibitemOpen
  \bibfield  {author} {\bibinfo {author} {\bibfnamefont {W.}~\bibnamefont
  {Tornow}}, \bibinfo {author} {\bibfnamefont {H.}~\bibnamefont {Wita\l{}a}}, \
  and\ \bibinfo {author} {\bibfnamefont {A.}~\bibnamefont {Kievsky}},\ }\href
  {\doibase 10.1103/PhysRevC.57.555} {\bibfield  {journal} {\bibinfo  {journal}
  {Phys. Rev. C}\ }\textbf {\bibinfo {volume} {57}},\ \bibinfo {pages} {555}
  (\bibinfo {year} {1998})}\BibitemShut {NoStop}%
\bibitem [{\citenamefont {Huber}\ and\ \citenamefont
  {Friar}(1998)}]{Huber:1998hu}%
  \BibitemOpen
  \bibfield  {author} {\bibinfo {author} {\bibfnamefont {D.}~\bibnamefont
  {Huber}}\ and\ \bibinfo {author} {\bibfnamefont {J.~L.}\ \bibnamefont
  {Friar}},\ }\href {\doibase 10.1103/PhysRevC.58.674} {\bibfield  {journal}
  {\bibinfo  {journal} {Phys. Rev. C}\ }\textbf {\bibinfo {volume} {58}},\
  \bibinfo {pages} {674} (\bibinfo {year} {1998})},\ \Eprint
  {http://arxiv.org/abs/nucl-th/9803038} {arXiv:nucl-th/9803038} \BibitemShut
  {NoStop}%
\bibitem [{\citenamefont {Anisovich}\ \emph {et~al.}(2000)\citenamefont
  {Anisovich}, \citenamefont {Anisovich},\ and\ \citenamefont
  {Sarantsev}}]{Anisovich:2000kxa}%
  \BibitemOpen
  \bibfield  {author} {\bibinfo {author} {\bibfnamefont {A.~V.}\ \bibnamefont
  {Anisovich}}, \bibinfo {author} {\bibfnamefont {V.~V.}\ \bibnamefont
  {Anisovich}}, \ and\ \bibinfo {author} {\bibfnamefont {A.~V.}\ \bibnamefont
  {Sarantsev}},\ }\href {\doibase 10.1103/PhysRevD.62.051502} {\bibfield
  {journal} {\bibinfo  {journal} {Phys. Rev. D}\ }\textbf {\bibinfo {volume}
  {62}},\ \bibinfo {pages} {051502} (\bibinfo {year} {2000})},\ \Eprint
  {http://arxiv.org/abs/hep-ph/0003113} {arXiv:hep-ph/0003113} \BibitemShut
  {NoStop}%
\bibitem [{\citenamefont {Ebert}\ \emph {et~al.}(2009)\citenamefont {Ebert},
  \citenamefont {Faustov},\ and\ \citenamefont {Galkin}}]{Ebert:2009ub}%
  \BibitemOpen
  \bibfield  {author} {\bibinfo {author} {\bibfnamefont {D.}~\bibnamefont
  {Ebert}}, \bibinfo {author} {\bibfnamefont {R.~N.}\ \bibnamefont {Faustov}},
  \ and\ \bibinfo {author} {\bibfnamefont {V.~O.}\ \bibnamefont {Galkin}},\
  }\href {\doibase 10.1103/PhysRevD.79.114029} {\bibfield  {journal} {\bibinfo
  {journal} {Phys. Rev. D}\ }\textbf {\bibinfo {volume} {79}},\ \bibinfo
  {pages} {114029} (\bibinfo {year} {2009})},\ \Eprint
  {http://arxiv.org/abs/0903.5183} {arXiv:0903.5183 [hep-ph]} \BibitemShut
  {NoStop}%
\bibitem [{\citenamefont {Meyer}\ and\ \citenamefont
  {Teper}(2005)}]{Meyer:2004jc}%
  \BibitemOpen
  \bibfield  {author} {\bibinfo {author} {\bibfnamefont {H.~B.}\ \bibnamefont
  {Meyer}}\ and\ \bibinfo {author} {\bibfnamefont {M.~J.}\ \bibnamefont
  {Teper}},\ }\href {\doibase 10.1016/j.physletb.2004.11.036} {\bibfield
  {journal} {\bibinfo  {journal} {Phys. Lett. B}\ }\textbf {\bibinfo {volume}
  {605}},\ \bibinfo {pages} {344} (\bibinfo {year} {2005})},\ \Eprint
  {http://arxiv.org/abs/hep-ph/0409183} {arXiv:hep-ph/0409183} \BibitemShut
  {NoStop}%
\bibitem [{\citenamefont {Hansen}\ and\ \citenamefont
  {Sharpe}(2019)}]{Hansen:2019nir}%
  \BibitemOpen
  \bibfield  {author} {\bibinfo {author} {\bibfnamefont {M.~T.}\ \bibnamefont
  {Hansen}}\ and\ \bibinfo {author} {\bibfnamefont {S.~R.}\ \bibnamefont
  {Sharpe}},\ }\href {\doibase 10.1146/annurev-nucl-101918-023723} {\bibfield
  {journal} {\bibinfo  {journal} {Ann. Rev. Nucl. Part. Sci.}\ }\textbf
  {\bibinfo {volume} {69}},\ \bibinfo {pages} {65} (\bibinfo {year} {2019})},\
  \Eprint {http://arxiv.org/abs/1901.00483} {arXiv:1901.00483 [hep-lat]}
  \BibitemShut {NoStop}%
\bibitem [{\citenamefont {Mai}\ \emph {et~al.}(2021)\citenamefont {Mai},
  \citenamefont {D\"oring},\ and\ \citenamefont {Rusetsky}}]{Mai:2021lwb}%
  \BibitemOpen
  \bibfield  {author} {\bibinfo {author} {\bibfnamefont {M.}~\bibnamefont
  {Mai}}, \bibinfo {author} {\bibfnamefont {M.}~\bibnamefont {D\"oring}}, \
  and\ \bibinfo {author} {\bibfnamefont {A.}~\bibnamefont {Rusetsky}},\ }\href
  {\doibase 10.1140/epjs/s11734-021-00146-5} {\bibfield  {journal} {\bibinfo
  {journal} {Eur. Phys. J. ST}\ }\textbf {\bibinfo {volume} {230}},\ \bibinfo
  {pages} {1623} (\bibinfo {year} {2021})},\ \Eprint
  {http://arxiv.org/abs/2103.00577} {arXiv:2103.00577 [hep-lat]} \BibitemShut
  {NoStop}%
\bibitem [{\citenamefont {Johnson}\ and\ \citenamefont
  {Dudek}(2021)}]{Johnson:2020ilc}%
  \BibitemOpen
  \bibfield  {author} {\bibinfo {author} {\bibfnamefont {C.~T.}\ \bibnamefont
  {Johnson}}\ and\ \bibinfo {author} {\bibfnamefont {J.~J.}\ \bibnamefont
  {Dudek}} (\bibinfo {collaboration} {Hadron Spectrum}),\ }\href {\doibase
  10.1103/PhysRevD.103.074502} {\bibfield  {journal} {\bibinfo  {journal}
  {Phys. Rev. D}\ }\textbf {\bibinfo {volume} {103}},\ \bibinfo {pages}
  {074502} (\bibinfo {year} {2021})},\ \Eprint
  {http://arxiv.org/abs/2012.00518} {arXiv:2012.00518 [hep-lat]} \BibitemShut
  {NoStop}%
\bibitem [{\citenamefont {L{\"u}scher}(1991)}]{Luscher:1990ux}%
  \BibitemOpen
  \bibfield  {author} {\bibinfo {author} {\bibfnamefont {M.}~\bibnamefont
  {L{\"u}scher}},\ }\href {\doibase 10.1016/0550-3213(91)90366-6} {\bibfield
  {journal} {\bibinfo  {journal} {Nucl. Phys.}\ }\textbf {\bibinfo {volume}
  {B354}},\ \bibinfo {pages} {531} (\bibinfo {year} {1991})}\BibitemShut
  {NoStop}%
\bibitem [{\citenamefont {L{\"u}scher}(1986)}]{Luscher:1986pf}%
  \BibitemOpen
  \bibfield  {author} {\bibinfo {author} {\bibfnamefont {M.}~\bibnamefont
  {L{\"u}scher}},\ }\href {\doibase 10.1007/BF01211097} {\bibfield  {journal}
  {\bibinfo  {journal} {Commun. Math. Phys.}\ }\textbf {\bibinfo {volume}
  {105}},\ \bibinfo {pages} {153} (\bibinfo {year} {1986})}\BibitemShut
  {NoStop}%
\bibitem [{\citenamefont {Rummukainen}\ and\ \citenamefont
  {Gottlieb}(1995)}]{Rummukainen:1995vs}%
  \BibitemOpen
  \bibfield  {author} {\bibinfo {author} {\bibfnamefont {K.}~\bibnamefont
  {Rummukainen}}\ and\ \bibinfo {author} {\bibfnamefont {S.~A.}\ \bibnamefont
  {Gottlieb}},\ }\href {\doibase 10.1016/0550-3213(95)00313-H} {\bibfield
  {journal} {\bibinfo  {journal} {Nucl. Phys. B}\ }\textbf {\bibinfo {volume}
  {450}},\ \bibinfo {pages} {397} (\bibinfo {year} {1995})},\ \Eprint
  {http://arxiv.org/abs/hep-lat/9503028} {arXiv:hep-lat/9503028} \BibitemShut
  {NoStop}%
\bibitem [{\citenamefont {Kim}\ \emph {et~al.}(2005)\citenamefont {Kim},
  \citenamefont {Sachrajda},\ and\ \citenamefont {Sharpe}}]{Kim:2005gf}%
  \BibitemOpen
  \bibfield  {author} {\bibinfo {author} {\bibfnamefont {C.~h.}\ \bibnamefont
  {Kim}}, \bibinfo {author} {\bibfnamefont {C.~T.}\ \bibnamefont {Sachrajda}},
  \ and\ \bibinfo {author} {\bibfnamefont {S.~R.}\ \bibnamefont {Sharpe}},\
  }\href {\doibase 10.1016/j.nuclphysb.2005.08.029} {\bibfield  {journal}
  {\bibinfo  {journal} {Nucl. Phys. B}\ }\textbf {\bibinfo {volume} {727}},\
  \bibinfo {pages} {218} (\bibinfo {year} {2005})},\ \Eprint
  {http://arxiv.org/abs/hep-lat/0507006} {arXiv:hep-lat/0507006} \BibitemShut
  {NoStop}%
\bibitem [{\citenamefont {Fu}(2012)}]{Fu:2011xz}%
  \BibitemOpen
  \bibfield  {author} {\bibinfo {author} {\bibfnamefont {Z.}~\bibnamefont
  {Fu}},\ }\href {\doibase 10.1103/PhysRevD.85.014506} {\bibfield  {journal}
  {\bibinfo  {journal} {Phys. Rev. D}\ }\textbf {\bibinfo {volume} {85}},\
  \bibinfo {pages} {014506} (\bibinfo {year} {2012})},\ \Eprint
  {http://arxiv.org/abs/1110.0319} {arXiv:1110.0319 [hep-lat]} \BibitemShut
  {NoStop}%
\bibitem [{\citenamefont {G{\"o}ckeler}\ \emph {et~al.}(2012)\citenamefont
  {G{\"o}ckeler}, \citenamefont {Horsley}, \citenamefont {Lage}, \citenamefont
  {Mei{\ss}ner}, \citenamefont {Rakow}, \citenamefont {Rusetsky}, \citenamefont
  {Schierholz},\ and\ \citenamefont {Zanotti}}]{Gockeler:2012yj}%
  \BibitemOpen
  \bibfield  {author} {\bibinfo {author} {\bibfnamefont {M.}~\bibnamefont
  {G{\"o}ckeler}}, \bibinfo {author} {\bibfnamefont {R.}~\bibnamefont
  {Horsley}}, \bibinfo {author} {\bibfnamefont {M.}~\bibnamefont {Lage}},
  \bibinfo {author} {\bibfnamefont {U.~G.}\ \bibnamefont {Mei{\ss}ner}},
  \bibinfo {author} {\bibfnamefont {P.~E.~L.}\ \bibnamefont {Rakow}}, \bibinfo
  {author} {\bibfnamefont {A.}~\bibnamefont {Rusetsky}}, \bibinfo {author}
  {\bibfnamefont {G.}~\bibnamefont {Schierholz}}, \ and\ \bibinfo {author}
  {\bibfnamefont {J.~M.}\ \bibnamefont {Zanotti}},\ }\href {\doibase
  10.1103/PhysRevD.86.094513} {\bibfield  {journal} {\bibinfo  {journal} {Phys.
  Rev.}\ }\textbf {\bibinfo {volume} {D86}},\ \bibinfo {pages} {094513}
  (\bibinfo {year} {2012})},\ \Eprint {http://arxiv.org/abs/1206.4141}
  {arXiv:1206.4141 [hep-lat]} \BibitemShut {NoStop}%
\bibitem [{\citenamefont {Guo}\ \emph {et~al.}(2013)\citenamefont {Guo},
  \citenamefont {Dudek}, \citenamefont {Edwards},\ and\ \citenamefont
  {Szczepaniak}}]{Guo:2012hv}%
  \BibitemOpen
  \bibfield  {author} {\bibinfo {author} {\bibfnamefont {P.}~\bibnamefont
  {Guo}}, \bibinfo {author} {\bibfnamefont {J.}~\bibnamefont {Dudek}}, \bibinfo
  {author} {\bibfnamefont {R.}~\bibnamefont {Edwards}}, \ and\ \bibinfo
  {author} {\bibfnamefont {A.~P.}\ \bibnamefont {Szczepaniak}},\ }\href
  {\doibase 10.1103/PhysRevD.88.014501} {\bibfield  {journal} {\bibinfo
  {journal} {Phys. Rev. D}\ }\textbf {\bibinfo {volume} {88}},\ \bibinfo
  {pages} {014501} (\bibinfo {year} {2013})},\ \Eprint
  {http://arxiv.org/abs/1211.0929} {arXiv:1211.0929 [hep-lat]} \BibitemShut
  {NoStop}%
\bibitem [{\citenamefont {Luu}\ and\ \citenamefont
  {Savage}(2011)}]{Luu:2011ep}%
  \BibitemOpen
  \bibfield  {author} {\bibinfo {author} {\bibfnamefont {T.}~\bibnamefont
  {Luu}}\ and\ \bibinfo {author} {\bibfnamefont {M.~J.}\ \bibnamefont
  {Savage}},\ }\href {\doibase 10.1103/PhysRevD.83.114508} {\bibfield
  {journal} {\bibinfo  {journal} {Phys. Rev.}\ }\textbf {\bibinfo {volume}
  {D83}},\ \bibinfo {pages} {114508} (\bibinfo {year} {2011})},\ \Eprint
  {http://arxiv.org/abs/1101.3347} {arXiv:1101.3347 [hep-lat]} \BibitemShut
  {NoStop}%
\bibitem [{\citenamefont {Leskovec}\ and\ \citenamefont
  {Prelovsek}(2012)}]{Leskovec:2012gb}%
  \BibitemOpen
  \bibfield  {author} {\bibinfo {author} {\bibfnamefont {L.}~\bibnamefont
  {Leskovec}}\ and\ \bibinfo {author} {\bibfnamefont {S.}~\bibnamefont
  {Prelovsek}},\ }\href {\doibase 10.1103/PhysRevD.85.114507} {\bibfield
  {journal} {\bibinfo  {journal} {Phys. Rev. D}\ }\textbf {\bibinfo {volume}
  {85}},\ \bibinfo {pages} {114507} (\bibinfo {year} {2012})},\ \Eprint
  {http://arxiv.org/abs/1202.2145} {arXiv:1202.2145 [hep-lat]} \BibitemShut
  {NoStop}%
\bibitem [{\citenamefont {{Brice\~no}}\ and\ \citenamefont
  {Davoudi}(2013)}]{Briceno:2012yi}%
  \BibitemOpen
  \bibfield  {author} {\bibinfo {author} {\bibfnamefont {R.~A.}\ \bibnamefont
  {{Brice\~no}}}\ and\ \bibinfo {author} {\bibfnamefont {Z.}~\bibnamefont
  {Davoudi}},\ }\href {\doibase 10.1103/PhysRevD.88.094507} {\bibfield
  {journal} {\bibinfo  {journal} {Phys. Rev. D}\ }\textbf {\bibinfo {volume}
  {88}},\ \bibinfo {pages} {094507} (\bibinfo {year} {2013})},\ \Eprint
  {http://arxiv.org/abs/1204.1110} {arXiv:1204.1110 [hep-lat]} \BibitemShut
  {NoStop}%
\bibitem [{\citenamefont {{Brice\~no}}(2014)}]{Briceno:2014oea}%
  \BibitemOpen
  \bibfield  {author} {\bibinfo {author} {\bibfnamefont {R.~A.}\ \bibnamefont
  {{Brice\~no}}},\ }\href {\doibase 10.1103/PhysRevD.89.074507} {\bibfield
  {journal} {\bibinfo  {journal} {Phys. Rev. D}\ }\textbf {\bibinfo {volume}
  {89}},\ \bibinfo {pages} {074507} (\bibinfo {year} {2014})},\ \Eprint
  {http://arxiv.org/abs/1401.3312} {arXiv:1401.3312 [hep-lat]} \BibitemShut
  {NoStop}%
\bibitem [{\citenamefont {Berkowitz}\ \emph {et~al.}(2017)\citenamefont
  {Berkowitz}, \citenamefont {Kurth}, \citenamefont {Nicholson}, \citenamefont
  {Joo}, \citenamefont {Rinaldi}, \citenamefont {Strother}, \citenamefont
  {Vranas},\ and\ \citenamefont {Walker-Loud}}]{Berkowitz:2015eaa}%
  \BibitemOpen
  \bibfield  {author} {\bibinfo {author} {\bibfnamefont {E.}~\bibnamefont
  {Berkowitz}}, \bibinfo {author} {\bibfnamefont {T.}~\bibnamefont {Kurth}},
  \bibinfo {author} {\bibfnamefont {A.}~\bibnamefont {Nicholson}}, \bibinfo
  {author} {\bibfnamefont {B.}~\bibnamefont {Joo}}, \bibinfo {author}
  {\bibfnamefont {E.}~\bibnamefont {Rinaldi}}, \bibinfo {author} {\bibfnamefont
  {M.}~\bibnamefont {Strother}}, \bibinfo {author} {\bibfnamefont {P.~M.}\
  \bibnamefont {Vranas}}, \ and\ \bibinfo {author} {\bibfnamefont
  {A.}~\bibnamefont {Walker-Loud}},\ }\href {\doibase
  10.1016/j.physletb.2016.12.024} {\bibfield  {journal} {\bibinfo  {journal}
  {Phys. Lett.}\ }\textbf {\bibinfo {volume} {B765}},\ \bibinfo {pages} {285}
  (\bibinfo {year} {2017})},\ \Eprint {http://arxiv.org/abs/1508.00886}
  {arXiv:1508.00886 [hep-lat]} \BibitemShut {NoStop}%
\bibitem [{\citenamefont {Wu}\ \emph {et~al.}(2016)\citenamefont {Wu},
  \citenamefont {Lee}, \citenamefont {Leinweber}, \citenamefont {Thomas},\ and\
  \citenamefont {Young}}]{Wu:2015evh}%
  \BibitemOpen
  \bibfield  {author} {\bibinfo {author} {\bibfnamefont {J.-J.}\ \bibnamefont
  {Wu}}, \bibinfo {author} {\bibfnamefont {T.-S.~H.}\ \bibnamefont {Lee}},
  \bibinfo {author} {\bibfnamefont {D.~B.}\ \bibnamefont {Leinweber}}, \bibinfo
  {author} {\bibfnamefont {A.~W.}\ \bibnamefont {Thomas}}, \ and\ \bibinfo
  {author} {\bibfnamefont {R.~D.}\ \bibnamefont {Young}},\ }\bibfield
  {booktitle} {\emph {\bibinfo {booktitle} {{Proceedings, 10th International
  Workshop on the Physics of Excited Nucleons (NSTAR 2015): Osaka, Japan, May
  25-28, 2015}}},\ }\href {\doibase 10.7566/JPSCP.10.062002} {\bibfield
  {journal} {\bibinfo  {journal} {JPS Conf. Proc.}\ }\textbf {\bibinfo {volume}
  {10}},\ \bibinfo {pages} {062002} (\bibinfo {year} {2016})},\ \Eprint
  {http://arxiv.org/abs/1512.02771} {arXiv:1512.02771 [hep-lat]} \BibitemShut
  {NoStop}%
\bibitem [{\citenamefont {Li}\ \emph {et~al.}(2020)\citenamefont {Li},
  \citenamefont {Wu}, \citenamefont {Abell}, \citenamefont {Leinweber},\ and\
  \citenamefont {Thomas}}]{Li:2019qvh}%
  \BibitemOpen
  \bibfield  {author} {\bibinfo {author} {\bibfnamefont {Y.}~\bibnamefont
  {Li}}, \bibinfo {author} {\bibfnamefont {J.-J.}\ \bibnamefont {Wu}}, \bibinfo
  {author} {\bibfnamefont {C.~D.}\ \bibnamefont {Abell}}, \bibinfo {author}
  {\bibfnamefont {D.~B.}\ \bibnamefont {Leinweber}}, \ and\ \bibinfo {author}
  {\bibfnamefont {A.~W.}\ \bibnamefont {Thomas}},\ }\href {\doibase
  10.1103/PhysRevD.101.114501} {\bibfield  {journal} {\bibinfo  {journal}
  {Phys. Rev. D}\ }\textbf {\bibinfo {volume} {101}},\ \bibinfo {pages}
  {114501} (\bibinfo {year} {2020})},\ \Eprint
  {http://arxiv.org/abs/1910.04973} {arXiv:1910.04973 [hep-lat]} \BibitemShut
  {NoStop}%
\bibitem [{\citenamefont {Li}\ \emph {et~al.}(2021)\citenamefont {Li},
  \citenamefont {Wu}, \citenamefont {Leinweber},\ and\ \citenamefont
  {Thomas}}]{Li:2021mob}%
  \BibitemOpen
  \bibfield  {author} {\bibinfo {author} {\bibfnamefont {Y.}~\bibnamefont
  {Li}}, \bibinfo {author} {\bibfnamefont {J.-J.}\ \bibnamefont {Wu}}, \bibinfo
  {author} {\bibfnamefont {D.~B.}\ \bibnamefont {Leinweber}}, \ and\ \bibinfo
  {author} {\bibfnamefont {A.~W.}\ \bibnamefont {Thomas}},\ }\href {\doibase
  10.1103/PhysRevD.103.094518} {\bibfield  {journal} {\bibinfo  {journal}
  {Phys. Rev. D}\ }\textbf {\bibinfo {volume} {103}},\ \bibinfo {pages}
  {094518} (\bibinfo {year} {2021})},\ \Eprint
  {http://arxiv.org/abs/2103.12260} {arXiv:2103.12260 [hep-lat]} \BibitemShut
  {NoStop}%
\bibitem [{\citenamefont {Amarasinghe}\ \emph {et~al.}(2021)\citenamefont
  {Amarasinghe}, \citenamefont {Baghdadi}, \citenamefont {Davoudi},
  \citenamefont {Detmold}, \citenamefont {Illa}, \citenamefont {{Parre\~no}},
  \citenamefont {Pochinsky}, \citenamefont {Shanahan},\ and\ \citenamefont
  {Wagman}}]{Amarasinghe:2021lqa}%
  \BibitemOpen
  \bibfield  {author} {\bibinfo {author} {\bibfnamefont {S.}~\bibnamefont
  {Amarasinghe}}, \bibinfo {author} {\bibfnamefont {R.}~\bibnamefont
  {Baghdadi}}, \bibinfo {author} {\bibfnamefont {Z.}~\bibnamefont {Davoudi}},
  \bibinfo {author} {\bibfnamefont {W.}~\bibnamefont {Detmold}}, \bibinfo
  {author} {\bibfnamefont {M.}~\bibnamefont {Illa}}, \bibinfo {author}
  {\bibfnamefont {A.}~\bibnamefont {{Parre\~no}}}, \bibinfo {author}
  {\bibfnamefont {A.~V.}\ \bibnamefont {Pochinsky}}, \bibinfo {author}
  {\bibfnamefont {P.~E.}\ \bibnamefont {Shanahan}}, \ and\ \bibinfo {author}
  {\bibfnamefont {M.~L.}\ \bibnamefont {Wagman}},\ }\href@noop {} {\  (\bibinfo
  {year} {2021})},\ \Eprint {http://arxiv.org/abs/2108.10835} {arXiv:2108.10835
  [hep-lat]} \BibitemShut {NoStop}%
\bibitem [{\citenamefont {Weissbluth}(2012)}]{weissbluth2012atoms}%
  \BibitemOpen
  \bibfield  {author} {\bibinfo {author} {\bibfnamefont {M.}~\bibnamefont
  {Weissbluth}},\ }\href {https://books.google.com.au/books?id=n1mbd1SafcQC}
  {\emph {\bibinfo {title} {Atoms and Molecules}}}\ (\bibinfo  {publisher}
  {Elsevier Science},\ \bibinfo {year} {2012})\BibitemShut {NoStop}%
\bibitem [{\citenamefont {D{\"o}ring}\ \emph {et~al.}(2012)\citenamefont
  {D{\"o}ring}, \citenamefont {Mei{\ss}ner}, \citenamefont {Oset},\ and\
  \citenamefont {Rusetsky}}]{Doring:2012eu}%
  \BibitemOpen
  \bibfield  {author} {\bibinfo {author} {\bibfnamefont {M.}~\bibnamefont
  {D{\"o}ring}}, \bibinfo {author} {\bibfnamefont {U.}~\bibnamefont
  {Mei{\ss}ner}}, \bibinfo {author} {\bibfnamefont {E.}~\bibnamefont {Oset}}, \
  and\ \bibinfo {author} {\bibfnamefont {A.}~\bibnamefont {Rusetsky}},\ }\href
  {\doibase 10.1140/epja/i2012-12114-6} {\bibfield  {journal} {\bibinfo
  {journal} {Eur. Phys. J. A}\ }\textbf {\bibinfo {volume} {48}},\ \bibinfo
  {pages} {114} (\bibinfo {year} {2012})},\ \Eprint
  {http://arxiv.org/abs/1205.4838} {arXiv:1205.4838 [hep-lat]} \BibitemShut
  {NoStop}%
\bibitem [{\citenamefont {Bali}\ \emph {et~al.}(2013)\citenamefont {Bali} \emph
  {et~al.}}]{Bali:2012qs}%
  \BibitemOpen
  \bibfield  {author} {\bibinfo {author} {\bibfnamefont {G.~S.}\ \bibnamefont
  {Bali}} \emph {et~al.},\ }\href {\doibase 10.1016/j.nuclphysb.2012.08.009}
  {\bibfield  {journal} {\bibinfo  {journal} {Nucl. Phys.}\ }\textbf {\bibinfo
  {volume} {B866}},\ \bibinfo {pages} {1} (\bibinfo {year} {2013})},\ \Eprint
  {http://arxiv.org/abs/1206.7034} {arXiv:1206.7034 [hep-lat]} \BibitemShut
  {NoStop}%
\bibitem [{\citenamefont {Beane}\ \emph {et~al.}(2008)\citenamefont {Beane},
  \citenamefont {Luu}, \citenamefont {Orginos}, \citenamefont {Parre\~no},
  \citenamefont {Savage}, \citenamefont {Torok},\ and\ \citenamefont
  {Walker-Loud}}]{Beane:2007xs}%
  \BibitemOpen
  \bibfield  {author} {\bibinfo {author} {\bibfnamefont {S.~R.}\ \bibnamefont
  {Beane}}, \bibinfo {author} {\bibfnamefont {T.~C.}\ \bibnamefont {Luu}},
  \bibinfo {author} {\bibfnamefont {K.}~\bibnamefont {Orginos}}, \bibinfo
  {author} {\bibfnamefont {A.}~\bibnamefont {Parre\~no}}, \bibinfo {author}
  {\bibfnamefont {M.~J.}\ \bibnamefont {Savage}}, \bibinfo {author}
  {\bibfnamefont {A.}~\bibnamefont {Torok}}, \ and\ \bibinfo {author}
  {\bibfnamefont {A.}~\bibnamefont {Walker-Loud}},\ }\href {\doibase
  10.1103/PhysRevD.77.014505} {\bibfield  {journal} {\bibinfo  {journal} {Phys.
  Rev. D}\ }\textbf {\bibinfo {volume} {77}},\ \bibinfo {pages} {014505}
  (\bibinfo {year} {2008})},\ \Eprint {http://arxiv.org/abs/0706.3026}
  {arXiv:0706.3026 [hep-lat]} \BibitemShut {NoStop}%
\bibitem [{\citenamefont {Beane}\ \emph {et~al.}(2009)\citenamefont {Beane},
  \citenamefont {Detmold}, \citenamefont {Luu}, \citenamefont {Orginos},
  \citenamefont {Parre\~no}, \citenamefont {Savage}, \citenamefont {Torok},\
  and\ \citenamefont {Walker-Loud}}]{Beane:2009kya}%
  \BibitemOpen
  \bibfield  {author} {\bibinfo {author} {\bibfnamefont {S.~R.}\ \bibnamefont
  {Beane}}, \bibinfo {author} {\bibfnamefont {W.}~\bibnamefont {Detmold}},
  \bibinfo {author} {\bibfnamefont {T.~C.}\ \bibnamefont {Luu}}, \bibinfo
  {author} {\bibfnamefont {K.}~\bibnamefont {Orginos}}, \bibinfo {author}
  {\bibfnamefont {A.}~\bibnamefont {Parre\~no}}, \bibinfo {author}
  {\bibfnamefont {M.~J.}\ \bibnamefont {Savage}}, \bibinfo {author}
  {\bibfnamefont {A.}~\bibnamefont {Torok}}, \ and\ \bibinfo {author}
  {\bibfnamefont {A.}~\bibnamefont {Walker-Loud}},\ }\href {\doibase
  10.1103/PhysRevD.79.114502} {\bibfield  {journal} {\bibinfo  {journal} {Phys.
  Rev. D}\ }\textbf {\bibinfo {volume} {79}},\ \bibinfo {pages} {114502}
  (\bibinfo {year} {2009})},\ \Eprint {http://arxiv.org/abs/0903.2990}
  {arXiv:0903.2990 [hep-lat]} \BibitemShut {NoStop}%
\bibitem [{\citenamefont {Feng}\ \emph {et~al.}(2010)\citenamefont {Feng},
  \citenamefont {Jansen},\ and\ \citenamefont {Renner}}]{Feng:2009ij}%
  \BibitemOpen
  \bibfield  {author} {\bibinfo {author} {\bibfnamefont {X.}~\bibnamefont
  {Feng}}, \bibinfo {author} {\bibfnamefont {K.}~\bibnamefont {Jansen}}, \ and\
  \bibinfo {author} {\bibfnamefont {D.~B.}\ \bibnamefont {Renner}},\ }\href
  {\doibase 10.1016/j.physletb.2010.01.018} {\bibfield  {journal} {\bibinfo
  {journal} {Phys. Lett. B}\ }\textbf {\bibinfo {volume} {684}},\ \bibinfo
  {pages} {268} (\bibinfo {year} {2010})},\ \Eprint
  {http://arxiv.org/abs/0909.3255} {arXiv:0909.3255 [hep-lat]} \BibitemShut
  {NoStop}%
\bibitem [{\citenamefont {Detmold}\ and\ \citenamefont
  {Smigielski}(2010)}]{Detmold:2010unv}%
  \BibitemOpen
  \bibfield  {author} {\bibinfo {author} {\bibfnamefont {W.}~\bibnamefont
  {Detmold}}\ and\ \bibinfo {author} {\bibfnamefont {B.}~\bibnamefont
  {Smigielski}},\ }\href {\doibase 10.22323/1.105.0100} {\bibfield  {journal}
  {\bibinfo  {journal} {PoS}\ }\textbf {\bibinfo {volume} {LATTICE2010}},\
  \bibinfo {pages} {100} (\bibinfo {year} {2010})},\ \Eprint
  {http://arxiv.org/abs/1101.2639} {arXiv:1101.2639 [hep-lat]} \BibitemShut
  {NoStop}%
\bibitem [{\citenamefont {Detmold}\ and\ \citenamefont
  {Smigielski}(2011)}]{Detmold:2011kw}%
  \BibitemOpen
  \bibfield  {author} {\bibinfo {author} {\bibfnamefont {W.}~\bibnamefont
  {Detmold}}\ and\ \bibinfo {author} {\bibfnamefont {B.}~\bibnamefont
  {Smigielski}},\ }\href {\doibase 10.1103/PhysRevD.84.014508} {\bibfield
  {journal} {\bibinfo  {journal} {Phys. Rev. D}\ }\textbf {\bibinfo {volume}
  {84}},\ \bibinfo {pages} {014508} (\bibinfo {year} {2011})},\ \Eprint
  {http://arxiv.org/abs/1103.4362} {arXiv:1103.4362 [hep-lat]} \BibitemShut
  {NoStop}%
\bibitem [{\citenamefont {Detmold}\ \emph {et~al.}(2012)\citenamefont
  {Detmold}, \citenamefont {Orginos},\ and\ \citenamefont
  {Shi}}]{Detmold:2012wc}%
  \BibitemOpen
  \bibfield  {author} {\bibinfo {author} {\bibfnamefont {W.}~\bibnamefont
  {Detmold}}, \bibinfo {author} {\bibfnamefont {K.}~\bibnamefont {Orginos}}, \
  and\ \bibinfo {author} {\bibfnamefont {Z.}~\bibnamefont {Shi}},\ }\href
  {\doibase 10.1103/PhysRevD.86.054507} {\bibfield  {journal} {\bibinfo
  {journal} {Phys. Rev. D}\ }\textbf {\bibinfo {volume} {86}},\ \bibinfo
  {pages} {054507} (\bibinfo {year} {2012})},\ \Eprint
  {http://arxiv.org/abs/1205.4224} {arXiv:1205.4224 [hep-lat]} \BibitemShut
  {NoStop}%
\bibitem [{\citenamefont {Culver}\ \emph {et~al.}(2019)\citenamefont {Culver},
  \citenamefont {Mai}, \citenamefont {Alexandru}, \citenamefont {D\"oring},\
  and\ \citenamefont {Lee}}]{Culver:2019qtx}%
  \BibitemOpen
  \bibfield  {author} {\bibinfo {author} {\bibfnamefont {C.}~\bibnamefont
  {Culver}}, \bibinfo {author} {\bibfnamefont {M.}~\bibnamefont {Mai}},
  \bibinfo {author} {\bibfnamefont {A.}~\bibnamefont {Alexandru}}, \bibinfo
  {author} {\bibfnamefont {M.}~\bibnamefont {D\"oring}}, \ and\ \bibinfo
  {author} {\bibfnamefont {F.~X.}\ \bibnamefont {Lee}},\ }\href {\doibase
  10.1103/PhysRevD.100.034509} {\bibfield  {journal} {\bibinfo  {journal}
  {Phys. Rev. D}\ }\textbf {\bibinfo {volume} {100}},\ \bibinfo {pages}
  {034509} (\bibinfo {year} {2019})},\ \Eprint
  {http://arxiv.org/abs/1905.10202} {arXiv:1905.10202 [hep-lat]} \BibitemShut
  {NoStop}%
\bibitem [{\citenamefont {Culver}\ \emph {et~al.}(2020)\citenamefont {Culver},
  \citenamefont {Mai}, \citenamefont {Brett}, \citenamefont {Alexandru},\ and\
  \citenamefont {D\"oring}}]{Culver:2019vvu}%
  \BibitemOpen
  \bibfield  {author} {\bibinfo {author} {\bibfnamefont {C.}~\bibnamefont
  {Culver}}, \bibinfo {author} {\bibfnamefont {M.}~\bibnamefont {Mai}},
  \bibinfo {author} {\bibfnamefont {R.}~\bibnamefont {Brett}}, \bibinfo
  {author} {\bibfnamefont {A.}~\bibnamefont {Alexandru}}, \ and\ \bibinfo
  {author} {\bibfnamefont {M.}~\bibnamefont {D\"oring}},\ }\href {\doibase
  10.1103/PhysRevD.101.114507} {\bibfield  {journal} {\bibinfo  {journal}
  {Phys. Rev. D}\ }\textbf {\bibinfo {volume} {101}},\ \bibinfo {pages}
  {114507} (\bibinfo {year} {2020})},\ \Eprint
  {http://arxiv.org/abs/1911.09047} {arXiv:1911.09047 [hep-lat]} \BibitemShut
  {NoStop}%
\bibitem [{\citenamefont {Mahbub}\ \emph {et~al.}(2013)\citenamefont {Mahbub},
  \citenamefont {Kamleh}, \citenamefont {Leinweber}, \citenamefont {Moran},\
  and\ \citenamefont {Williams}}]{Mahbub:2013ala}%
  \BibitemOpen
  \bibfield  {author} {\bibinfo {author} {\bibfnamefont {M.~S.}\ \bibnamefont
  {Mahbub}}, \bibinfo {author} {\bibfnamefont {W.}~\bibnamefont {Kamleh}},
  \bibinfo {author} {\bibfnamefont {D.~B.}\ \bibnamefont {Leinweber}}, \bibinfo
  {author} {\bibfnamefont {P.~J.}\ \bibnamefont {Moran}}, \ and\ \bibinfo
  {author} {\bibfnamefont {A.~G.}\ \bibnamefont {Williams}},\ }\href {\doibase
  10.1103/PhysRevD.87.094506} {\bibfield  {journal} {\bibinfo  {journal} {Phys.
  Rev. D}\ }\textbf {\bibinfo {volume} {87}},\ \bibinfo {pages} {094506}
  (\bibinfo {year} {2013})},\ \Eprint {http://arxiv.org/abs/1302.2987}
  {arXiv:1302.2987 [hep-lat]} \BibitemShut {NoStop}%
\bibitem [{\citenamefont {Stokes}\ \emph {et~al.}(2015)\citenamefont {Stokes},
  \citenamefont {Kamleh}, \citenamefont {Leinweber}, \citenamefont {Mahbub},
  \citenamefont {Menadue},\ and\ \citenamefont {Owen}}]{Stokes:2013fgw}%
  \BibitemOpen
  \bibfield  {author} {\bibinfo {author} {\bibfnamefont {F.~M.}\ \bibnamefont
  {Stokes}}, \bibinfo {author} {\bibfnamefont {W.}~\bibnamefont {Kamleh}},
  \bibinfo {author} {\bibfnamefont {D.~B.}\ \bibnamefont {Leinweber}}, \bibinfo
  {author} {\bibfnamefont {M.~S.}\ \bibnamefont {Mahbub}}, \bibinfo {author}
  {\bibfnamefont {B.~J.}\ \bibnamefont {Menadue}}, \ and\ \bibinfo {author}
  {\bibfnamefont {B.~J.}\ \bibnamefont {Owen}},\ }\href {\doibase
  10.1103/PhysRevD.92.114506} {\bibfield  {journal} {\bibinfo  {journal} {Phys.
  Rev. D}\ }\textbf {\bibinfo {volume} {92}},\ \bibinfo {pages} {114506}
  (\bibinfo {year} {2015})},\ \Eprint {http://arxiv.org/abs/1302.4152}
  {arXiv:1302.4152 [hep-lat]} \BibitemShut {NoStop}%
\bibitem [{\citenamefont {Edwards}\ and\ \citenamefont
  {Joo}(2005)}]{Edwards:2004sx}%
  \BibitemOpen
  \bibfield  {author} {\bibinfo {author} {\bibfnamefont {R.~G.}\ \bibnamefont
  {Edwards}}\ and\ \bibinfo {author} {\bibfnamefont {B.}~\bibnamefont {Joo}}
  (\bibinfo {collaboration} {SciDAC, LHPC, UKQCD}),\ }\href {\doibase
  10.1016/j.nuclphysbps.2004.11.254} {\bibfield  {journal} {\bibinfo  {journal}
  {Nucl. Phys. B Proc. Suppl.}\ }\textbf {\bibinfo {volume} {140}},\ \bibinfo
  {pages} {832} (\bibinfo {year} {2005})},\ \Eprint
  {http://arxiv.org/abs/hep-lat/0409003} {arXiv:hep-lat/0409003} \BibitemShut
  {NoStop}%
\end{thebibliography}%


\end{document}